\documentclass[twocolumn]{aastex631}

\newcommand{\teff}{$T_{\rm eff}$}
\newcommand{\tint}{$T_{\rm int}$} 
\newcommand{\teq}{$T_{\rm eq}$}
\newcommand{\co}{CO}
\newcommand{\meth}{CH$_4$}
\newcommand{\sotwo}{SO$_2$}
\newcommand{\amon}{NH$_3$}
\newcommand{\cotwo}{CO$_2$} 
\newcommand{\ntwo}{N$_2$} 
\newcommand{\water}{H$_2$O}

\newcommand{\cms}{cm$^2$s$^{-1}$}

\newcommand{\tp}{$T(P)$}

\newcommand{\kzz}{$K_{zz}$}

\newcommand{\RNum}[1]{\uppercase\expandafter{\romannumeral #1\relax}}

\newcommand{\orcid}[1]{\href{https://orcid.org/#1}{\includegraphics[width=10pt]{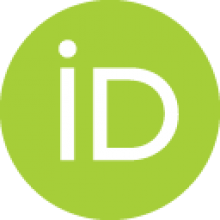}}}

\usepackage{amsmath}
\begin{document}

\title{Effects of Planetary Parameters on Disequilibrium Chemistry in Irradiated Planetary Atmospheres: From Gas Giants to Sub-Neptunes}

\email{samukher@ucsc.edu}

\author{Sagnick Mukherjee \orcid{0000-0003-1622-1302}}
\affiliation{Department of Astronomy and Astrophysics, University of California, Santa Cruz, CA 95064, USA \\ }
\affiliation{Department of Physics and Astronomy, Johns Hopkins University, Baltimore, MD, USA \\ }
\author{Jonathan J. Fortney\orcid{0000-0002-9843-4354}}
\affiliation{Department of Astronomy and Astrophysics, University of California, Santa Cruz, CA 95064, USA \\ }

\author{Nicholas F. Wogan}
\affiliation{Space Science Division, NASA Ames Research Center, Moffett Field, CA 94035, USA \\ }

\author{David K. Sing}
\affiliation{Department of Earth and Planetary Sciences, Johns Hopkins University, Baltimore, MD, USA \\}
\affiliation{Department of Physics and Astronomy, Johns Hopkins University, Baltimore, MD, USA \\ }

\author{Kazumasa Ohno \orcid{0000-0003-3290-6758}}
\affiliation{Division of Science, National Astronomical Observatory of Japan, 2-21-1 Osawa, Mitaka-shi, Tokyo, Japan\\ }



\begin{abstract}

A primary goal of characterizing exoplanet atmospheres is to constrain planetary bulk properties, such as their metallicity, C/O ratio, and intrinsic heat. However, there are significant uncertainties in many aspects of atmospheric physics, such as the strength of vertical mixing. Here we use \texttt{PICASO} and the \texttt{photochem} model to explore how atmospheric chemistry is influenced by planetary properties like metallicity, C/O ratio, {\tint}, {\teq}, and {\kzz} in hydrogen-dominated atmospheres. We vary the {\teq} of the planets between 400 K- 1600 K, across ``cold",``warm," and `hot" objects. We also explore an extensive range of {\tint} values between 30-500 K, representing sub-Neptunes to massive gas giants. We find that gases like {\co} and {\cotwo} show a drastically different dependence on {\kzz} and C/O for planets with cold interiors (e.g., sub-Neptunes) compared to planets with hotter interiors (e.g., Jupiter mass planets), for the same {\teq}. We also find that gases like CS and CS$_2$ can carry a significant portion of the S- inventory in the upper atmosphere near {\teq} $\le$ 600 K, below which {\sotwo} ceases to be abundant. For solar C/O, we show that the {\co}/{\meth} ratio in the upper atmospheres of planets can become $\le$1 for planets with low {\teq}, but only if their interiors are cold ({\tint}$\le$100 K). We find that photochemical haze precursor molecules in the upper atmosphere show very complex dependence on C/O, {\kzz}, {\teq}, and {\tint} for planets with cold interiors (e.g., sub-Neptunes).  We also briefly explore fully coupling \texttt{PICASO} and \texttt{photochem} to generate self-consistent radiative-convective-photochemical-equilibrium models. 

\end{abstract}



\section{Introduction} \label{sec:intro}
Major goals of characterizing exoplanet atmospheres are to understand the formation and evolutionary history of planets. This process involves observing exoplanetary atmospheres and intrepreting these observations to estimate some key bulk planetary parameters, like metal enrichment of the atmosphere, which can be used to understand their formation scenario and evolutionary history. However, our interpretation of observational data of exoplanetary atmospheres requires a robust understanding of the various physical and chemical processes ongoing in the planet's atmosphere. In this work, we explore how a series of atmospheric processes can affect our interpretation of exoplanet atmospheric observations. We explore how these atmospheric processes specifically impact atmospheric chemistry, which is a crucial link between understanding exoplanetary atmospheres and constraining their formation and evolutionary processes.

There has been a considerable amount of modeling work revolving around the idea that planets retain some amount of information about their formation scenario and subsequent evolution in their current state \citep[e.g.,][]{oberg11,molliere22,mordasini16}. This information can be retained within important planetary parameters like the planetary radius, bulk metal enrichment, elemental abundance ratios, internal heat flux of the planet, etc. \citep[e.g.,][]{madhu12,fortney2007planetary,oberg11,molliere22}. For transiting planets, parameters like planetary mass and radius are directly measurable. But other critical parameters like bulk metal-enrichment, bulk elemental abundance ratios, and internal heat flux of planets need to be inferred by observing and understanding their atmospheres. As planetary atmospheres are quite complex, drawing inferences about their bulk properties requires a detailed understanding of the multiple interconnected physical processes at play.

{\it JWST} observations have opened a new era of precision chemical analysis of transiting planet atmospheres \citep[e.g.,][]{prism22,nircam22,g395h_22,niriss22,tsai23_S,kirk24,thao24,welbanks24,sing24,beatty2024,bell23,grant23,fu24,radica23,august,schlawin2024,xue24,inglis24,bean23,gagnebin24}. This is especially interesting because the atmospheric chemistry of  planets with H/He atmospheres (from giant planets down to sub-Neptunes) is sensitive to three key planetary parameters -- 1) bulk metal-enrichment (or metallicity), 2) bulk elemental abundance ratios, and 3) internal heat flux of planets \citep[e.g.,][]{fortney20,molliere15}. 

The metallicity of a planet's atmosphere is defined as  (X/H)$_{\rm planet}$/(X/H)$_{\rm sun}$, where X represents the number of atoms of a certain element (like C-, O-, S-, etc) in the object and H is the number of hydrogen atoms. Therefore a planet with higher metallicity will have significantly different atmospheric chemical composition than one with lower metallicity. Two planets with the same metallicity might have the ``metals" distributed differently among various elements. For example, one planet can have a higher C/H ratio and lower O/H ratio than the other and yet maintain the same bulk metallicity \citep[e.g.,][]{madhu12,moses2013}. These abundance ratios are important potential markers of the formation location of planets because of the presence of various ice lines in propoplanetary disks \citep[e.g.,][]{oberg11,molliere22,mordasini16}. It is clear and well-studied how both of these bulk parameters can affect atmospheric chemistry especially if the atmosphere is in thermochemical equilibrium throughout \citep[e.g.,][]{visscher05,visscher06,goyal18,molliere15,lodders04,lodders02}. However, studies of solar system giant planets and brown dwarf atmospheres have shown that disequilibrium chemical processes in addition to thermochemistry can play a significant role across planetary atmospheres \citep[e.g.,][]{zhang20,hubeny07,karilidi21,Mukherjee22,Miles20,Philips20,tsai17,moses11,mukherjee24,venot12,drummond16}.

The chemical structure of a planet's atmosphere is shaped by a number of other physical processes too, namely -- 1) atmospheric mixing, 2) photochemistry, 3) condensation, and 4) molecular diffusion. As a result, the photospheric abundance of a gas like {\meth} is not only controlled by the metallicity, C/O ratio, and heat flux from the deep interior of the planet  but also by parameters like the stellar UV flux incident on the planet along with atmospheric properties like strength of vertical and horizontal dynamics in the atmosphere \citep[e.g.,][]{tsai21,tsai23,moses11,fortney20,tsai23daynight}. Molecular diffusion is an important process controlling the chemical nature of atmospheres at very low pressures but the importance of this process is often overshadowed by photochemical processes, especially in highly irradiated giant exoplanets \citep[e.g.,][]{tsai21}. Atmospheric mixing, photochemistry, condensation, and internal heat flux of exoplanets remain the least understood/constrained processes/parameters in exoplanet atmospheres. We briefly explain these processes here.

{\it Atmospheric Mixing} -- Atmospheres of planets are dynamic, not static, in nature. This dynamics can play out in three dimensions causing both vertical and horizontal bulk transport of gases and aerosols in atmospheres \citep[e.g.,][]{showman09,menou09,steinruck19,roth24}. As a result, photospheric abundances of gases such as {\meth}, {\co}, {\amon} can be significantly altered depending on the strength of dynamics in the atmosphere \citep{cooper06,drummond16,steinruck19,mendonca18,drummond18,2020drummond,zamyatina23,lee23}. 3D models of tidally locked close-in giant exoplanets have revealed three major types of transport operational in their atmospheres -- 1) vertical transport, 2) day-to-night transport, and 3) pole-to-equator transport. Vertical transport can lift gases and aerosols from the deeper parts of the atmosphere across several pressure scale-heights, while the day-to-night winds can transport gases from the hotter day-side of the planet to their colder night-side. Clearly, a full understanding of the role of atmospheric dynamics on planet-wide chemistry requires the use of 3D models to explore vast swathes of the multidimensional parameter space involved. This vast parameter space includes dimensions like the stellar flux incident on the planet, the planet's gravity, atmospheric composition, rotational period, interior properties such as interior heat flux, cloud composition, etc. The computational run times involved in 3D atmospheric models often prohibit a full scale parameter space exploration. 

However, 1D models can somewhat capture the vertical transport operating in these atmospheres by simplifying the transport as a diffusive process, parametrized with the 1D eddy diffusion coefficient-- {\kzz}. {\kzz} is defined as $v_{\rm mix}L_{\rm mix}$, where $v_{\rm mix}$ is the typical velocity of transport and $L_{\rm mix}$ is the typical length-scale at which this bulk transport is operational \citep{allen81}. But, observational constraints on  {\kzz} are very weak. Recently, several studies have constrained {\kzz} to a certain extent in brown dwarf atmospheres with ground-based and space-based data \citep[e.g.,][]{Miles20,Mukherjee22a,madurowicz23,beiler23,mukherjee24,kothari24} but we have only started to constrain it for irradiated planets \citep[e.g.,][]{sing24,welbanks24,kawashima21}. Various 3D and 2D modeling efforts have found a large range of {\kzz} values in model atmospheres \citep[e.g.,][]{tan22,parmentier13,komacek19,tsai23,freytag10,menou19}, keeping the parameter uncertain by several orders of magnitude in irradiated planets. It is important to especially mitigate this uncertainty due to the influence of {\kzz} on key gas abundances that are used to infer bulk properties of planets such as C/O or metallicity.

{\it Internal Heat Flux} -- As planets evolve, they slowly lose their gravitational heat of formation to space. This energy is carried from the deep interior through convection to the deep atmosphere up to the point where the atmosphere is opaque at all wavelengths \citep[e.g.,][]{fortney2007planetary}. Beyond that, the heat is lost to space via radiation in the partially or fully optically thin radiative atmosphere. Therefore, detailed properties of the atmosphere, such as its composition and thermal structure, control the rate of cooling of the whole planet over its entire lifetime. This rate of cooling, in turn, also controls the contraction of the planet over time. Additionally, external energy sources like tidal heating can also deposit energy to the deep interior of the planet, the extent of which depends on various orbital and bulk planet properties. 

The internal heat flux cannot be measured directly through transmission and emission spectroscopy of close-in exoplanet atmospheres because it typically only alters the temperature structure of the planet where its atmosphere is optically thick, leading to little or no affect on its photospheric temperatures. However, internal heat flux can indirectly affect the photospheric chemical abundances due to the vertical mixing phenomenon discussed previously \citep[e.g.,][]{fortney20,agundez14}. If gases are mixed up to the photospheres of the planets from the deep optically thick atmosphere, where internal heat flux affects chemistry, then they can alter the observable photospheric chemistry of the planet and let us infer the internal heat flux of the planet as well. This works for  gases which are stable for long times and not chemically destroyed in the photosphere immediately after being transported. Gases such as {\co} and {\ntwo} fit into this category as destroying them chemically at low temperature photospheres and converting them to gases like {\meth} and {\amon} is generally energetically prohibitive. Therefore, photospheric abundances of such gases are strongly influenced by the internal heat flux of the planet. Recently, {\it JWST} transmission spectroscopy observations of WASP-107b were used to constrain its internal heat flux by  \citet{welbanks24} and \citet{sing24}. Both of these analysis revealed a high {\tint} ($\ge$345 K), which was interpreted as evidence for tidal heating of the planet's interior. Similar constraints on other transiting planets have also been obtained using {\it HST} \citep[e.g.,][]{barat24}. These studies are an excellent example of how understanding atmospheric chemistry can reveal details of close-in giant planet evolution. 

The internal heat flux of a planet at a given age depends on the mass of the planet. For example, at a similar age, a more massive planet is expected to have a hotter interior than a less massive planet, if there are no external perturbations on the planet such as tidal heating. However, interior temperatures also depend on aspects of atmospheric physics. For example, the strength of vertical mixing plays a major role in affecting both atmospheric clouds and chemistry. Both of these things can delay/accelerate the cooling of planets over time \citep[e.g.,][]{diamondback,saumonmarley08}. 

{\it Photochemistry} -- Planets can have significant X-ray and UV photons incident on them from their host stars. These photons can cause the photolysis of weakly bonded molecules such as {\meth}, {\amon}, H$_2$S, etc. in the upper atmosphere \citep[e.g.,][]{moses11,tsai21,tsai23,tsai23daynight,wogan23,hu21,venot12,moses16}. This can trigger a whole new set of chemical reactions altering photospheric composition. The first concrete evidence of a photochemical by-product was confirmed with {\it JWST} in WASP-39b by the presence of photochemically produced SO$_2$ in its upper atmosphere \citep{tsai23,nircam22,g395h_22,prism22}. While significant modeling work on understanding photochemistry in close-in  planets has been done previously \citep[e.g.,][]{moses11,tsai21,tsai23,tsai23daynight,crossfield23,zahnle09,venot12}, this \emph{JWST} work  demonstrated that photochemistry cannot be neglected while modeling the photospheric chemical nature of planets and using abundances to infer planetary properties.

Additionally, photolysis of gases like {\meth} and {\amon} can lead to the production of C- and N- bearing molecules like C$_2$H$_2$ and HCN. These gases in the upper atmospheres of H$_2$/He-rich atmospheres can act as precursors for further reactions to form photochemical hazes \citep[e.g.,][]{morley13,morley15,fortney13,Gao20aerosol}. The optical properties of these hazes are such that they can play a significant role in shaping the observable spectra of exoplanet atmospheres \citep[e.g.,][]{lavvas17,kawashima18,ohno20,steinruek23,arfaux22,morley15}.

The discussion above shows that the observational data from transiting planets today is at a stage where its chemistry can be leveraged to not only understand atmospheric elemental compositions of planets but also should be leveraged to understand exoplanetary atmospheric processes better as well. These include constraining key atmospheric processes such as atmospheric dynamics and parameters like internal heat flux of planets. Therefore, it is necessary to explore how atmospheric dynamics and internal heat flux can influence the photospheric abundance of gasses in exoplanets with varying bulk properties such as equilibrium temperature, C/O ratio, metallicity, and varying amounts of incident UV flux. It is also important to identify the key molecules that can produce the best constraints on very uncertain parameters like {\kzz} in different parts of this vast parameter space.

This paper presents a broad parameter space study to address these questions. Our work is applicable to a wide variety of  planets. In terms of internal heat flux, this includes sub-Neptune mass planets with lower internal heat fluxes, to  Jupiter mass giant planets with much hotter interiors. In terms of planetary temperatures, we explore planets between cold giant planets ({\teq}$\sim$ 400 K) to hot Jupiters ({\teq}$\sim$ 1600 K). We also explore the atmospheric chemistry of planets with varying C- and O- abundances. We also investigate the effects of a wide range of the atmospheric mixing parameter {\kzz} and atmospheric metallicity on the  chemistry of these planets.

We describe our modeling setup in\S\ref{sec:model} followed by the results presented in \S\ref{sec:results}. We discuss the limitations and caveats in this analysis in \S\ref{sec:disc} followed by summarizing our key conclusions in \S\ref{sec:conc}. 

\section{Modeling}\label{sec:model}

The aim of the modeling presented in this work is to explore a significant parameter space of properties that can influence atmospheric chemistry of transiting planets with H/He atmospheres. The parameters we consider in this study are the equilibrium temperature (\teq), interior heat flux through the parameter {\tint}, strength of vertical mixing parameterized with {\kzz}, C/O ratio, and metallicity. The modeling setup to achieve this parameter space exploration can be divided into two different components -- 1) 1D radiative--convective atmospheric modeling, and 2) chemical kinetics modeling. We describe each component in detail below.
\subsection{1D Radiative--Convective Equilibrium Models}
\begin{figure*}
  \centering
  \includegraphics[width=1\textwidth]{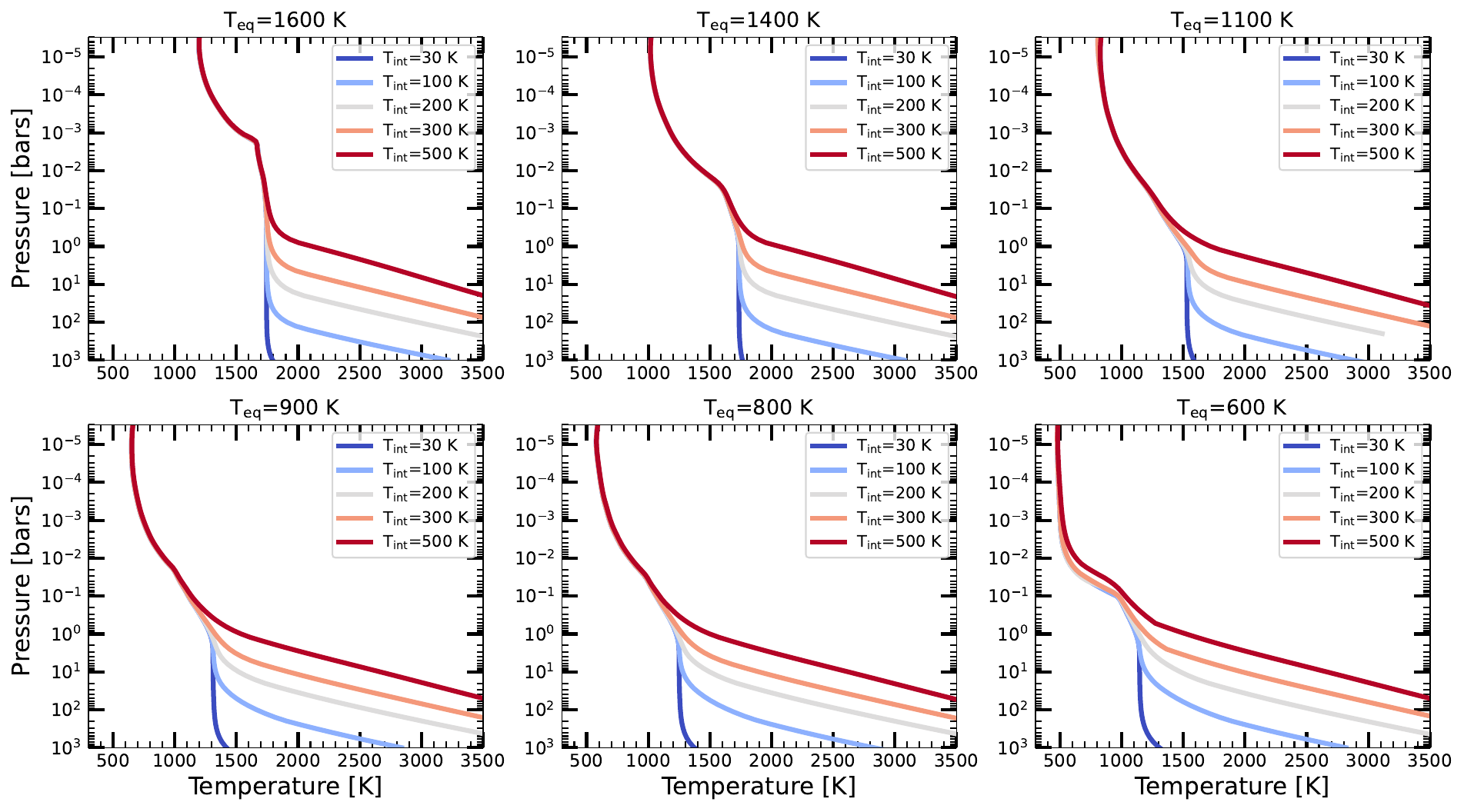}
  \caption{{\tp} profiles computed using 1D radiative--convective--equilibrium modeling at various {\teq} and {\tint} values are shown. Each panel shows the model {\tp} profiles computed for a {\teq} value at five different {\tint} values between 30 K and 500 K. The six panels correspond to {\teq}= 1600 K, 1400 K, 1100 K, 900 K, 800 K, and 600 K.  The kink at 1 mbar and 1700 K in the hottest models ({\teq}=1600 K) is due to a small amount of TiO opacity. All the {\tp} profiles have been computed for a planet with gravity=4.5 ms$^{-2}$ around a star with {\teff}= 5326.6 K, $log(g)$=4.38, and Radius=0.93$R_{\odot}$. }
\label{fig:TP_profiles}
\end{figure*}

We use the 1D radiative--convective equilibrium model \texttt{PICASO} \citep{Mukherjee22,batalha19} to generate atmospheric temperature--pressure (\tp) profiles for irradiated planets. We generate the {\tp} profiles for a planet with gravity= 4.5 ms$^{-2}$ around a star with {\teff}=5327 K, $log(g)$=4.38, [M/H]= -0.03, and Radius= 0.932 R$_{\odot}$. These system parameters are chosen to be the same as WASP-39b. We divide the planet atmosphere into 91 plane-parallel levels (90 layers) logarithmically spaced in pressure between 10$^{-6}$ to 100/5000 bars. We generate the {\tp} profiles using the heat recirculation factor \texttt{rfacv}= 0.5, which corresponds to the case of full heat redistribution across the planet. We place the planet at different distances from the star such that the {\teq} of the planet (assuming 0 albedo) are 1600 K, 1400 K, 1100 K, 900 K, 800 K, 600 K, and 400 K.  For each of these cases, we generate 5 different cases with different {\tint} values of 30 K, 100 K, 200 K, 300 K, and 500 K.  The lowest value, 30 K, is representative of Gyr+ old sub-Neptunes \citep{lopez14}.  Jupiter's value today is 100 K. More massive, younger, or inflated gas giants (such as by tidal heating or other factors) would have larger \tint\ values \citep{fortney20}. The equilibrium chemistry calculation and gaseous opacities used for computing these \tp\ profiles are the same as used in \citet{mukherjee24}. Figure \ref{fig:TP_profiles} shows these generated {\tp} profiles where each panel corresponds to a different {\teq} case. Note that these radiative--convective--thermochemical--equilibrium (RCTE) models were calculated by including TiO and VO opacities, which are known to cause thermal inversions in atmospheres which have {\teq}$\ge$ 1600 K. We generate these {\tp} profiles at 10$\times$solar metallicity and solar C/O ratio (C/O= 0.458 \citep{lodders09}) assuming thermochemical equilibrium throughout the atmosphere.

\begin{figure*}
  \centering
  \includegraphics[width=1\textwidth]{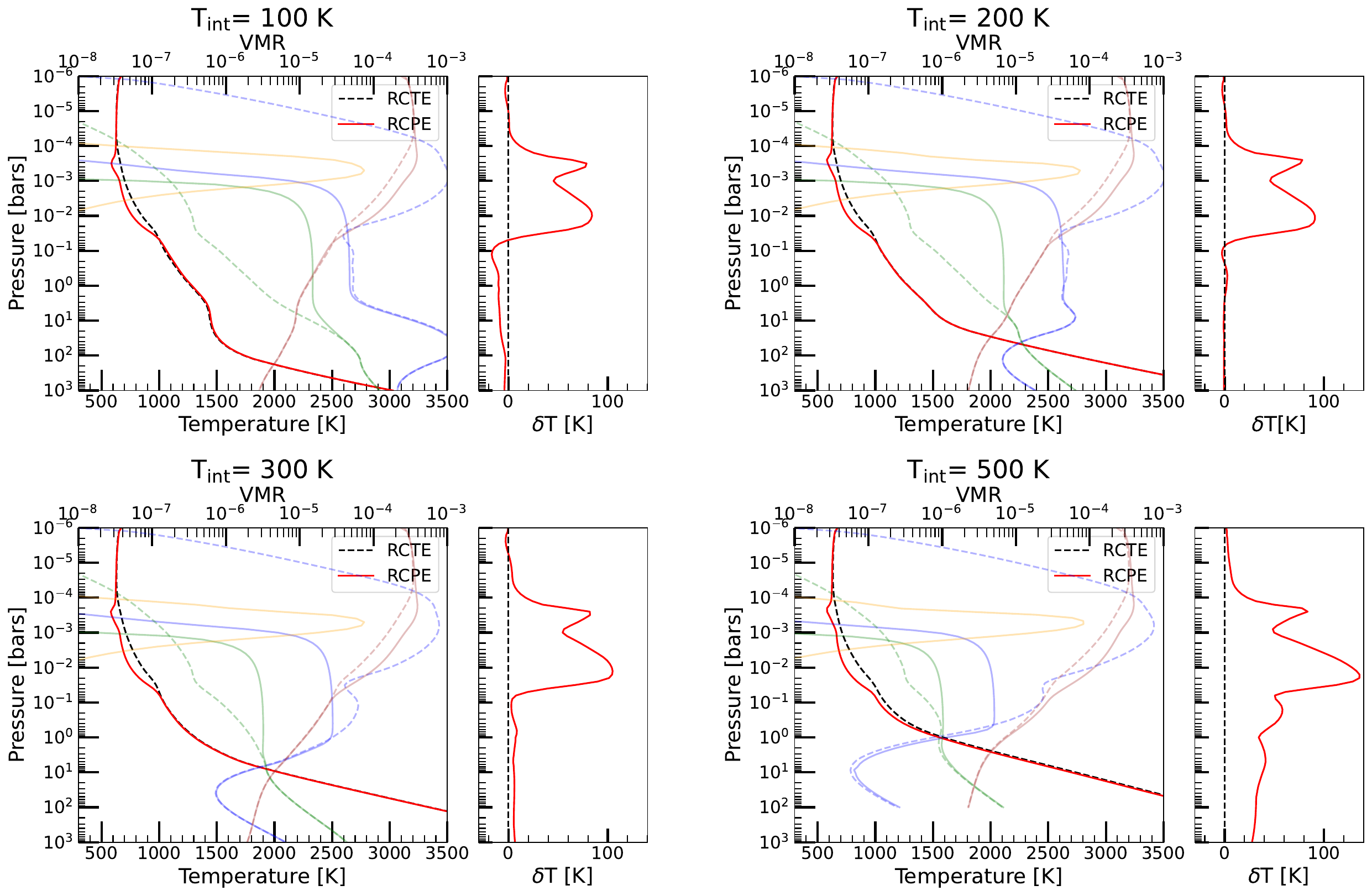}
  \caption{{\tp} profiles computed assuming radiative--convective--thermochemical equilibrium (RCTE) and radiative--convective--photochemical equilibrium (RCPE) are shown with dashed black lines and solid red lines, respectively. The assumed {\teq}, [M/H], and C/O of the model is 800 K, +1.0, and 1$\times$solar, respectively. Each panel corresponds to a different {\tint} from 100 K to 500 K. The faded lines show the volume mixing ratios (top x-axis) of {\meth}, {\amon}, {\sotwo}, and {\cotwo} in blue, green, orange, and brown, respectively. The dashed VMR profiles are from the RCTE models whereas the solid VMR profiles correspond to the RCPE models. The narrower panel accompanying each of the four panels shows the different in the {\tp} profile between the two modeling assumptions in each case as a function of pressure.}
\label{fig:rcpe}
\end{figure*}

\subsection{Chemical Kinetics Modeling}

To simulate disequilibrium chemistry within our model atmospheres, we use the python wrapped {\it Photochem}\footnote{https://github.com/Nicholaswogan/photochem} chemical kinetics model \citep{wogan23,wogan24}. We use the {\tp} profiles computed using the method described in the previous section as inputs in the {\it Photochem} package. In addition to the {\it Photochem} package, we use the {\it Cantera} software \citep{cantera} to simulate thermochemical equilibrium in the very deep atmosphere of the model planet atmospheres. For each model, we use the quench time approximation with chemical timescales from \citet{Zahnle14} to determine the quench pressures of key gases like {\meth}, {\amon}, {\cotwo}, etc. At atmospheric levels with pressures much greater than the highest quench pressures among these gases (5-8 atmospheric levels deeper than the highest quench pressure), we use {\it Cantera} to calculate the thermochemical equilibrium state of the atmospheres. At pressures smaller than this pressure level we use {\it Photochem} to simulate chemical disequilibrium with its kinetics calculations. The convergence criteria used in our photochemistry modeling is presented in \S\ref{sec:convergence}. This modeling setup helps us in calculating photochemical models significantly quicker than using the chemical kinetics calculation for all the pressure layers in the atmosphere including the very high pressure and high temperature deep atmosphere where chemical reactions are expected to be very fast. The setup saves computational time by preventing the kinetics model from taking extremely small time steps to model the very rapid chemical reactions in the very deep atmosphere, which is bound to remain in thermochemical equilibrium anyway. To maintain sufficient accuracy in the 1D chemical kinetics calculations, we interpolate the atmospheric {\tp} profiles with 91 pressure levels to a finer grid of pressures with 180 pressure levels before ingesting them within {\it Photochem}. 

For a given metallicity, we scale the C/H, O/H, N/H, S/H from the solar values with the metallicity factor in our model. To change the C/O ratio for a given metallicity, we change both the C/H and O/H such that (C+O)/H remains the same. This is crucial to maintain the same atmospheric metallicity while changing the C/O ratio of the atmosphere. To account for O- atoms locked up in various expected condensates such as silicates, we remove 20\% of the O from the atmospheric gas phase.

\texttt{PICASO} and {\it Photochem} can be fully and iteratively coupled, instead of \texttt{PICASO} using pre-computed equilibrium chemistry tables \citep[e.g.,][]{marley21,fortney2007planetary,morley13,Mukherjee22,morley14,fortney2005comp,fortney2008unified}, or using on-the-fly non-equilibrium chemistry due to mixing implemented using quench time approximation \citep[e.g.,][]{Mukherjee22a,mukherjee24}. Figure \ref{fig:rcpe} illustrates the difference in the {\tp} profile between a fully self-consistent treatment of photochemistry within the climate model and a profile computed assuming radiative--convective--thermochemical equilibrium (RCTE) for the same set of physical parameters -- {\teq}= 800 K, [M/H]=+1.0, and C/O=1$\times$solar. The radiative--convective--photochemical equilibrium (RCPE) \citep[e.g.,][]{bell23,welbanks24} model was computed by combining the {\it Photochem} kinetics model \citep{wogan23} calculations within the \texttt{PICASO} climate calculation framework. Each panel of Figure \ref{fig:rcpe} illustrates the difference between the {\tp} profiles for four different {\tint} values of 100 K, 200 K, 300 K, and 500 K with {\kzz}=10$^{9}${\cms}. The RCTE (dashed) and RCPE (solid) chemical profiles for {\meth} (blue), {\amon} (green), and {\sotwo} (orange) are also shown in each panel. It is clear that for lower values of {\tint}, the maximum difference in the {\tp} profile is typically around $\sim$ 100 K in the 0.1 mbar to 0.1 bar range. Whereas as the {\tint} increases, small differences of the order of $\sim$ 20 K remain in the profiles, even at deeper pressures till around 100 bars. This shows that the corrections in the {\tp} profiles due to radiative feedback from the changed atmospheric chemistry due to vertical mixing or photochemistry is an important effect, especially for high {\tint} values. This will be especially relevant for interpreting emission spectra of planets. Even though this work focuses on modeling disequilibrium chemistry effects, we choose not to use the RCPE approach for our {\tp} profile calculations mainly because this can vastly increase the computational time involved in running a 1D model. We make this choice because we focus on transmission spectroscopy in this paper where temperature has a relatively small effect compared to emission spectroscopy. Moreover, in our parameter space exploration, we vary both metallicity and C/O ratio but we do not self-consistently include the variation of the {\tp} profile with metallicity and C/O in our models. We discuss the effect of these approximations further in \S\ref{sec:disc}.

Our parameter space exploration (\S\ref{sec:cbyo}) is focused on investigating roles of C/O and {\kzz} across planets with different {\teq} and {\tint}. For this, we run photochemical models for 13 different C/O values between a range of C/O ratios starting from 0.0458 to 1.125 with a step size of 0.0916, exploring the very O- rich to very C- rich atmospheres at 10$\times$solar metallicity. We also vary the {\kzz} parameter between 10$^{6}$ to 10$^{13}$ cm$^2$s$^{-1}$ with steps of 1 in the log$_{10}$(\kzz) space. This range reflects the huge range of uncertainty on this parameter covering cases of very slow vertical mixing to rapid mixing. Therefore, in the first parameter space exploration we compute chemical kinetics models for each {\tp} profile shown in Figure \ref{fig:TP_profiles} for a range of C/O and {\kzz}. For each model, we use the X-ray/UV spectra for WASP-39b used in \citet{tsai23}. Note that we {\it do not} scale the UV flux incident on the planet as we move the planet closer or further away from the star to change its {\teq}. We make this choice as we want to explore the chemical trends in atmospheres as a function of varying \teq\ with a uniform X-ray/UV flux incident on the planet. This is to avoid changing this additional parameter while we sweep the \teq\ parameter space. Our second part of the parameter space exploration (\S\ref{sec:met}) is focused on exploring the role of metallicity, C/O, {\kzz}, and {\tint} on the photospheric abundances but at a fixed {\teq}. We present the trends in photospheric abundance with varying {\teq} in \S\ref{sec:teq} and the sensitivity of the transmission spectra to various planetary parameters in \S\ref{sec:spectra}. We also present trends in our estimates of photochemical precursors to haze/soot across the parameter space in \S\ref{sec:haze}.

\section{Results}\label{sec:results}
\subsection{Exploring how abundances depend on {\kzz}, C/O, and {\tint} across different {\teq}}\label{sec:cbyo}

\begin{figure*}
  \centering
  \includegraphics[width=1\textwidth]{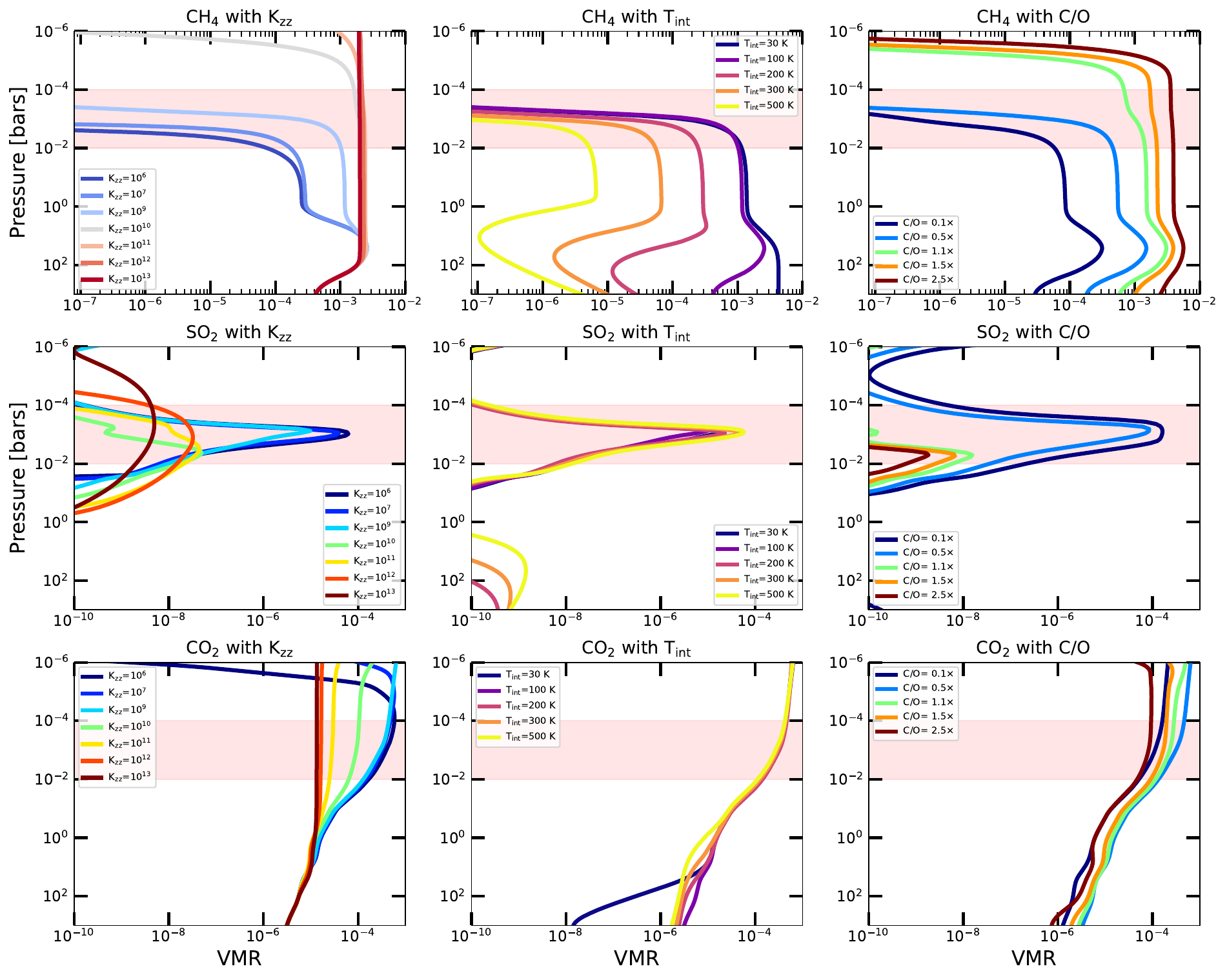}
  \caption{The dependence of chemical abundance profiles on {\kzz}, {\tint}, and C/O are shown in the left, middle, and right columns, respectively. The upper row shows this dependence for {\meth} while the middle and lower rows show this dependence for {\sotwo} and {\cotwo}, respectively. The pressure values typically probed by transmission spectra are shown with the red shaded region. }
\label{fig:chemical_profiles}
\end{figure*}
In this section, we present the variation of key molecular abundances at the pressures typically probed by transmission spectroscopy for transiting planets. The top three panels of Figure \ref{fig:chemical_profiles} shows an example of how {\meth} abundance profiles change with changing {\kzz}, {\tint}, and C/O, respectively for a planet with {\teq}= 800 K. The middle and bottom three panels of Figure \ref{fig:chemical_profiles} shows the same but for {\sotwo} and {\cotwo}, respectively. The approximate pressure range probed by transmission spectroscopy is shown with the red shaded region in each panel. It is clear that {\tint}, {\kzz}, and C/O can cause significant changes to the photospheric abundance of gases like {\meth} but Figure \ref{fig:chemical_profiles} also shows that not all gases are equally sensitive to all of these parameters.

 In order to present the trends in photospheric abundance of gases with variations in these parameters, we use a box-shaped kernel between 10 mbar and 0.1 mbar which is also the region shaded in Figure \ref{fig:chemical_profiles}. For each gas abundance profile, we calculate the logarithm of the average photospheric abundance using,
\begin{equation}\label{eq:xw}
    log_{10}(X_{\rm w}) = \dfrac{{\int_{P_1}^{P_2}\log_{10}(X(p))dp}}{{\int_{P_1}^{P_2}dp}}
\end{equation}
where $X(P)$ is the volume mixing ratio profile of gas $X$, $p_1$ is 10 mbar, and $p_2$ is 0.1 mbar. This range covers a typical pressure range probed by transmission spectroscopy. We present the trends in $X_{\rm w}$ for various gases in the following sections.

\subsubsection{Trends in {\meth}, CO, {\water}, and {\cotwo}}
\begin{center}
\item{}
\paragraph{{\meth}}
\end{center}

\begin{figure*}
  \centering
  \includegraphics[width=1\textwidth]{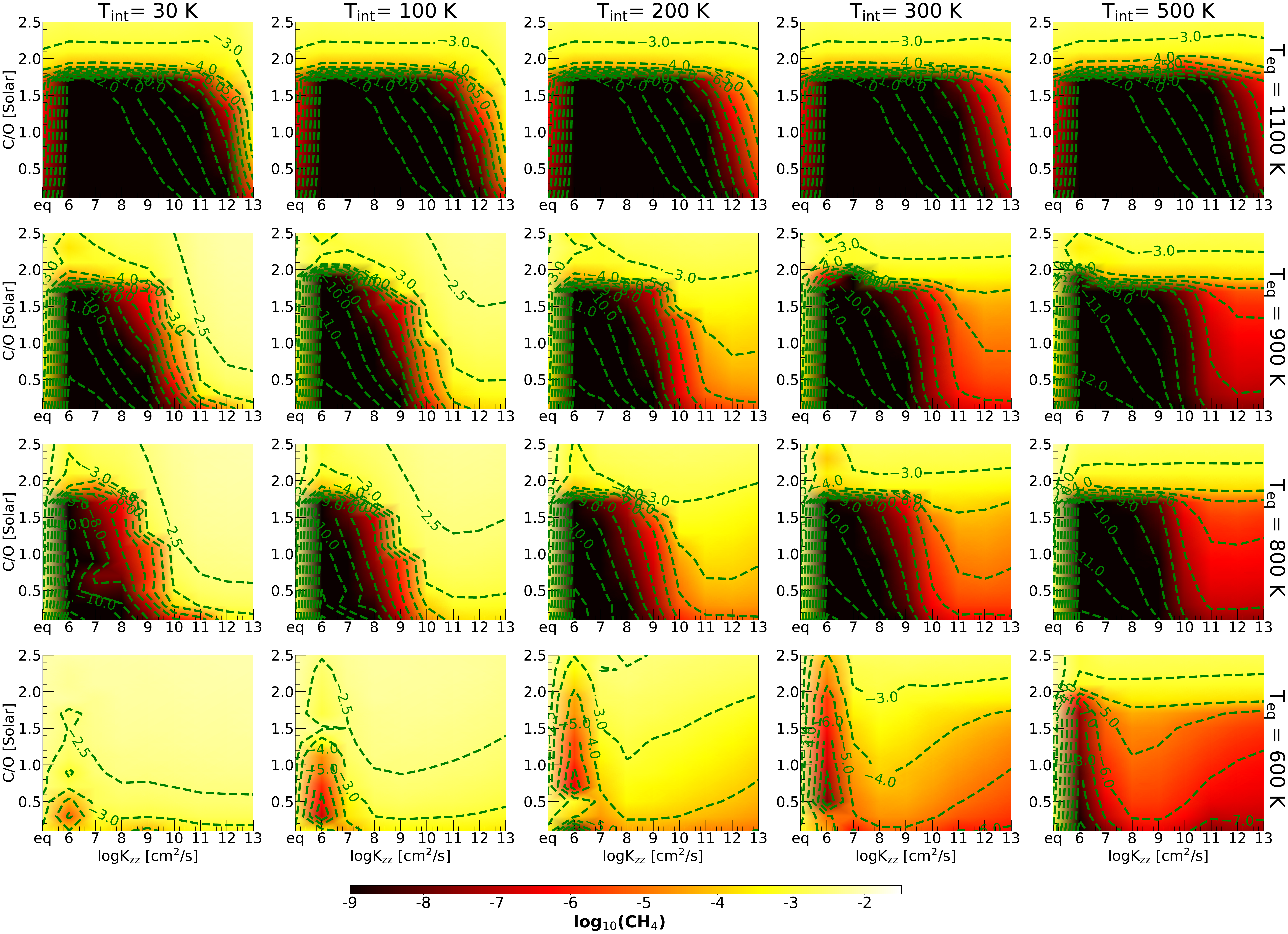}
  \caption{The weighted abundance of {\meth} calculated using Equation \ref{eq:xw} as a function of C/O and {\kzz} in each panel. Each row corresponds to a different {\teq} value from 1100 K to 600 K from top to bottom. Each column correspond to a different {\tint} value between 30 K and 500 K from left to right. Note that the C/O is shown relative to an assumed solar value of 0.458 in this plot.}
\label{fig:ch4_profiles}
\end{figure*}

{\meth} is one of the major C-bearing gases in giant planet atmospheres. Figure \ref{fig:ch4_profiles} shows the quantity $X_{\rm w}$ for {\meth} with a heat map as  a function of {\kzz} and C/O at various {\teq} and {\tint} values. Each row in Figure \ref{fig:ch4_profiles} corresponds to a different {\teq} from 1100 K to 600 K, while each column corresponds to a different {\tint} from 30 K to 500 K. We do not show the heat maps computed for {\teq} of 1400 and 1600 K here as they are qualitatively similar to the 1100 K case. The left most part of each sub-figure shows the trend in {\meth} $X_{\rm w}$ in the case of thermochemical equilibrium. 

Figure \ref{fig:ch4_profiles} shows that for {\teq} $\ge$ 1100 K, {\meth} only becomes abundant in the photosphere when the atmosphere is very C- rich with C/O $>$ 2$\times$solar or when {\kzz} $\ge$ 10$^{12}$ {\cms}. \citet{madhu12} found very similar C/O dependence for {\meth}, albeit their calculations assumed thermochemical equilibrium. The amount of {\meth} in the photosphere for these hotter planets shows some {\tint} dependence when the vertical mixing is extremely vigorous with {\kzz} $\ge$ 10$^{12}$ {\cms}. In these cases, a colder interior planet (e.g., {\tint}= 30 K or 100 K) shows higher amount of photospheric {\meth} than planets with much hotter interior (e.g., {\tint}= 500 K).  This shows that the {\tint} parameter, which mostly affects the {\tp} and chemistry in the deep atmosphere, already starts to affect the {\meth} abundance at much smaller pressures at {\teq}=1100 K, when mixing is vigorous.

Figure \ref{fig:ch4_profiles} shows that for a given C/O, {\tint}, or {\kzz}, the {\meth} abundance is much higher in the {\teq} $\le$ 900 K planets than the {\teq} $\ge$ 1100 K planets. For {\teq}= 900 K, Figure \ref{fig:ch4_profiles} shows that the {\meth} abundance varies very strongly with {\kzz}. For a given C/O, the photospheric {\meth} abundance shows 6-7 orders of magnitude increase with the {\kzz} increasing by 7 orders of magnitude. The abundance of photospheric {\meth} is much lower than the expectations from thermochemical equilibrium for {\kzz} $\le$ 10$^{11}${\cms} when {\tint}$\le$ 200 K. For stronger vertical mixing, the amount of {\meth} rises to thermochemical equilibrium expectations for these colder interior planets.  Whereas, the {\meth} abundance is always lower than the expected amount from thermochemical equilibrium for hotter interiors for all the {\kzz} values explored in Figure \ref{fig:ch4_profiles}. Equal abundance contours plotted on Figure \ref{fig:ch4_profiles} {\teq}=900 K panels are nearly vertical which suggests that the variation of {\meth} abundance with changing {\kzz} is stronger than its variation with C/O. For a given C/O and {\kzz}, the {\meth} abundance typically decreases by 3-4 orders of magnitude with the {\tint} increasing from 30 K to 500 K for {\teq}=900 K. Figure \ref{fig:ch4_profiles} shows that the qualitative behavior shown by the {\meth} abundance is very similar between {\teq}=900 K and {\teq}=800 K panels except the {\meth} abundances are higher in the {\teq}=800 K panels than the {\teq}=900 K panels for a given C/O, {\tint}, and {\kzz} value.

The last row in Figure \ref{fig:ch4_profiles}
 shows {\meth} abundances for {\teq}= 600 K. For these cold {\teq} values, {\meth} abundance shows a significant increase compared to the {\teq}= 800 K cases. This increase is especially high for {\kzz}$\le$10$^{10}${\cms}. Also, the equal abundance contours for {\teq}= 600 K, are nearly horizontal which shows that the {\meth} abundance is much more sensitive to C/O than {\kzz} for these cold planets. At {\teq}= 600 K, for a given C/O and {\kzz} value, the {\meth} abundance shows a 2-3 orders of magnitude decrease with {\tint} increasing from 30 K to 500 K. The equal abundance contours shown in Figure \ref{fig:ch4_profiles} show that while interpreting the C/O of a planet from its measured {\meth} abundance, one should be particularly careful in parts of the parameter space where the contours in Figure \ref{fig:ch4_profiles} are not horizontal. This means that the same photospheric {\meth} abundance can be a result of a wide range C/O in these parts of the parameter space. This also suggests that the {\meth} abundance, if detected, can be a good diagnostic of {\kzz} in parts of the parameter space where the equal abundance contours are nearly vertical. Figure \ref{fig:ch4_profiles} suggests that {\meth} becomes a favorable marker of vertical transport when {\teq} is between 800--900 K, which is slightly colder than the `sweet-spot' identified in the 3D models presented in \citet{zamyatina23}. Note that we have shown these trends at a fixed metallicity (i.e., 10$\times$solar).

\begin{center}
\item{}
\paragraph{{\co}}
\end{center}

\begin{figure*}
  \centering
  \includegraphics[width=1\textwidth]{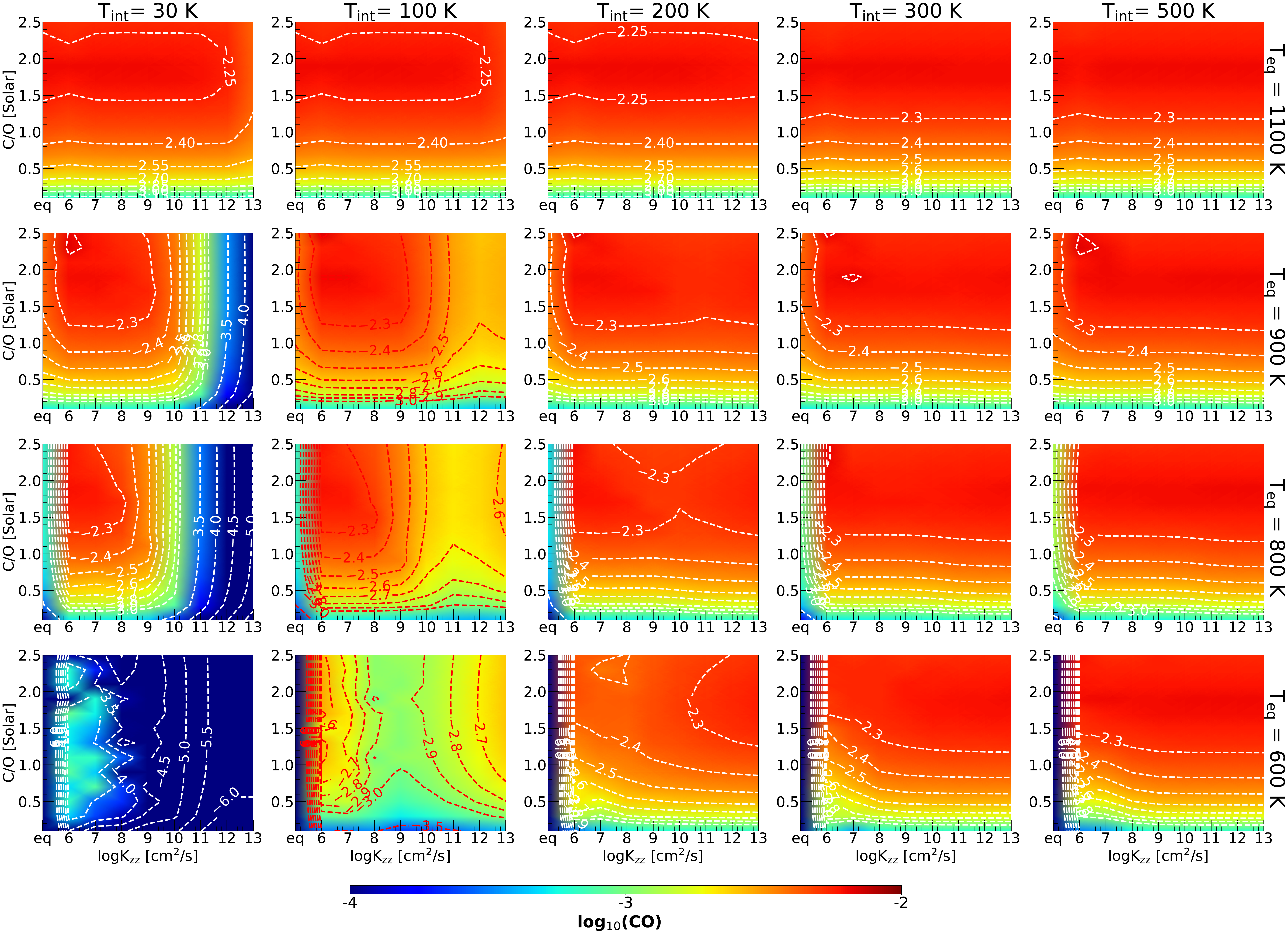}
  \caption{The weighted abundance of {\co} calculated using Equation \ref{eq:xw} has been shown as a heat map as a function of C/O and {\kzz} in each panel. Each row corresponds to a different {\teq} value from 1100 K to 600 K from top to bottom. Each column correspond to a different {\tint} value between 30 K and 500 K from left to right. Note that the C/O has been shown relative to the assumed solar value of 0.458 in this plot.}
\label{fig:co_profiles}
\end{figure*}

\begin{figure*}
  \centering
  \includegraphics[width=1\textwidth]{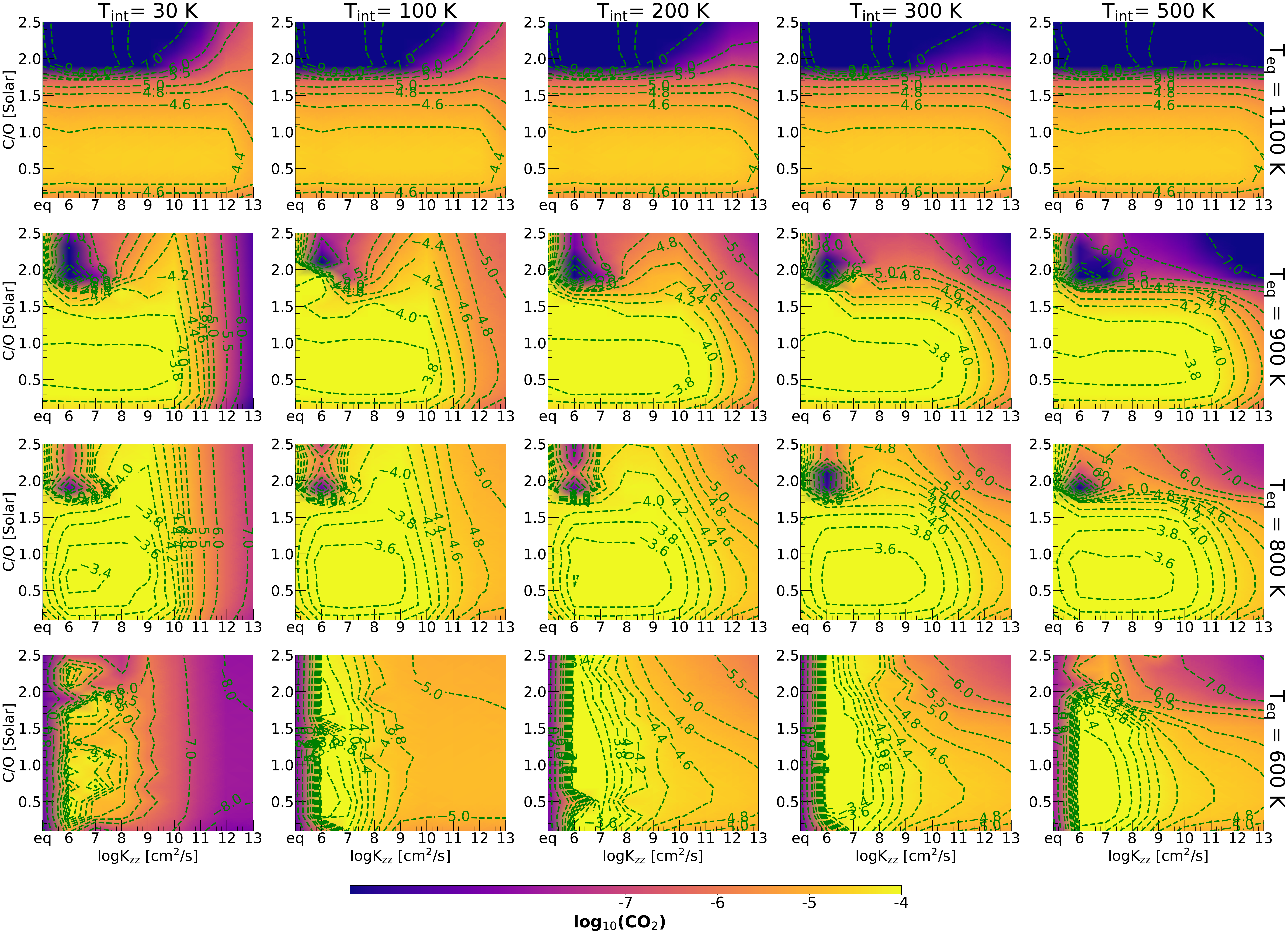}
  \caption{The weighted abundance of {\cotwo} calculated using Equation \ref{eq:xw} has been shown as a heat map as a function of C/O and {\kzz} in each panel. Each row corresponds to a different {\teq} value from 1100 K to 600 K from top to bottom. Each column correspond to a different {\tint} value between 30 K and 500 K from left to right. Note that the C/O has been shown relative to the assumed solar value of 0.458 in this plot.}
\label{fig:cotwo_profiles}
\end{figure*}

Figure \ref{fig:co_profiles} presents the $X_w$ for CO across different C/O ratios and {\kzz}. Again, each panel corresponds to a  different {\teq} and {\tint} combination. For {\teq}= 1100 K, Figure \ref{fig:co_profiles} shows that the CO abundance is only sensitive to C/O showing a rapid increase with increasing C/O around C/O$\sim$0.3$\times$solar. Figure \ref{fig:co_profiles} suggests that the {\co} gas can be a great tracer of C/O in planets with {\teq} $\ge$ 1100 K irrespective of the strength of {\kzz}. The {\co} abundane for a given C/O also shows almost no {\tint} dependence for {\teq}$\ge$1100 K. The difference in CO abundance between the thermochemical equilibrium and disequilibrium chemistry at {\teq}= 1100 K and 900 K is almost negligible. However, some {\kzz} dependence of {\co} abundance appears at {\teq}= 900 K, when {\tint}$\le$ 100 K. For these cases, {\co} shows a decrease with increasing {\kzz} when {\kzz}$\ge$ 10$^{9}${\cms}. This decrease of {\co} abundance with increasing {\kzz} beyond 10$^{9}${\cms} is much more pronounced when {\tint}=30 K than the case of {\tint}= 100 K. Similar behavior is also shown by the {\teq}= 800 K models, where the {\kzz} dependence of {\co}  abundance for cold interior planets ({\tint}$\le$ 100 K) sets in at a slightly lower {\kzz} of 10$^{8}${\cms}. This suggests that for planets with {\teq} at 800 K or 900 K and hot interiors (\tint\ $\ge$ 200 K), {\co} can be a great tracer of C/O, irrespective of {\kzz} and with minimal dependence on {\tint} as well. But for such planets with cold interiors {\tint\ $\le$ 100 K}, whether {\co} will trace the C/O of the atmosphere depends largely on the strength of {\kzz}.

The same behavior is not only shown but enhanced further in {\teq}= 600 K planets shown in the last row of Figure \ref{fig:co_profiles}. The {\co} abundance mainly depends on {\kzz} only in these cases for \tint\ $\le$ 100 K and is dependant only on C/O for \tint\ $\ge$ 200 K. This suggests that for planets with colder {\teq} ($\le$ 900 K) and cold interiors (\tint\ $\le$ 100 K), {\co} abundance can be a tracer of {\kzz}, whereas for all the other cases of hotter planets or hotter interiors, {\co} abundance is a great tracer of atmospheric C/O. 

\begin{figure*}
  \centering
  \includegraphics[width=0.45\textwidth]{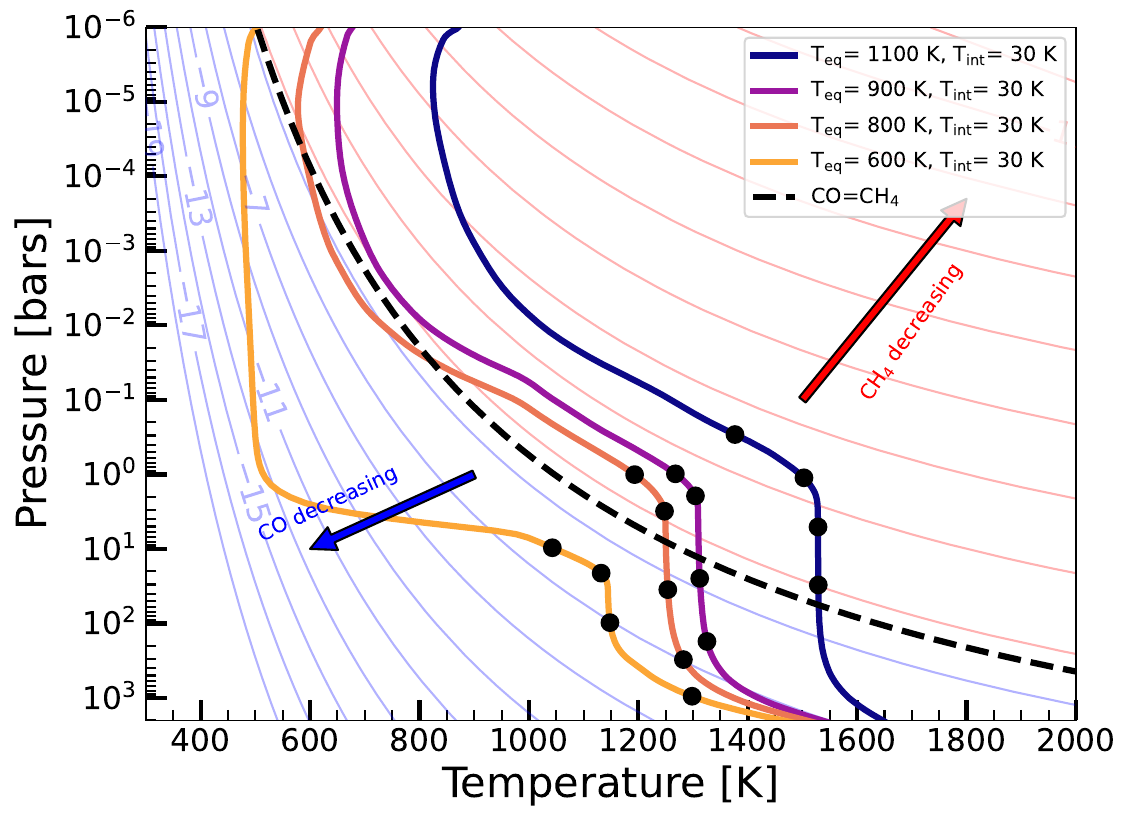}
  \includegraphics[width=0.45\textwidth]{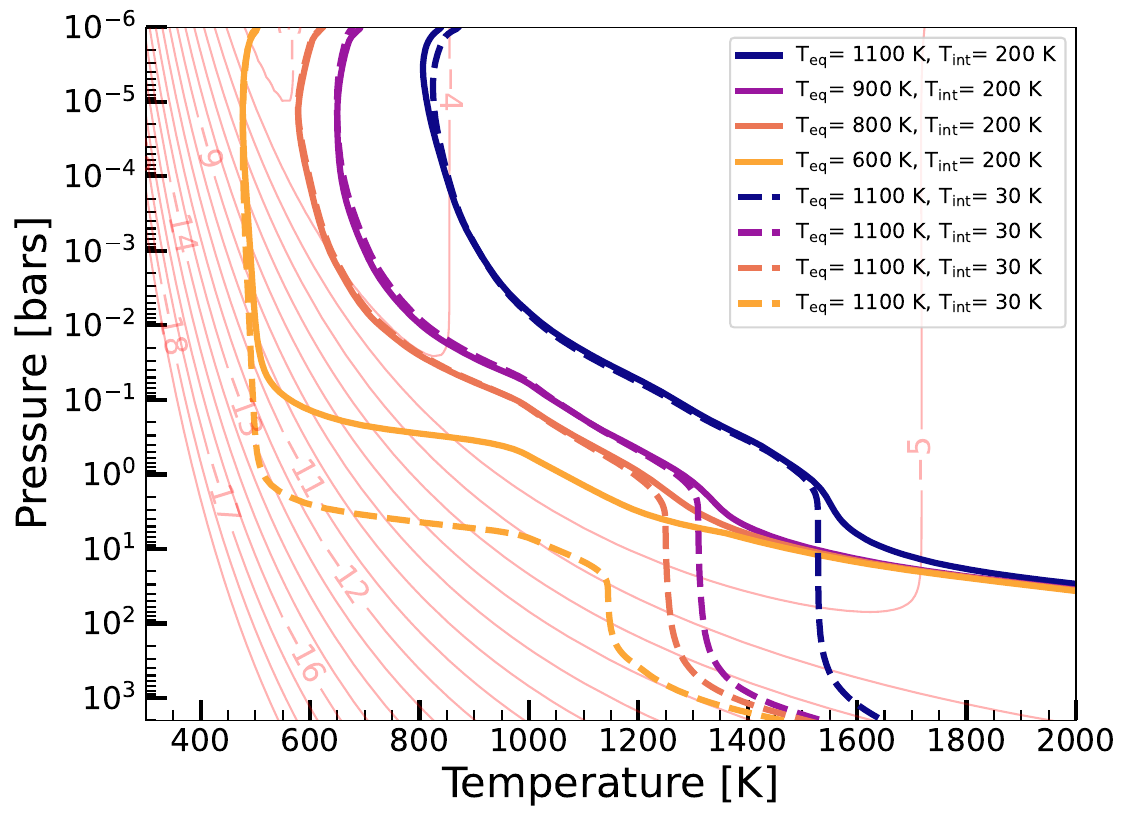}
  \caption{{\bf Left panel:} The {\tp} profiles of planets with {\tint}= 30 K and {\teq} between 1100 and 600 K are shown with the four solid lines. The black dashed line shows the {\co}={\meth} equal abundance curve from chemical equilibrium calculations at [M/H]=+1.0 and C/O=1$\times$solar. Equal abundance contours for {\co} and {\meth} are shown in blue and red, respectively. The four black markers on each {\tp} profile show the quench pressures of {\co} for 10$^6$, 10$^8$, 10$^{10}$, and 10$^{12}$ {\cms}. {\bf Right panel:} The {\tp} profiles of planets with {\tint}= 200 K and {\teq} between 1100 and 600 K are shown with the four solid lines, whereas the dashed lines show the models for {\tint}= 30 K. Equal abundance contours for {\cotwo} from thermochemical equilibrium are shown in red.}
\label{fig:co_behavior}
\end{figure*}

The CO abundance in planets with cold interiors and colder {\teq} values shows a decrease with increasing {\kzz}, beyond a certain {\kzz} value, because their {\tp} profiles straddle the {\co}={\meth} equal abundance curve \citep{fortney20}. This has been shown in Figure \ref{fig:co_behavior} right panel, where the black dashed curve shows the {\co}={\meth} equal abundance curve and the solid {\tp} profiles represent {\tint}=30 K models with {\teq} lying between 1100 K and 600 K. {\co} abundance decreases in the bottom left direction of the figure and {\meth} abundance decreases towards the top right direction of the plot. The black points on each {\tp} profile show the quench points for {\co}, when the {\kzz} is 10$^6$, 10$^8$, 10$^{10}$, and 10$^{12}$ {\cms}. The {\co} abundance remains almost constant with changing {\kzz} if the quench pressure lie in the {\co} dominated region, which is the top right part above the {\co}={\meth} equal abundance curve. If quenching happens in this region, the quenched {\co} abundance is expected to be insensitive to {\kzz}. However, if {\co} is quenched in the lower left half, where {\co} abundance is rapidly decreasing, then the {\co} abundance shows a rapid decrease with increasing {\kzz}. This decrease is sharper because of the isothermal nature of the {\tp} profile, which are far from being parallel to the equal abundance contours of {\co} in this part of the $T-P$ space.

Another important behavior reflected in Figure \ref{fig:co_profiles} is that the photospheric {\co} abundance remains the same between thermochemical equilibrium and chemical disequilibrium for {\teq} $\ge$ 900 K, but for colder planets, the {\co} abundance in the thermochemical equilibrium cases are smaller than the models with low {\kzz} values (\kzz\ $\le$ 10$^7${\cms}), irrespective of {\tint}. This suggests that even though the {\co} abundance can act as a tracer of C/O (without any {\kzz} dependence) in planets with hot interiors, disequilibrium chemistry calculations must still be used to interpret the correct C/O of a planet from its measured {\co} abundance. This behavior is a direct effect of the shape of the {\tp} profiles of these planets, when their {\teq} is $\le$ 800 K and can also be understood with the left panel of Figure \ref{fig:co_behavior}. The {\tp} profiles of planets with {\teq}$\le$800 K transition from the CO-dominated to the {\meth} dominated regions in the $T-P$ space, as shown in Figure \ref{fig:co_behavior}. This leads to a sharp decrease in photospheric CO in these objects. However, even for low {\kzz} values (e.g., 10$^6$ {\cms}), the CO gets quenched at much deeper pressures, where their {\tp} profiles cross equal abundance contours corresponding to higher CO abundance than the low photospheric CO abundance expected from thermochemical equilibrium. As the {\co} abundance becomes roughly constant on the top right side of the {\co}={\meth} line, this doesn't happen when {\teq}$\ge$900 K.


\begin{center}
\item{}
\paragraph{{\cotwo}}
\end{center}

Figure \ref{fig:cotwo_profiles} shows the $X_w$ for {\cotwo}. At {\teq}= 1100 K, the {\cotwo} abundance shows a sharp decrease when C/O $\ge$ 1.9$\times$solar. This behaviour is unlike {\co}, which doesn't show such a sharp change in its abundance when C/O is $\ge$ 1.9$\times$solar as shown in the top panels of Figure \ref{fig:co_profiles}. {\cotwo} shows this sharp decline beyond C/O $\ge$ 1.9$\times$solar due to the unavailability of enough O- atoms to form {\cotwo} when the C/O is approaching 1. At this {\teq}, {\cotwo} remains practically independent of {\kzz} across all {\tint} values, except for some slight {\kzz} sensitivity for {\kzz $\ge$ 10$^{12}${\cms}. However, Figure \ref{fig:cotwo_profiles} also shows that the photospheric {\cotwo} abundance doesn't show much change with changing C/O between 0.3$\times$solar$\le$C/O$\le$1.2$\times$solar across all {\tint} values. This might limit the utility of {\cotwo} as a C/O diagnostic for this range of C/O values. This suggests the {\cotwo} can be a good tracer for C/O in these {\teq} $\ge$ 1100 K planets, except for very high strengths of vertical mixing or for 0.3$\times$solar$\le$C/O$\le$1.2$\times$solar. {\cotwo} also doesn't show much dependence on {\tint} at this {\teq}.

{\cotwo} has a larger {\kzz} dependence for planets with {\teq}$\le$ 900 K. For {\teq}= 900 K and {\tint}= 30 K, the {\cotwo} remains insensitive to {\kzz} for {\kzz}$\le$10$^9${\cms}. However, the {\cotwo} shows very strong sensitivity to {\kzz} for higher values of {\kzz} causing the {\cotwo} abundance to drop by two orders of magnitude with {\kzz} increasing by four orders of magnitude. This behavior is present for hotter {\tint} models as well, albeit with a slower decrease of {\cotwo} abundance with increasing {\kzz}. With increasing {\tint}, the {\kzz} value at which {\cotwo} transitions from being insensitive to sensitive to {\kzz} also shifts slowly towards higher value. The {\teq}= 800 K models shown in the third row of Figure \ref{fig:cotwo_profiles}, also show the same qualitative behavior as the {\teq} =900 K models, but with an even stronger dependence of {\cotwo} abundance on {\kzz} once the {\kzz} value is higher than the ``transition" {\kzz} value. This behavior can be readily understood with the right panel of Figure \ref{fig:co_behavior}, which shows the {\tp} profiles for {\tint}= 200 K and 30 K models with solid and dashed lines, respectively. Each profile corresponds to a different {\teq}. The equal abundance contours for {\cotwo} from thermochemical equilibrium are shown with the red lines. For planets with hotter {\teq} or hotter {\tint}, most of the {\tp} profile lie in parts of the $T-P$ space where the {\cotwo} equal abundance contours appear less dense. This suggests that a large change in quench pressure of {\cotwo} due to changing {\kzz} will only cause a small change in the quenched {\cotwo} abundance for such {\tp} profiles. However, for planets with colder {\teq} or colder {\tint}, the {\tp} profiles traverse a denser region of equal abundance contours of {\cotwo}. For these models, a change in quench pressure of {\cotwo} due to changing {\kzz} will produce a much more pronounced effect on the quenched {\cotwo} abundance. These models suggest that for {\teq}=800 and 900 K planets, irrespective of {\tint}, the {\cotwo} abundance remains sensitive to {\kzz} to varying degrees if {\kzz} is higher than a ``transition" value which is typically somewhere between 10$^{9}$-10$^{10}${\cms}. Other than the coldest interior case of {\tint}=30 K, the {\cotwo} abundance doesn't show very strong dependence on {\tint}.

Although for {\teq}$\ge$800 K, the photospheric {\cotwo} abundances between thermochemical equilibrium and low {\kzz} values are nearly the same, this breaks down for {\teq}= 600 K. The {\cotwo} abundance shows a sharp increase between thermochemical equilibrium models and {\kzz}=10$^6${\cms} for these models. This effect can be understood with the interplay between the `U' shaped {\cotwo} equal abundance contours and the {\tp} profiles shown in the right panel of Figure \ref{fig:co_behavior}. Most of the {\tp} profiles of planets with {\teq}$\ge$800 K remain in the $T-P$ space where the {\cotwo} contours are less dense and nearly vertical. As a result, the equilibrium abundance of {\cotwo} monotonically and slowly increases with decreasing pressure in these objects. Therefore, there is generally not a large difference between the quenched abundance and chemical equilibrium abundance of {\cotwo} in these objects. However, the {\teq}= 600 K model has a complex shape with a nearly horizontal region traversing the densely packed contours, where the chemical equilibrium {\cotwo} abundance decreases rapidly. If the {\cotwo} abundance is quenched in this region, the quenched abundance will be significantly higher than expected equilibrium chemistry abundance of {\cotwo} in the photosphere. Moreover, at {\teq}=600 K, {\cotwo} depends almost solely depends on {\kzz}, and not C/O, for {\tint}$\le$100 K. Some C/O dependence of the photospheric {\cotwo} abundance sets in for higher {\tint} values. This suggests that near {\teq}=600 K, the {\cotwo} abundance can be a good tracer of {\kzz} for planets with cold interiors (\tint\ $\le$ 100 K) but for planets with hotter interiors, it can be quite degenerate between {\kzz} and C/O.

\begin{center}
\item{}
    \paragraph{{\water}}
\end{center}

The $X_w$ for {\water} is shown in Figure \ref{fig:h2o_profiles}. The range of variation in $X_w$ for {\water} is much lower when compared with other gases shown in Figures \ref{fig:ch4_profiles}, \ref{fig:co_profiles}, and \ref{fig:cotwo_profiles}. {\water} shows strong sensitivity to C/O when {\teq}= 1100 K without much dependence on {\kzz}. At {\teq}= 900 K, when the {\tint}$\ge$ 200 K, this behaviour of {\water} abundance being only dependant on C/O continues. The dependence of {\water}
 abundance on {\tint} is also extremely weak in these cases. {\water} starts to show some {\kzz} dependence when {\kzz} $\ge$ 10$^{10}${\cms} for {\tint} $\le$ 100 K at {\teq}= 900 K. This behaviour continues for {\tint}$\le$ 100 K at {\teq}= 800 K as well, except the {\kzz} dependence of {\water} abundance sets in a lower value of {\kzz}$\ge$10$^{9}${\cms}. The {\water} abundance shows an increase with {\kzz} beyond this value which is also accompanied by a sharp decrease in {\co} and {\cotwo} abundance seen in this part of the parameter space in Figures \ref{fig:co_profiles} and \ref{fig:cotwo_profiles}, especially at {\tint}= 30 K. The {\water} abundance shows a decrease between thermochemical equilibrium and the {\kzz}=10$^6${\cms} chemical disequilibrium calculations for models with {\teq}$\le$800 K. This decrease is particularly strong for C/O$\ge$1.7$\times$solar for {\teq}= 800 K models. This decrease in {\water} abundance between thermochemical equilibrium and {\kzz}=10$^6${\cms} is also seen for  {\teq}=600 K models. However, the amount of decrease seen in these models increases with increasing {\tint} at this {\teq}. This highlights that for these hotter {\tint} values (\tint\ $\ge$ 200 K) at {\teq} $\le$ 900 K, {\water} can still be a good tracer of C/O but chemical disequilibrium calculations must still be carried out to interpret the C/O from {\water} abundance accurately. The {\teq}=600 K models shown in the last row of Figure \ref{fig:h2o_profiles} show much less sensitivity to {\kzz} compared to the {\teq}=900 K and 800 K models. This behavior is similar to the smaller sensitivity shown by {\meth} abundance to {\kzz} in Figure \ref{fig:ch4_profiles} at {\teq}= 600 K.

\begin{figure*}
  \centering
  \includegraphics[width=1\textwidth]{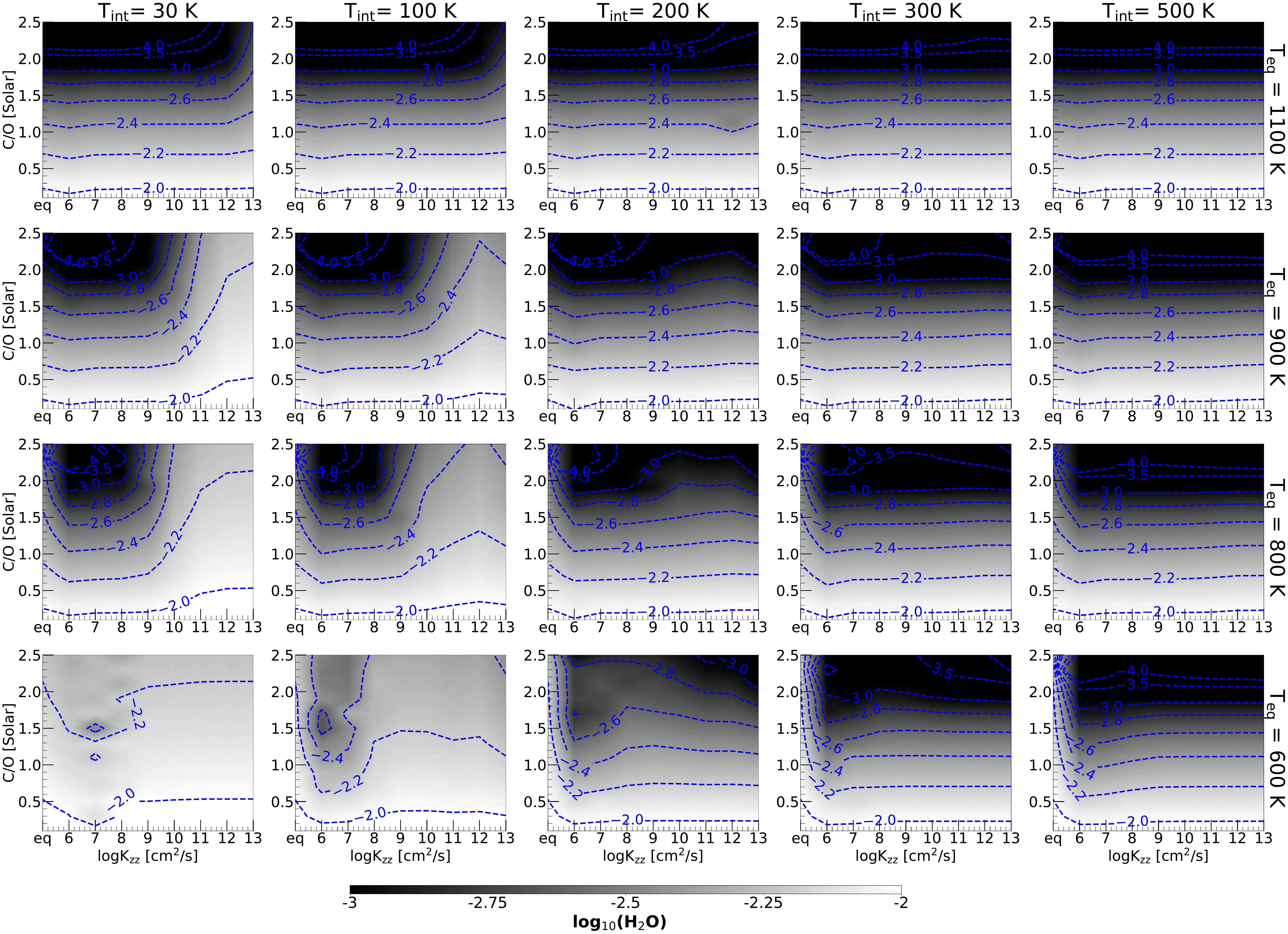}
  \caption{The weighted abundance of {\water} calculated using Equation \ref{eq:xw} has been shown as a heat map as a function of C/O and {\kzz} in each panel. Each row corresponds to a different {\teq} value from 1100 K to 600 K from top to bottom. Each column correspond to a different {\tint} value between 30 K and 500 K from left to right. Note that the C/O has been shown relative to the assumed solar value of 0.458 in this plot.}
\label{fig:h2o_profiles}
\end{figure*}

\subsubsection{Trends in NH$_3$ and HCN}
\begin{center}
\item{}
    \paragraph{{\amon}}
\end{center}

Figure \ref{fig:nh3_profiles} shows the $X_w$ metric for {\amon}. {\amon} neither contains C- or O- and therefore is expected to not depend strongly on C/O. All panels of Figure \ref{fig:nh3_profiles} reflect this behavior. At {\teq}= 1100 K, the photospheres are very {\amon} poor.  For {\teq}= 900 K, the photospheric {\amon} abundance slowly starts to rise, much more for the colder {\tint} models. The vertical equal abundance contours in Figure \ref{fig:nh3_profiles} show that {\amon} is very sensitive to {\kzz} in addition to high sensitivity to {\tint}. Similar to the {\meth} behavior seen in the previous section, {\amon} abundance monotonically increases with {\kzz}. For the coldest interior of {\tint}= 30 K at {\teq}= 900 K or 800 K, {\amon} abundance increases by 7 orders of magnitude between {\kzz} of 10$^{6}$}{\cms} and 10$^{13}${\cms}. The equal abundance curves for {\amon} are known to be nearly parallel to the adiabats for H$_2$/He gas mixture in the deep atmosphere \citep{Zahnle14,fortney20,ohno232}. This means that if {\amon} is quenched in the deep convective atmosphere, its photospheric abundance will not be very sensitive to {\kzz}. {\amon} can become more sensitive to {\kzz} if it is quenched in the radiative atmosphere in irradiated planets, where the lapse rate of the {\tp} profile is very different than the adiabatic lapse rate. The photodissociation of {\amon} due to the incident UV radiation also depends on {\kzz}. A low {\kzz} in the upper atmosphere causes the {\amon} abundance to be depleted very rapidly with decreasing pressure. The depletion of {\amon} with decreasing pressure is much less if the {\kzz} in the upper atmosphere is high. Both of these effects shape the {\kzz} dependence of the {\amon} abundance shown in Figure \ref{fig:nh3_profiles} in the pressures typically probed by transmission spectroscopy. 

At {\teq}$\le$900 K, the {\amon} abundance also declines by several orders of magnitude as {\tint} rises from 30 K to 500 K. Similar to the findings of \citet{ohno232}, we also find that when {\tint} $\le$ 200 K, the photospheric {\amon} abundance shows a very sharp rise with {\teq} below {\teq}=800 K when vertical mixing is slow ({\kzz}$\le$10$^{8}$cm$^2$s$^{-1}$). But when vertical mixing is faster, the photospheric {\amon} abundance shows a much more gradual rise with {\teq}, as was also found by \citet{ohno232}.  Even though detecting {\amon} has proven to be a difficult task so far with {\it JWST} \citep{welbanks24}, its abundance can be of great use for reducing the degeneracies discussed in the previous section arising from just fitting {\meth} abundances and measuring {\kzz} and {\tint}, especially for warm planets.
\begin{center}
\item{}
    \paragraph{{HCN}}
\end{center}

\begin{figure*}
  \centering
  \includegraphics[width=1\textwidth]{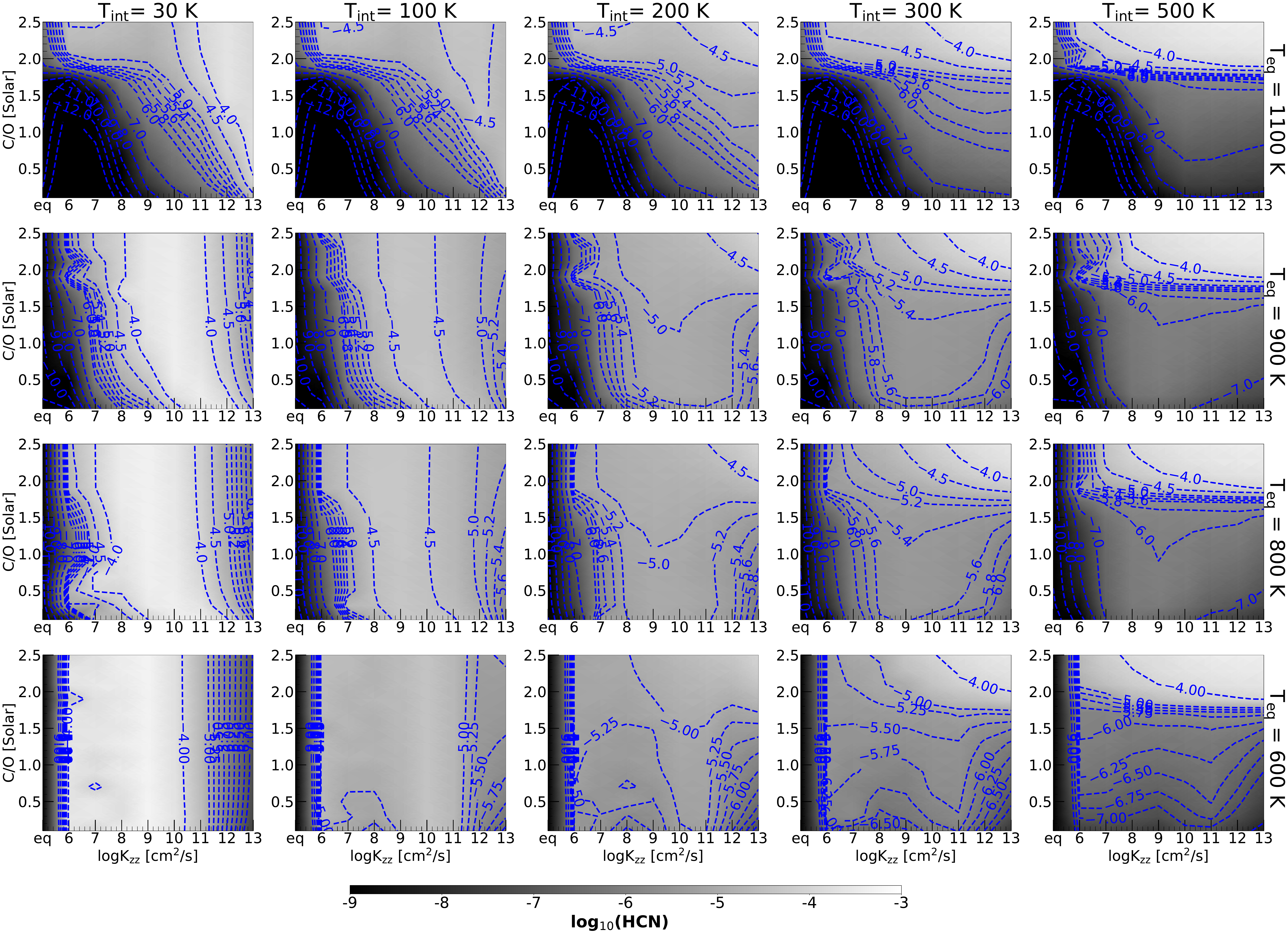}
  \caption{The weighted abundance of {HCN} calculated using Equation \ref{eq:xw} has been shown as a heat map as a function of C/O and {\kzz} in each panel. Each row corresponds to a different {\teq} value from 1100 K to 600 K from top to bottom. Each column correspond to a different {\tint} value between 30 K and 500 K from left to right. Note that the C/O has been shown relative to the assumed solar value of 0.458 in this plot.}
\label{fig:hcn_profiles}
\end{figure*}

HCN is a molecule which is quenched due to {\kzz} as well as produced photochemically. Moreover, unlike {\amon}, it has C- and therefore should be sensitive to C/O. Figure \ref{fig:hcn_profiles} shows the $X_w$ calculated for HCN. As expected, photospheric HCN shows sensitivity to both C/O and {\kzz}, and also depends on {\tint}. Figure \ref{fig:hcn_profiles} shows that the disequilibrium chemistry abundance of HCN is much higher than the expected thermochemical equilibrium abundance. The equal abundance contours in the top row for {\teq}= 1100 K run diagonally which shows that same HCN abundance can be very degenerate corresponding to a wide range of C/O ratios and {\kzz} for these hotter planets. At {\teq}= 900 K ,800 K and 600 K, the HCN abundance is mostly sensitive to {\kzz} and not much to C/O if {\kzz} is either low or very high. The HCN abundance is not sensitive to either {\kzz} or C/O for intermediate values of {\kzz}.

\begin{figure*}
  \centering
  \includegraphics[width=1\textwidth]{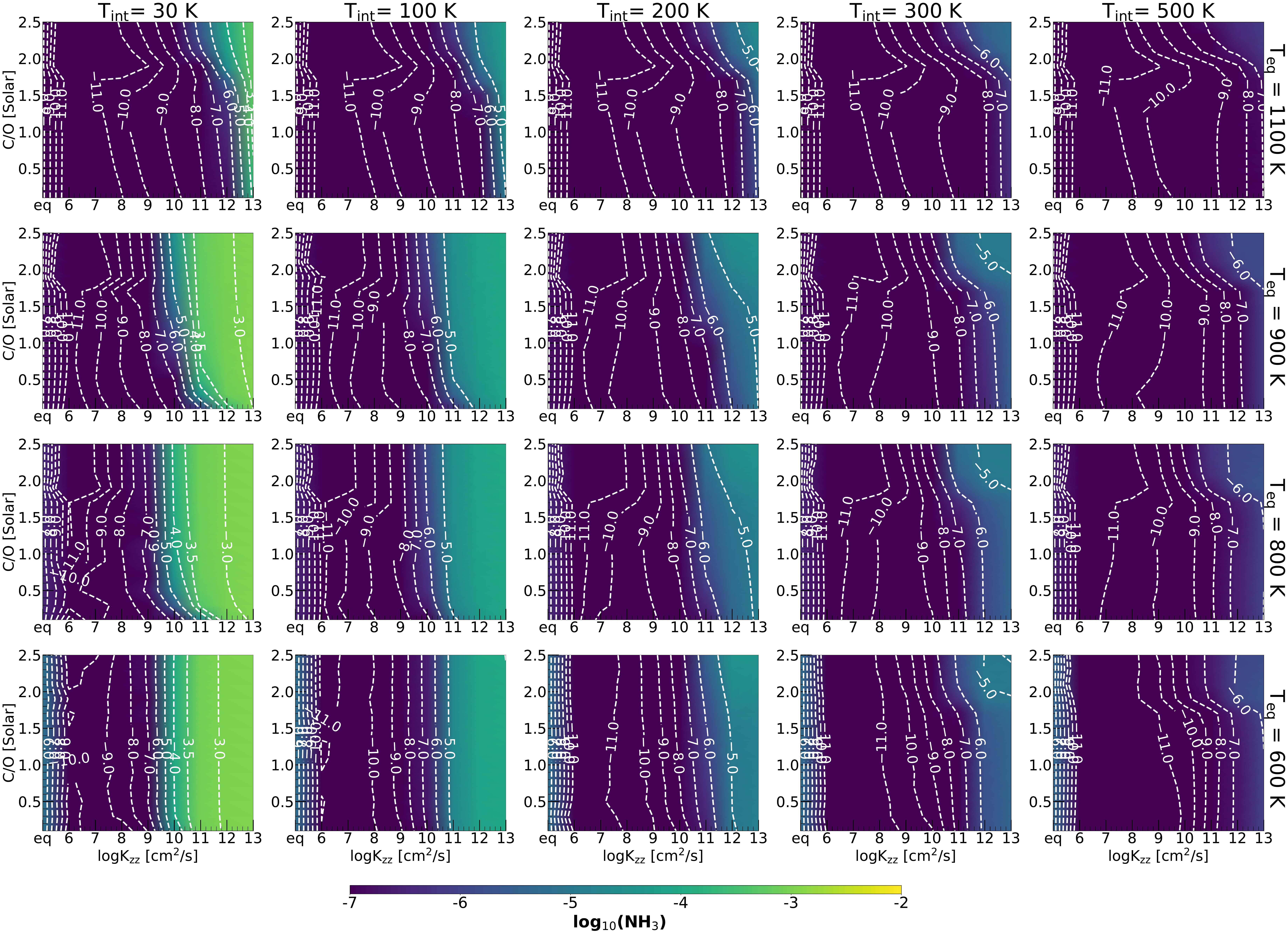}
  \caption{The weighted abundance of {\amon} calculated using Equation \ref{eq:xw} has been shown as a heat map as a function of C/O and {\kzz} in each panel. Each row corresponds to a different {\teq} value from 1100 K to 600 K from top to bottom. Each column correspond to a different {\tint} value between 30 K and 500 K from left to right. Note that the C/O has been shown relative to the assumed solar value of 0.458 in this plot.}
\label{fig:nh3_profiles}
\end{figure*}

\subsubsection{Trends in Sulfur Species}

\begin{figure*}
  \centering
  \includegraphics[width=1\textwidth]{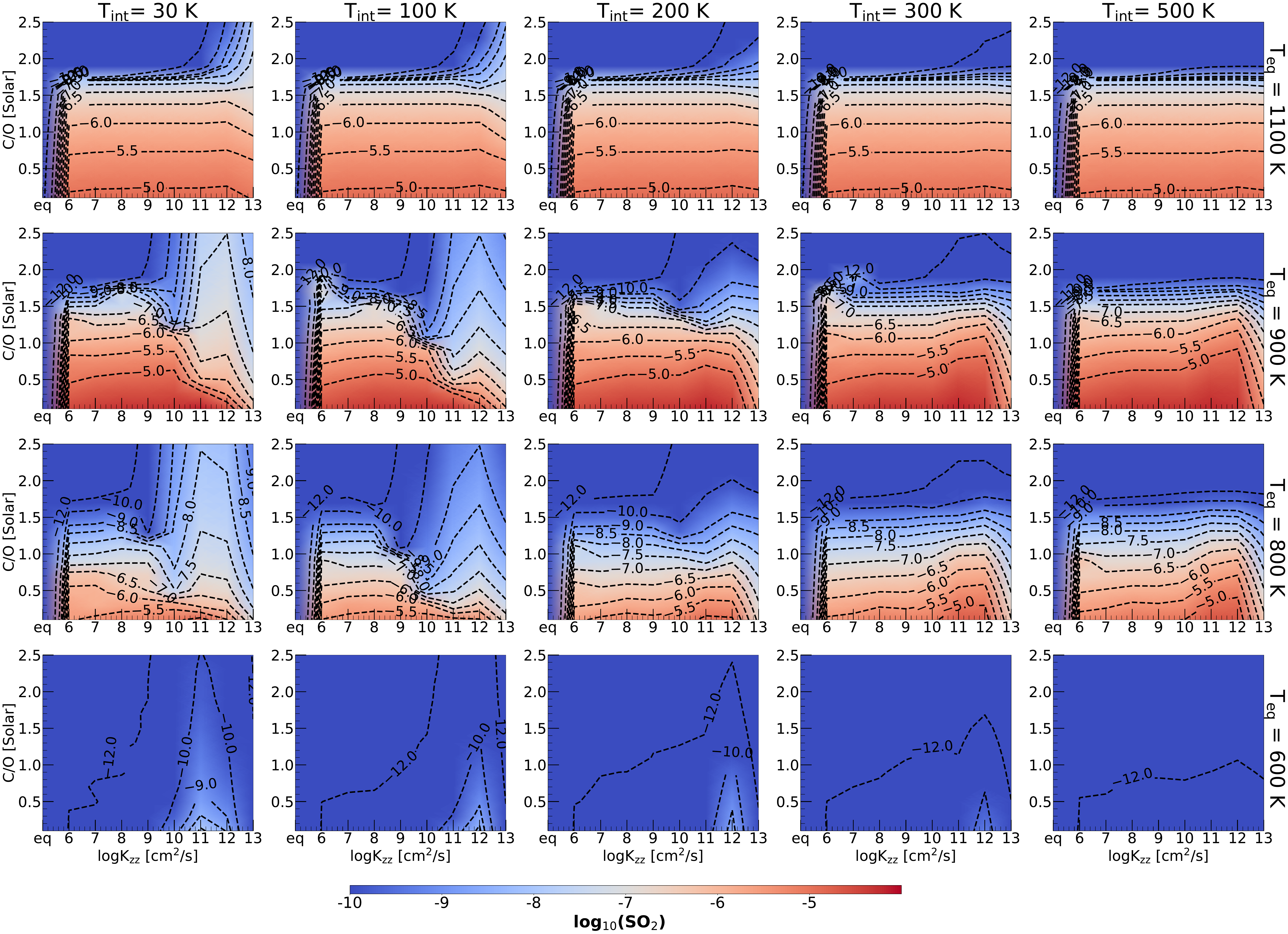}
  \caption{The weighted abundance of {\sotwo} calculated using Equation \ref{eq:xw} has been shown as a heat map as a function of C/O and {\kzz} in each panel. Each row corresponds to a different {\teq} value from 1100 K to 600 K from top to bottom. Each column correspond to a different {\tint} value between 30 K and 500 K from left to right. Note that the C/O has been shown relative to the assumed solar value of 0.458 in this plot.}
\label{fig:so2_profiles}
\end{figure*}

\begin{figure*}
  \centering
  \includegraphics[width=0.8\textwidth]{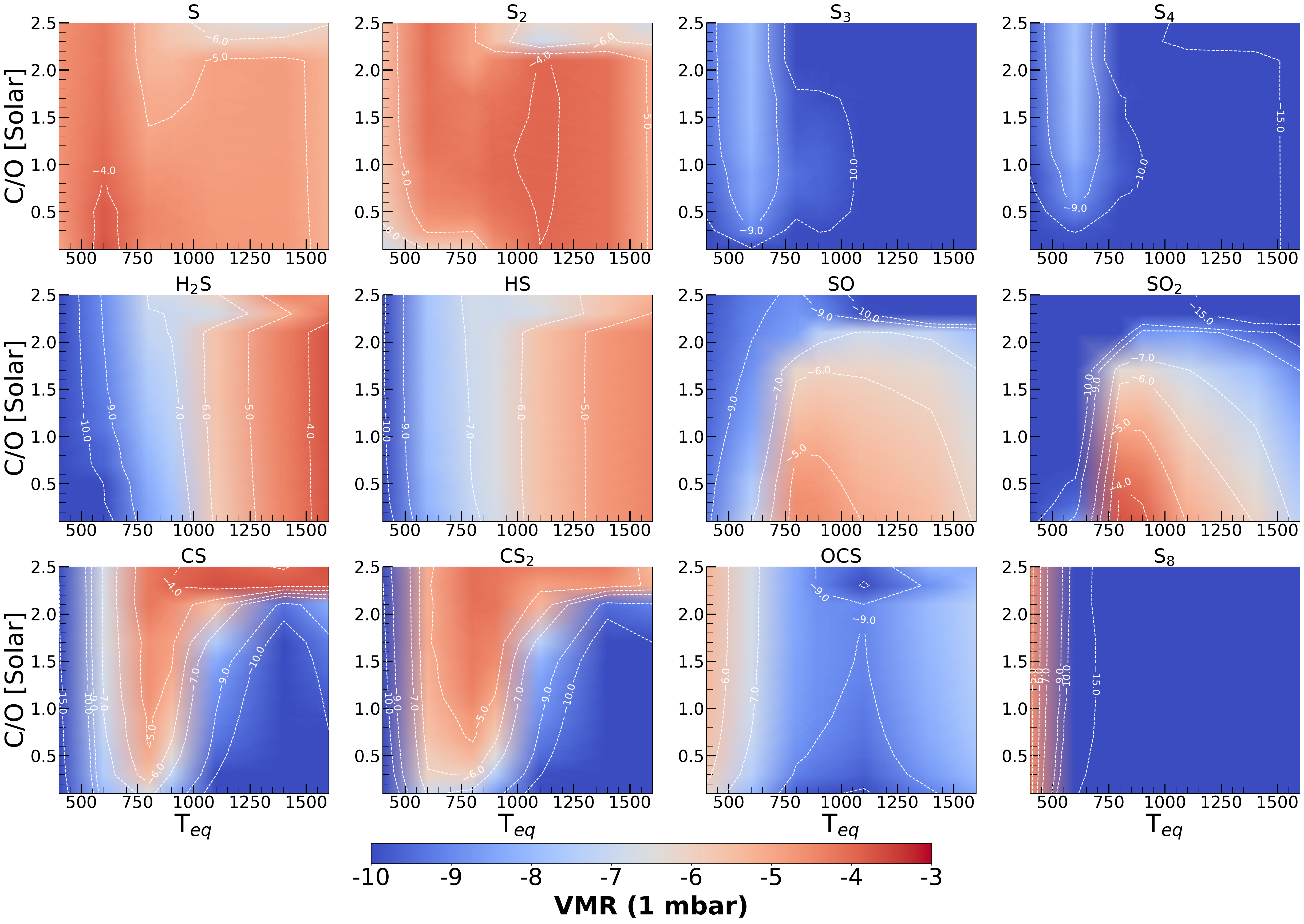}
  \includegraphics[width=0.8\textwidth]{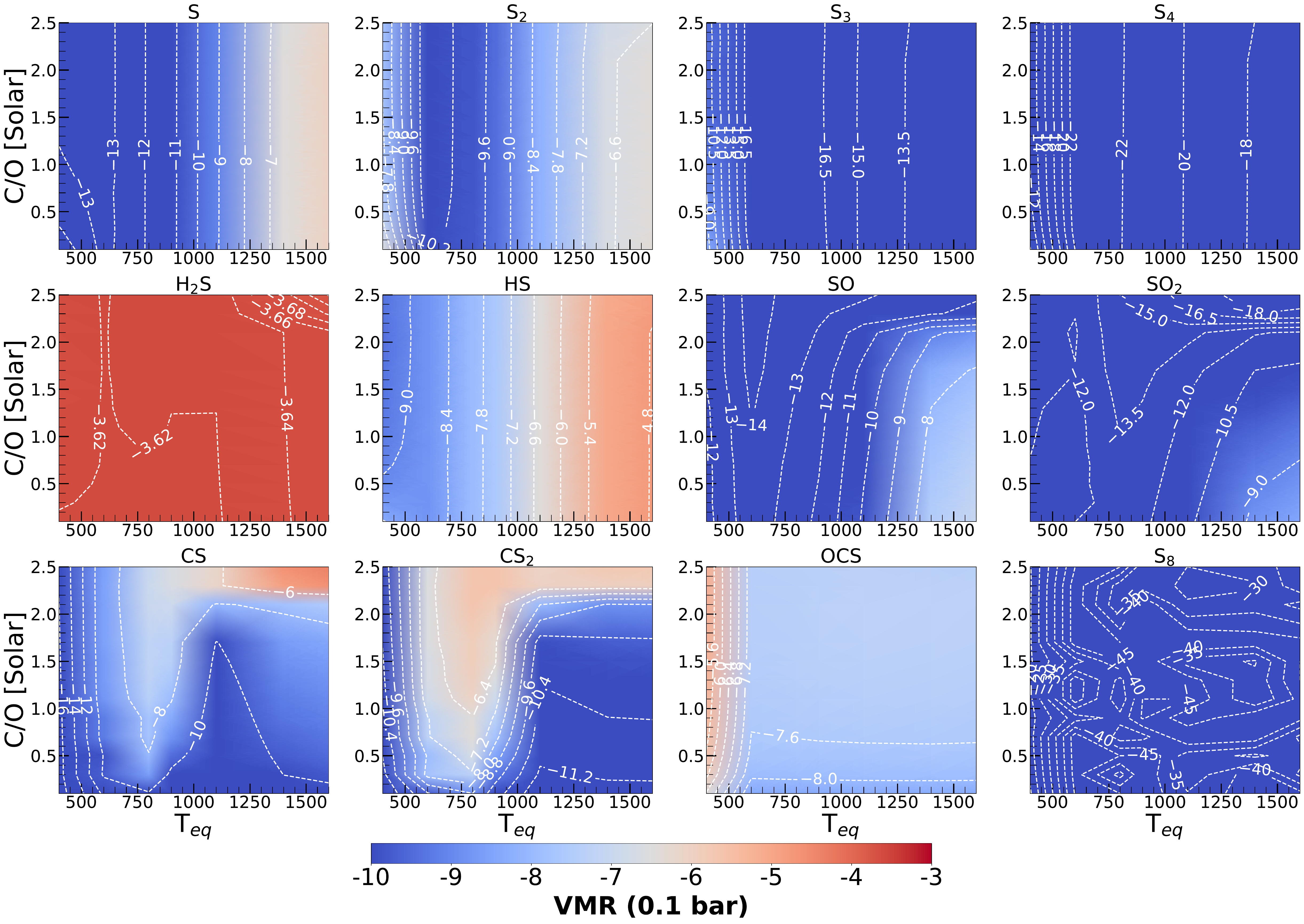}
  \caption{The top set of panels show the abundance of  various Sulfur species at 1 mbar as a heat map as a function of C/O and {\teq} in each panel. Each panel corresponds to a different S- bearing gas. All models shown here have {\tint}=200 K and {\kzz}=10$^9$cm$^2$s$^{-1}$. The lower panels show the abundance of the same species but at 0.1 bar instead. This is to show the abundance of S- bearing gases both at pressures probed by transmission and emission spectroscopy, respetively.}
\label{fig:s_profiles}
\end{figure*}

The detection of {\sotwo} in WASP-39b has garnered great interest in the community due to its photochemical origin. Figure \ref{fig:chemical_profiles} shows how {\sotwo} is produced in a narrow pressure region in the upper atmosphere. But as transmission spectra of gas giant planets probes this particular pressure region as well, {\sotwo} features can be easily observed in transmission spectroscopy. Figure \ref{fig:so2_profiles} shows the $X_w$ for {\sotwo}. At all {\teq} values except for 600 K, {\sotwo} shows a sharp increase from thermochemical equilibrium to disequilibrium chemistry cases. For {\teq}= 1100 K, the {\sotwo} abundance only depends on C/O and shows a small dependence on {\kzz}. For lower C/O values than 1.7$\times$solar, the {\sotwo} abundance increases very slowly with increasing {\kzz} for {\kzz\ $\le$ 10$^{12}${\cms}} followed by a small decrease for higher {\kzz} values. The nearly horizontal equal abundance contours in the top row of Figure \ref{fig:so2_profiles} shows that the {\sotwo} abundance is much more sensitive to C/O than {\kzz} at {\teq}=1100 K. The {\sotwo} abundance shows a gradual decrease with increasing C/O, which is expected as the {\sotwo} molecule contains two O- atoms and no C- atoms. The decrease of {\sotwo} with increasing C/O is particularly rapid when C/O$\ge$1.7$\times$solar for these hot planets, similar to {\cotwo}. This is again because of the lack of enough O- atoms to form {\sotwo} in these very C- rich atmospheres. For these hotter planets, {\sotwo} also remains independent of {\tint}. 

For 800 K$\le${\teq}$\le$ 900 K, the {\sotwo} abundance is overall higher than the {\teq}=1100 K. The {\sotwo} abundance starts to show strong dependence on {\kzz} at this {\teq}, especially for high {\kzz} values. This dependence particularly sets in when the {\kzz} is higher than another ``transition" value of {\kzz}. Below this transition value, the {\sotwo} shows little to no dependence on {\kzz}, but once the {\kzz} is higher than this value, {\sotwo} abundance shows a rapid decline with increasing {\kzz}. Figure \ref{fig:so2_profiles} second row shows that this transition {\kzz} depends on both C/O and {\tint}. For example, at {\teq}=900 K, this ``transition" {\kzz} can be identified to be around 10$^{10}${\cms} when {\tint}=30 K. But this ``transition" {\kzz} is higher at {\kzz=10$^{12}${\cms}} for {\tint}=300 K. Qualitatively similar behavior is also exhibited in the {\teq}=800 K models. {\sotwo} shows another interesting behavior for {\tint}$\le$200 K where  the {\sotwo} shows a rapid increase with {\kzz} when C/O$\ge$1.5$\times$solar and {\kzz} is between 10$^{10}$-10$^{12}${\cms}. The {\sotwo} abundance then declines rapidly when {\kzz} is increased further, which is the effect discussed just above. This rapid increase of {\sotwo} with increasing {\kzz} before showing a decline occurs for C/O$\ge$1.0$\times$solar in the {\teq}=800 K models shown in the third row of Figure \ref{fig:so2_profiles}.

The {\sotwo} abundance in the photosphere of the {\teq}= 600 K models declines by several orders of magnitude than the {\teq}=800 K models. This decline is close to 6-7 orders of magnitude for values of {\kzz} lower or higher than a narrow range of {\kzz} between 10$^{10}$$\le${\kzz}$\le$10$^{12}${\cms}, when {\tint}=30 K. For this narrow range of {\kzz} values, the decline is only about by $\sim$ 1-2 order of magnitudes. For higher {\tint} values, this narrow range of {\kzz} where {\sotwo} is still relatively more abundant than the rest of the parameter space, still exists. However, this region of the parameter space shrinks progressively along both the C/O and {\kzz} direction with increasing {\tint}. This shows that below {\teq}$\le$600 K, it is very difficult to photochemically produce a detectable amount of {\sotwo}, unless the atmospheric metallicity is much higher than 10$\times$ solar. The depletion of {\sotwo} at cool $T_{\rm eq}$ is consistent with previous work and is attributed to the depletion of OH radicals that are needed to form {\sotwo} \citep{tsai23_S}

{\sotwo} is not the only important S- bearing molecule in these atmospheres. Figure \ref{fig:s_profiles}, upper set of panels, shows the volume mixing ratio of various S-bearing molecules as a function of {\teq} and C/O at a fixed {\kzz}=10$^9$cm$^2$s$^{-1}$ and {\tint}=200 K. The volume mixing ratios at 1 mbar are shown. The right-most panel in the middle row of Figure \ref{fig:s_profiles} shows that the {\sotwo} abundance in the photosphere declines sharply when {\teq}$\le$ 600 K or C/O $\ge$2$\times$solar. The same behaviour is also shown by SO. Figure \ref{fig:s_profiles} also shows that SO is an almost equally abundant product of photochemistry expected in the same part of the parameter space as {\sotwo}. Comparing the SO and {\sotwo} panels in Figure \ref{fig:s_profiles} also shows that SO declines slower with increasing {\teq} compared to {\sotwo}. As both SO and {\sotwo} decline at high C/O$\ge$2$\times$solar, most of the S- in the photosphere goes into CS instead. With the decline of SO and {\sotwo} at {\teq}$\le$ 600 K, CS and CS$_2$ also start to carry a large fraction of the S- inventory. This is particularly interesting in the context of the recent tentative detection of CS$_2$ in the atmosphere of TOI-270 d with {\it JWST} \citep{benneke24,holmberg24}, which has a {\teq} between 350-380 K but is inferred to be much more metal rich than 10$\times$solar.

Among the other S$_x$ kind of molecules S$_2$ seems the most abundant. Molecules like S$_3$ and S$_4$ are at least 4-6 orders of magnitude less abundant than S$_2$. Interestingly, the abundance of S$_8$ is almost negligible for {\teq}$\ge$600 K, but Figure \ref{fig:s_profiles} shows that S$_8$ and OCS shows a very sharp increase in abundance for {\teq}$\le$400 K.  Figure \ref{fig:s_profiles} shows that among the H$_{x}$S gases, H$_2$S is more abundant than HS at 1 mbar, especially for {\teq}$\ge$ 1200 K. We note that S- polymerization kinetics is still not very well understood \citep[e.g.,][]{zahnle16} and there is also significant uncertainty in the OCS recombination rates \cite[e.g.,][]{ranjan20,tsai21}.

Figure \ref{fig:s_profiles}, lower set of panels, shows the abundance of the same gases but at 0.1 bar instead. It is clear that almost all of the S- is present in the form of H$_2$S with a minor amount in the form of HS in the deeper atmospheres. The abundance of gases like SO, {\sotwo}, and CS, which are the dominant S- bearers near the pressures probed by transmission spectroscopy, are much lower at these deeper pressures. This is expected as these gasses are photochemically enhanced at the smaller pressures. Interestingly, even though CS is more abundant than CS$_2$ in the upper atmosphere probed by transmission spectroscopy, CS$_2$ appears to be more abundant molecule in the deeper atmosphere probed by emission spectroscopy. The most abundant S- carrying gas after H$_2$S in the deeper atmosphere is HS for {\teq}$\ge$1000 K. But below {\teq}$\le$1000 K, CS$_2$ becomes the second-most abundant S- carrying gas after H$_2$S in the deep atmosphere. For {\teq} lower than 500 K, OCS replaces CS$_2$ as the second most S- carrying gas after H$_2$S.

\subsection{Exploring how abundances depend on {\kzz}, C/O, and {\tint} across different metallicities}\label{sec:met}

\begin{figure*}
  \centering
  \includegraphics[width=0.85\textwidth]{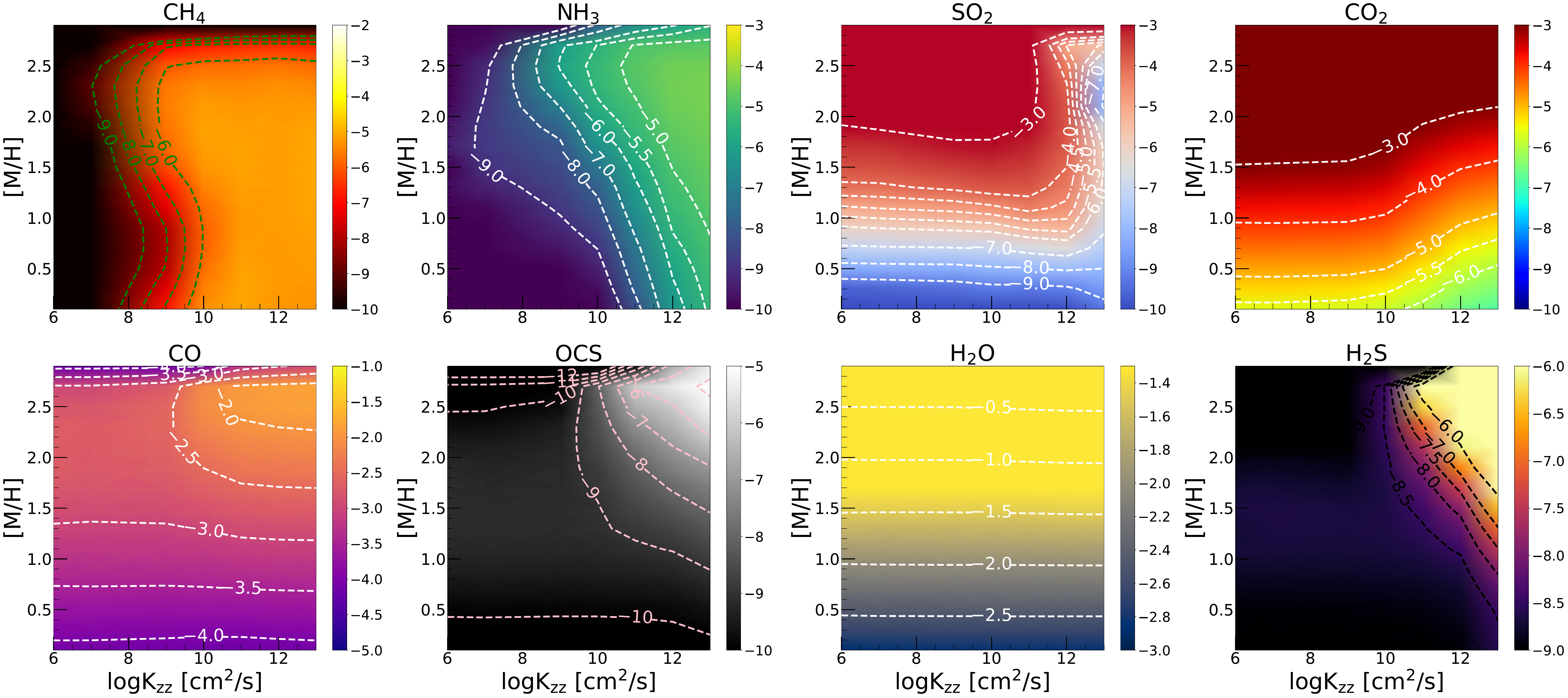}

  \caption{Each panel shows how $X_w$ for a different gas varies with [M/H] and {\kzz} as a heat map. All the models shown here have  {\teq}= 800 K, {\tint}= 200 K, and C/O= 0.1$\times$solar.}
\label{fig:met1}
\end{figure*}

\begin{figure*}
  \centering
  \includegraphics[width=0.85\textwidth]{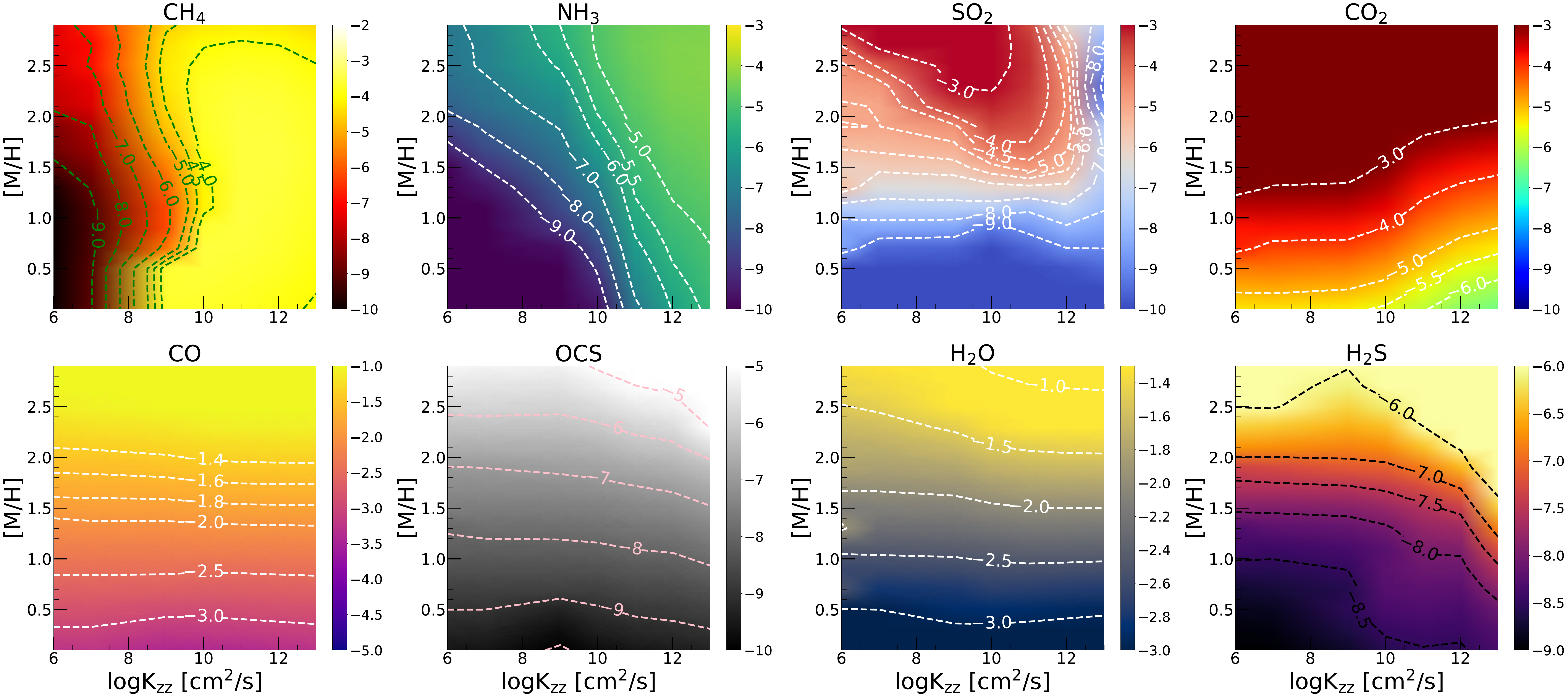}

  \caption{Each panel shows how $X_w$ for a different gas varies with [M/H] and {\kzz} as a heat map. All the models shown here have  {\teq}= 800 K, {\tint}= 200 K, and C/O= 1$\times$solar.}
\label{fig:met2}
\end{figure*}

Metallicity scales the abundance of C-, O-, N-, and S- relative to H in our modeling framework. \S\ref{sec:cbyo} showed that C/O can be degenerate with {\kzz} and {\tint} for most gases at small or big parts of the parameter space. Here, we explore how atmospheric metallicity influences photospheric abundance of gases across the same parameter space. We use models for a planet with {\teq}= 800 K for this purpose with a {\tint}=200 K.

Figure \ref{fig:met1} shows $X_w$ for {\meth}, {\amon}, {\sotwo}, {\cotwo}, {\co}, OCS, {\water}, and H$_2$S as a function of [M/H] and {\kzz} in the different panels. These models are shown for a very O- rich atmosphere with C/O=0.1$\times$ solar whereas Figure \ref{fig:met2} shows the same quantities but for C/O=1$\times$solar. The {\meth} panel of Figure \ref{fig:met1} shows that at C/O=0.1$\times$solar, increasing {\kzz} leads to increase in {\meth} when {\kzz}$\le$10$^{10}${\cms}. This holds true at C/O=1$\times$solar in Figure \ref{fig:met2} too. The overall {\meth} abundances are much lower in the C/O=0.1$\times$solar than the C/O=1$\times$solar, as expected. The equal abundance contours for {\meth} in the C/O=0.1$\times$solar case (Figure \ref{fig:met1}) are nearly vertical which suggests that the photospheric {\meth} abundance varies a lot with changing {\kzz}, but not much with [M/H]. This holds true at C/O=1$\times$solar as well (Figure \ref{fig:met2}) but only for [M/H]$\le$+0.5 and [M/H]$\ge$+1.0. For intermediate metallicities, Figure \ref{fig:met2} shows that {\meth} abundance can depend both on [M/H] and {\kzz}, increasing as both of these parameters are increased. For metallicities higher than [M/H]=+2.5, {\meth} abundance starts to decrease rapidly with increasing metallicity for the very O- rich case shown in Figure \ref{fig:met1}. This behavior is also present but to a smaller extent in the solar C/O models in Figure \ref{fig:met2}. Note that our models do not account for graphite saturation \citep{moses2013b}, which could draw down the {\meth} concentration and impact other carbon-bearing species in Figure \ref{fig:met2} for high metallicities ([M/H] $> 2$). 

{\amon} abundance in Figures \ref{fig:met1} and \ref{fig:met2} also show increase with increasing {\kzz}. The equal abundance contours of {\amon} are along the diagonal direction, which suggests that the {\amon} abundance shows a dependence on both [M/H] and {\kzz}. For a given {\kzz}, higher [M/H] values produces more photospheric {\amon} than lower [M/H] values, except for [M/H]$\ge$+2.5 models with C/O=0.1$\times$solar. Both {\sotwo} and {\cotwo} show much stronger [M/H] dependence than {\kzz} dependence, unlike {\meth} and {\amon}. This is expected as both {\sotwo} and {\cotwo} require one C- or S- and two O- atoms per molecule. It is interesting that for very O- rich atmospheres (Figure \ref{fig:met1}), the rate of increase of {\sotwo} with increasing [M/H] slows down considerably for high [M/H] values. This behavior also happens in the C/O=1$\times$solar atmospheres in Figure \ref{fig:met2}, but not to the extent present in Figure \ref{fig:met1}. The metallicity dependence of {\sotwo} only dissappears when the {\kzz} is higher than a threshold value. Beyond this value, the {\sotwo} abundance shows a sharp decline with increasing {\kzz} without much [M/H] dependence. In Figure \ref{fig:met1}, this threshold appears to be close to 10$^{12}${\cms}, whereas in Figure \ref{fig:met2}, the threshold is near 10$^{11}${\cms}. The {\cotwo} abundance also shows some {\kzz} dependence for high {\kzz} values, but not to the extent of the sharp change shown by {\sotwo}. For a given [M/H], the {\cotwo} abundance shows a slight increase with increasing {\kzz} when the {\kzz} is higher than 10$^{11}${\cms}. For lower {\kzz} values, the {\cotwo} abundance solely depends on [M/H] (for a fixed C/O).

Like {\cotwo}, {\co} shows little to no dependence on {\kzz} and strong dependence on metallicity. The main reason behind this is the relatively higher {\tint}=200 K of these models. Interestingly, {\co} abundances in Figure 
\ref{fig:met1} are generally lower than the {\co} abundances in Figure \ref{fig:met2}. This is the case as {\co} contains equal numbers of C- and O- atoms, so its abundance is diminished in a very O- rich and C- poor atmosphere. {\co} also shows some minor dependance on {\kzz} in the very O- rich atmospheres (Figure \ref{fig:met1}) when [M/H]$\ge$+2.0. OCS also shows a similar behavior as {\co}, where its abundance is diminished in the very O- rich/C- poor atmospheres compared to solar C/O atmospheres. The OCS abundance seems to primarily depend on atmospheric [M/H] unless {\kzz} is high. For high {\kzz}, the OCS abundance starts to increase with increasing {\kzz} and this increase is particularly rapid for significantly metal-enriched atmospheres. This behavior is also mimicked by H$_2$S, where its much less abundant in the O- rich/C- poor atmosphere than the solar C/O atmosphere. The photospheric abundance of H$_2$S also shows a rapid increase with increasing {\kzz} beyond a threshold value, particularly in high [M/H] atmospheres. Figures \ref{fig:met1} and \ref{fig:met2} show that {\water} abundance in the photosphere only depends on [M/H], with almost no {\kzz} dependence at all. 

\subsection{Chemical Transitions with {\teq}}\label{sec:teq}
\begin{figure*}
  \centering
  \includegraphics[width=1\textwidth]{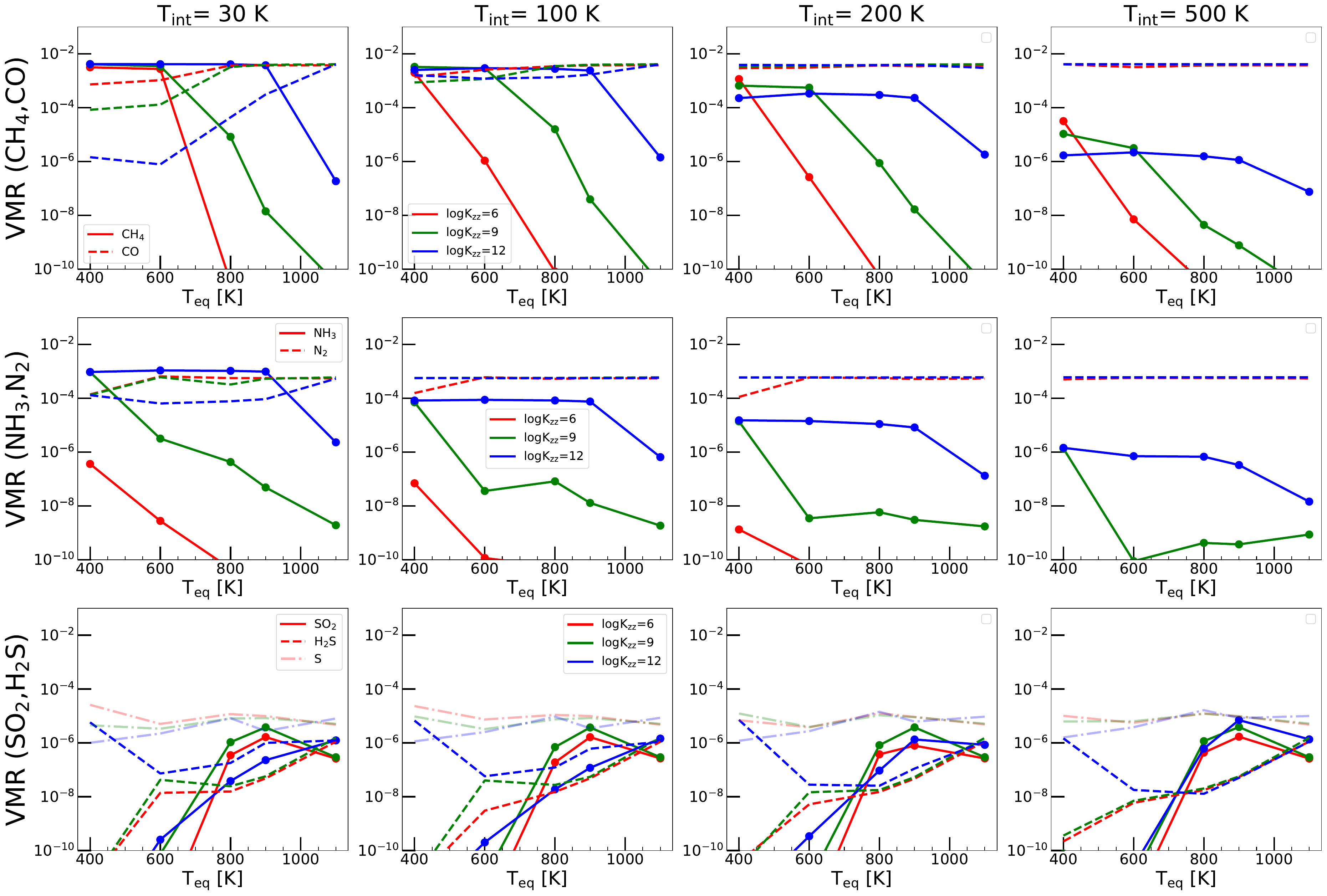}
  \caption{The variation of photospheric abundance of key atmospheric gases as a function of {\teq} is shown here. Each column shows the variation for a different value of {\tint}. The top row shows the evolution of {\meth} and {\co} abundance with {\teq}. The middle and last row shows the same for {\ntwo}-{\amon}, and H$_2$S-{\sotwo}-S, respectively. In each panel, the three colored lines shows the abundances for a three different values of {\kzz}. The C/O has been fixed to be 1$\times$solar in these models.  }
\label{fig:teqabun}
\end{figure*}

\citet{fortney20} and \citet{ohno232} presented how fundamental chemical transitions in the atmosphere, like from being {\co} rich to {\meth} rich, can depend on both {\tint} and {\kzz} in addition to {\teq}. Recently, \citet{bell23} reported the discovery of {\meth} in the atmosphere of WASP-80b detected in both transmission and emission spectroscopy of the planet with {\it JWST}. There have been efforts to measure the {\teq} value below which {\meth} first starts to appear in observations of exoplanetary atmospheres. But this can be a complex function of {\tint}, {\kzz}, C/O, and [M/H] \citep[e.g.,][]{fortney20}. 

Figure \ref{fig:teqabun} presents these transitions/behavior from our photochemical 1D chemical kinetics grid for three pairs of molecules -- {\co}-{\meth}, {\ntwo}-{\amon}, and H$_2$S-{\sotwo}-S at [M/H]=+1.0. The top row panels in Figure \ref{fig:teqabun} shows how {\co} and {\meth} photospheric abundance change with decreasing {\teq} for {\tint}=30 K, 100 K, 200 K, and 500 K from left to right, respectively. Each panel in Figure \ref{fig:teqabun} shows three colored solid lines denoting the {\meth} abundance for three different {\kzz} values, whereas the dotted lines indicate the {\co} abundance in each case. It is clear that the upper atmosphere probed by transmission spectroscopy can only have {\meth}$\ge${\co} in planets with cold interiors {\tint} $\sim$ 100 K or lower. For hotter interiors, {\co} always remains the majority C- carrier molecule with higher abundance than {\meth}. This transition was seen to be happening at or below {\tint}=150 K too by \citet{fortney20}, even though they used quench-time approximation for approximating the effect of {\kzz} on chemistry and also did not treat photochemistry in their models. We note that this threshold of \tint$=$150 K was found at 3$\times$solar metallicity in \citet{fortney20} instead of the 10$\times$solar metallicity assumed in our analysis. Figure \ref{fig:teqabun} also shows that the {\teq} value where this transition from {\co} rich to {\meth} rich occurs is a very strong function of {\kzz}. When {\kzz}=10$^{12}${\cms}, this transition happens near {\teq}= 1000 K for atmospheres with {\tint}$\le$100 K. But, for {\kzz}=10$^{9}${\cms} or {\kzz}=10$^{6}${\cms} , atmospheres can remain {\co} rich for {\teq} above 600 K. When {\tint}= 500 K, shown in the right most column of Figure \ref{fig:teqabun}, the {\meth} abundance is still an order of magnitude lower than {\co} even for {\teq}= 400 K.

The second row of Figure \ref{fig:teqabun} shows the same quantities as the first row but for {\amon} (solid lines) and {\ntwo} (dashed lines) instead. Figure \ref{fig:teqabun} second row shows that the {\amon} can only become the major N- carrier if the {\tint} is 30 K. Even for such cold interior, the mixing needs to be high to make {\amon} carry most of the N- atoms in the photosphere. For example, the {\amon} in the photosphere still remains almost two orders of magnitude lower than that of {\ntwo} abundance if {\kzz}=10$^{6}${\cms}, even if {\tint}=30 K. This result is also consistent with the findings of both \citet{fortney20}
and \citet{ohno232}. \citet{ohno23} found that {\amon} can become the dominant carrier of N- atoms in giant planets only in a small parameter space where the planet has a low mass and is very old, such that the interior is sufficiently cool and vertical mixing can dredge the {\amon} up to the upper atmosphere.

The last row of Figure \ref{fig:teqabun} shows the behavior of the abundance of three major S- carrying species -- H$_2$S (dashed lines), {\sotwo} (solid lines), and S (faded dash-dotted lines) as a function of {\teq}. The notable thing in Figure \ref{fig:teqabun} is that neither {\sotwo} or H$_2$S carries majority of the S- inventory of the atmosphere, not atleast at [M/H]=+1.0. Instead, neutral S- atoms form a much more substantial fraction of the total S- inventory in the upper atmosphere than {\sotwo}. These are shown with the faded dash dotted lines in all the panels and the same phenomenon is also shown in Figure \ref{fig:s_profiles}. This was seen in previous work \citep[e.g.,][]{tsai23,tsai23_S} as well. Figure \ref{fig:teqabun} also shows how the {\sotwo} abundance peaks near {\teq}= 900 K but gradually decreases when the {\teq} increases further. But the decrease of {\sotwo} with decreasing {\teq} is much more rapid as was also seen in Figure \ref{fig:s_profiles}.

\subsection{Sensitivity of Transmission Spectrum to {\kzz} and {\tint}}\label{sec:spectra}

\begin{figure*}
  \centering
  \includegraphics[width=0.8\textwidth]{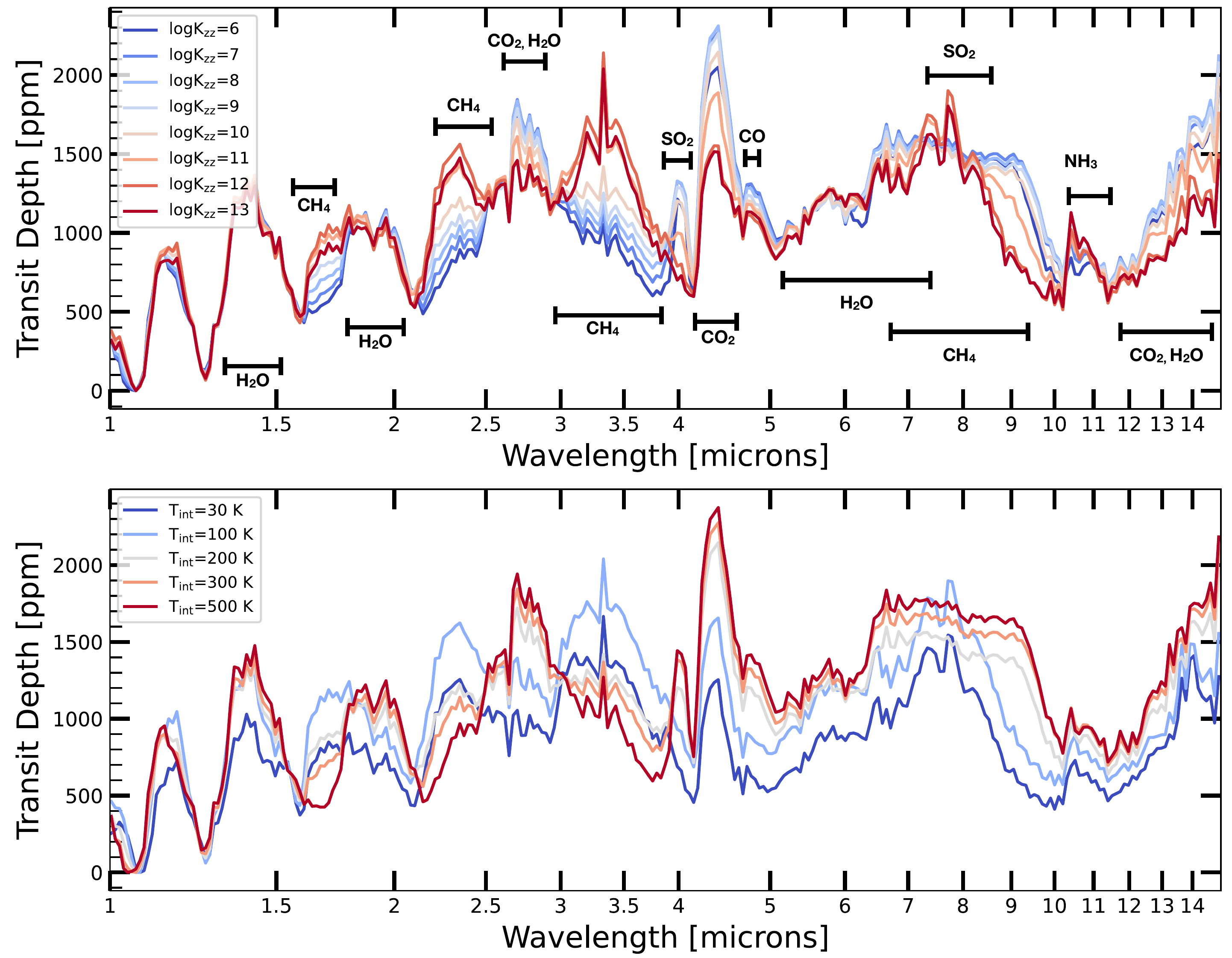}
  \caption{The top panel shows the variation of transmission spectra with changing {\kzz} while the bottom panel shows the variation of the transmission spectra of the planet with changing {\tint}. The models shown here are for {\teq}=800 K, [M/H]=+1.0, and C/O=1$\times$solar. The {\tint} has been fixed at 200 K for the top panel while the {\kzz} has been fixed at 10$^{10}${\cms} in the bottom panel. Note that the transmission spectra have been renormalized in such a manner that the minimum transit depth of each model is set at zero.}
\label{fig:TD_ex}
\end{figure*}

While we have presented our results on the detailed chemical abundances at different parts of the parameter space till now, here we present how this chemistry affects the observable -- the transmission spectra. The top panel of Figure \ref{fig:TD_ex} shows the transmission spectra computed from the {\teq}= 800 K, {\tint}= 200 K, C/O=1.1$\times$solar, and [M/H]=+1.0 models for a range of different {\kzz} values. It is clear that certain parts of the transmission spectra are quite sensitive to {\kzz} while others are not as much. For example, the {\meth} band between 3-3.5 $\mu$m, and the {\sotwo} band at 4$\mu$m and 7-10 $\mu$m are quite sensitive to {\kzz}. On the other hand, the {\water} bands between 1-1.5 $\mu$m does not show as much sensitivity to {\kzz}.

The lower panel in Figure \ref{fig:TD_ex} shows the sensitivity of the transmission spectrum to {\tint} with a fixed {\kzz}=10$^{10}${\cms}. The {\tint} has been varied from 30 K to 500 K. It is clear that the spectra is very {\meth} dominated for cold interiors with {\tint}=30 K and 100 K. As the {\tint} increases further, the effect of {\meth} on the spectra slowly decreases. Looking at Figure \ref{fig:TD_ex}, it is not exactly clear at how the transmission spectrum responds to changing {\kzz} and {\tint} in different parts of the vast parameter space explored in this work. Therefore, we present those trends in the upcoming section by computing the differential of the transmission spectra at each wavelength bin with changing {\kzz} and {\tint}.

\subsubsection{Dependence of spectra on {\kzz}}
\begin{figure*}
  \centering
  \includegraphics[width=1\textwidth]{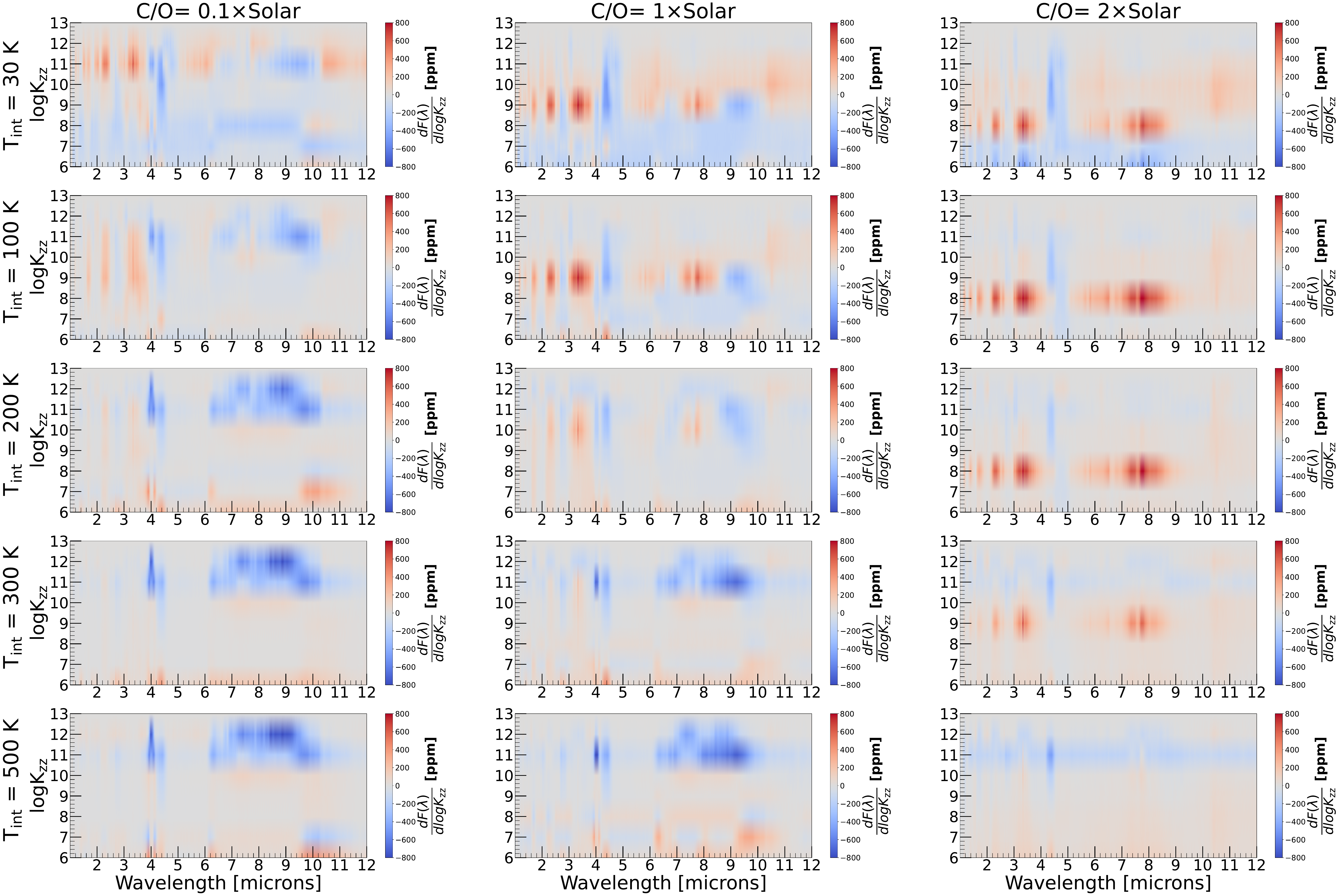}
  \caption{ Sensitivity of the transmission spectra to changing {\kzz} is shown here as a function of {\kzz} and wavelength. Each column corresponds to a different C/O value whereas each row shows models for different {\tint} values. All models shown here have {\teq}=800 K. A brighter red or a brighter blue color indicates that the transmission spectra is very sensitive to changing {\kzz} at that {\kzz} and wavelength value.}
\label{fig:TD_kzz_teq800_cto}
\end{figure*}

\begin{figure*}
  \centering
  \includegraphics[width=1\textwidth]{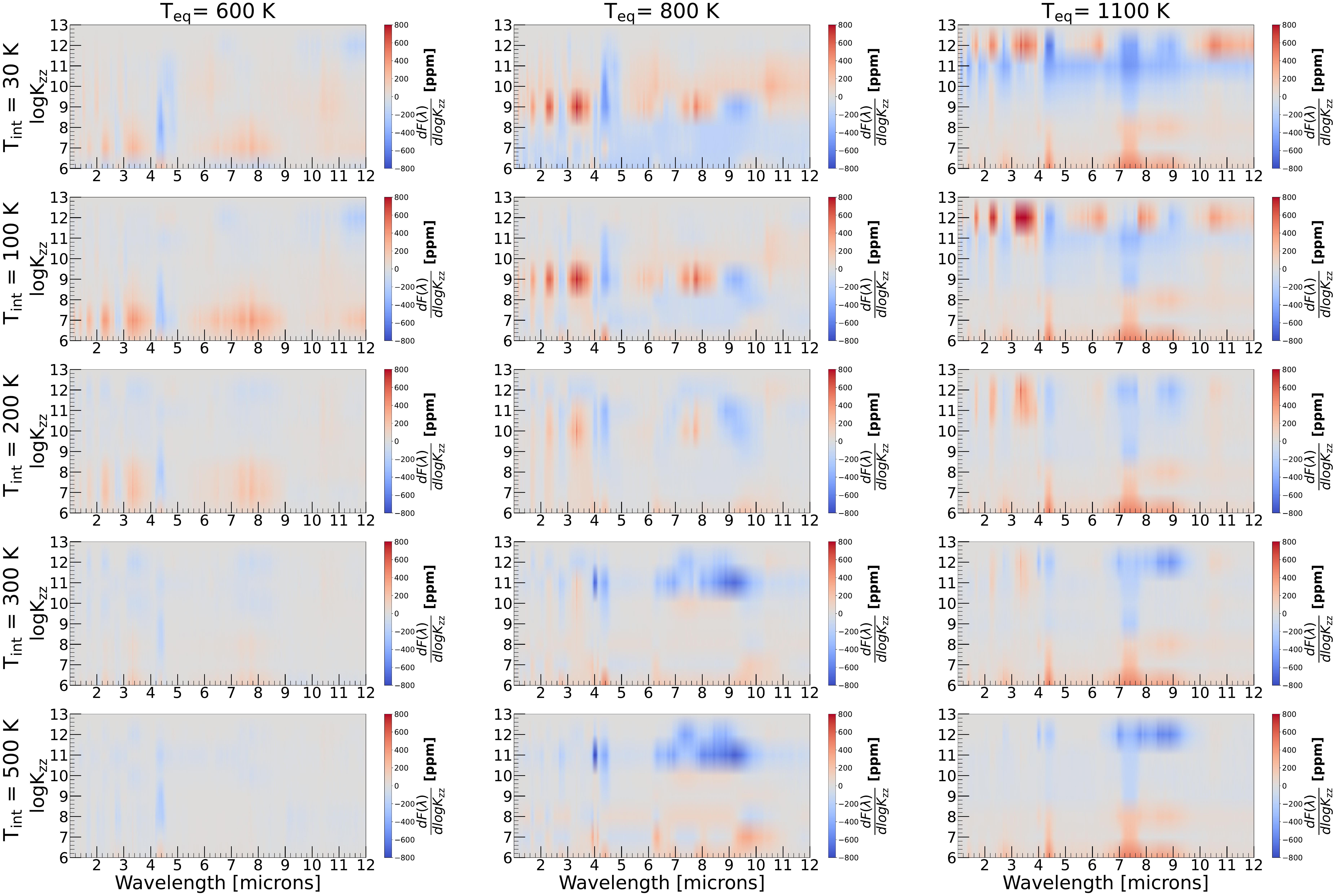}
  \caption{Sensitivity of the transmission spectra to changing {\kzz} is shown here as a function of {\kzz} and wavelength. Each column corresponds to a different {\teq} value whereas each row shows models for different {\tint} values. All models shown here have C/O=1$\times$solar. A brighter red or a brighter blue color indicates that the transmission spectra is very sensitive to changing {\kzz} at that {\kzz} and wavelength value.}
\label{fig:TD_kzz_teq_cto1pt1}
\end{figure*}

To highlight the wavelength ranges where the transmission spectra changes the most when a parameter like {\kzz} is changed, we first renormalise all the transmission spectra across our grid such that the minimum transit depth of each transmission spectra is zero. Then, we calculate the change in the renormalised transit depth within each wavelength bin from one grid point to the next grid point. For example, to highlight which part of the transmission spectra changes the most with a change in {\kzz} near {\tint}=30 K, {\teq}=800 K, C/O=0.1$\times$solar, [M/H]=+1.0, and {\kzz}=10$^{8}${\cms}, we calculate $dF$($\lambda$)/$dlog${\kzz}= ($F$($\lambda$,{\kzz}=10$^{8}$)- $F$($\lambda$,{\kzz}=10$^{7}$))/($log_{10}(10^8)$-$log_{10}(10^7)$), where $F$ denotes the renormalized transmission spectra at the above mentioned grid point parameters. We plot this quantity $dF$($\lambda$)/$dlog${\kzz} in the top left panel of Figure \ref{fig:TD_kzz_teq800_cto} as a function of {\kzz} with a color map. Each row in Figure \ref{fig:TD_kzz_teq800_cto} shows the same quantity for a different {\tint} value while each column corresponds to a different C/O for a fixed {\teq}= 800 K. The $dF$($\lambda$)/$dlog${\kzz} quantity shows how the transit depth in a given wavelength change in parts-per-million (ppm) in a particular part of the parameter space if {\kzz} is changed by one order of magnitude there. We note that the $dF$($\lambda$)/$dlog${\kzz} quantity is sparsely sampled in the {\kzz} direction, and as a result the trends shown in Figure \ref{fig:TD_kzz_teq800_cto} and \ref{fig:TD_kzz_teq_cto1pt1} may appear smoother than expected. This quantity helps us to identify which are the best wavelength range or molecular features to constrain {\kzz} in a particular part of the parameter space.

Figure \ref{fig:TD_kzz_teq800_cto} top row shows that if the interior of the planet has {\tint}=30 K, the {\meth} and {\cotwo} bands near 3.3 $\mu$m and 4.2-4.3$\mu$m, respectively, show the most sensitivity to changing {\kzz}. However, it is clear that the maximum change on the spectra due to changing {\kzz} appears at different {\kzz} values depending on the C/O. For example, in the C/O=0.1$\times$solar panel on the top left, the maximum change in the {\meth} feature appears at high {\kzz} values near 10$^{10}${\cms}. This region of maximum sensitivity of the {\meth} band shifts to lower {\kzz}=10$^{8}${\cms} value when the C/O is 1$\times$solar and to even lower {\kzz}= 10$^{7}${\cms} when the C/O is 2$\times$solar in the top row. However, Figure \ref{fig:TD_kzz_teq800_cto} top row shows that the {\cotwo} band near 4.2-4.3 $\mu$m remains quite sensitive to {\kzz} for almost all {\kzz} values in this cold interior scenario.

As the {\tint} is increased from 30 K to 500 K (moving from the top towards the bottom row in Figure \ref{fig:TD_kzz_teq800_cto}), the sensitivity of the {\meth} bands to {\kzz} slowly diminishes while the {\sotwo} bands near 4$\mu$m and 7-9 $\mu$m become increasingly sensitive to changing {\kzz}. This happens only in the C/O = 0.1$\times$ and 1$\times$solar columns. The sensitivity of the {\sotwo} bands set in particularly for high values of {\kzz} above 10$^{10}${\cms}. The rest of the transmission spectra still remains sensitive to {\kzz} but to a smaller degree. When C/O=2$\times$solar, the {\meth} bands remain the most sensitive parts of the transmission spectra till {\tint}$\le$300 K. The sensitivity of the 4.2-4.3 $\mu$m {\cotwo} slowly diminishes in this case as the interior warms up from a {\tint} of 30 K to 500 K.

Next, we present the trends in $dF$($\lambda$)/$dlog${\kzz} at three different {\teq} values in Figure \ref{fig:TD_kzz_teq_cto1pt1}, instead of the three different C/O values shown in Figure \ref{fig:TD_kzz_teq800_cto}. Each column in Figure \ref{fig:TD_kzz_teq_cto1pt1} shows the trends for {\teq}= 600 K, 800 K, and 1100 K while the C/O has been fixed at 1$\times$solar. Figure \ref{fig:TD_kzz_teq_cto1pt1} shows that transit depths at {\teq}= 600 K are much less sensitive to {\kzz} than those at {\teq}= 800 K and {\teq}= 1100 K. At {\teq}= 600 K, most of the sensitivity of the spectrum arises in the {\cotwo} bands (4.2-4.3 $\mu$m) and {\meth} bands (3-4 $\mu$m and others) at low values of {\tint}. The {\meth} bands show some sensitivity but only when {\kzz} is low whereas the {\cotwo} band shows sensitivity for a larger range of {\kzz} values. When the {\tint} $\ge$ 300 K, there is reduced sensitivity to {\kzz} across the transmission spectrum at {\teq}=600 K. 

The trends in the middle column of Figure \ref{fig:TD_kzz_teq_cto1pt1}
and the middle panel of Figure \ref{fig:TD_kzz_teq800_cto} represent the same set of models, so we don't describe them again here. The {\teq}=1100 K models shown in the right columns of Figure \ref{fig:TD_kzz_teq_cto1pt1} show that the {\meth} bands (between 3-4$\mu$m and others) are particularly sensitive to {\kzz}, especially at high {\kzz} values but this sensitivity diminishes slowly with increasing {\tint} value. The {\cotwo} feature (4.2-4.3 $\mu$m) shows high sensitivity to {\kzz} across all {\kzz} values, but this sensitivity slowly diminishes with increasing {\tint} at {\teq}=1100 K. While the 4 $\mu$m {\sotwo} feature doesn't show much sensitivity to {\kzz} for this {\teq} value, but the 7-9 $\mu$m {\sotwo} is very sensitive to {\kzz} throughout all {\tint} and {\kzz} values for the {\teq}=1100 K case. As it is not possible to present these trends for the whole suite of models within this paper in a wavelength resolved manner, we now move on to showing sensitivity of the transmission spectra to changing {\tint} instead of changing {\kzz}.

\subsubsection{Dependence of spectra on {\tint}}

\begin{figure*}
  \centering
  \includegraphics[width=1\textwidth]{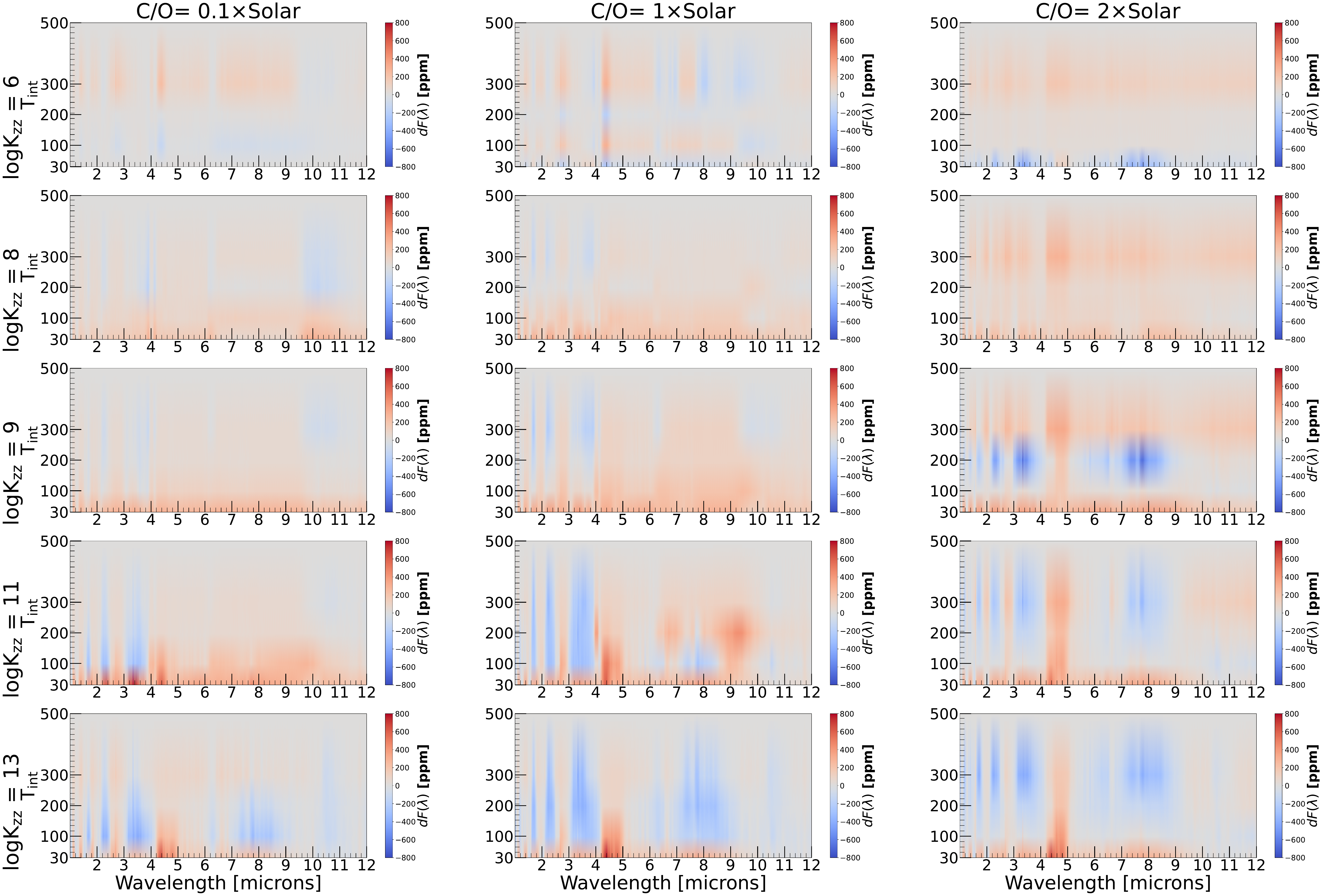}
  \caption{Similar to Figure \ref{fig:TD_kzz_teq800_cto} but the sensitivity of transmission spectra to changing {\tint} is shown here as a function of {\tint} and wavelength, instead of {\kzz} in Figure \ref{fig:TD_kzz_teq800_cto}. Each column corresponds to a different C/O value whereas each row shows models for different {\kzz} values. All models shown here have {\teq}=800 K. A brighter red or a brighter blue color indicates that the transmission spectra is very sensitive to changing {\tint} at that {\tint} and wavelength value.}
\label{fig:TD_tint_teq800_cto}
\end{figure*}

\begin{figure*}
  \centering
  \includegraphics[width=1\textwidth]{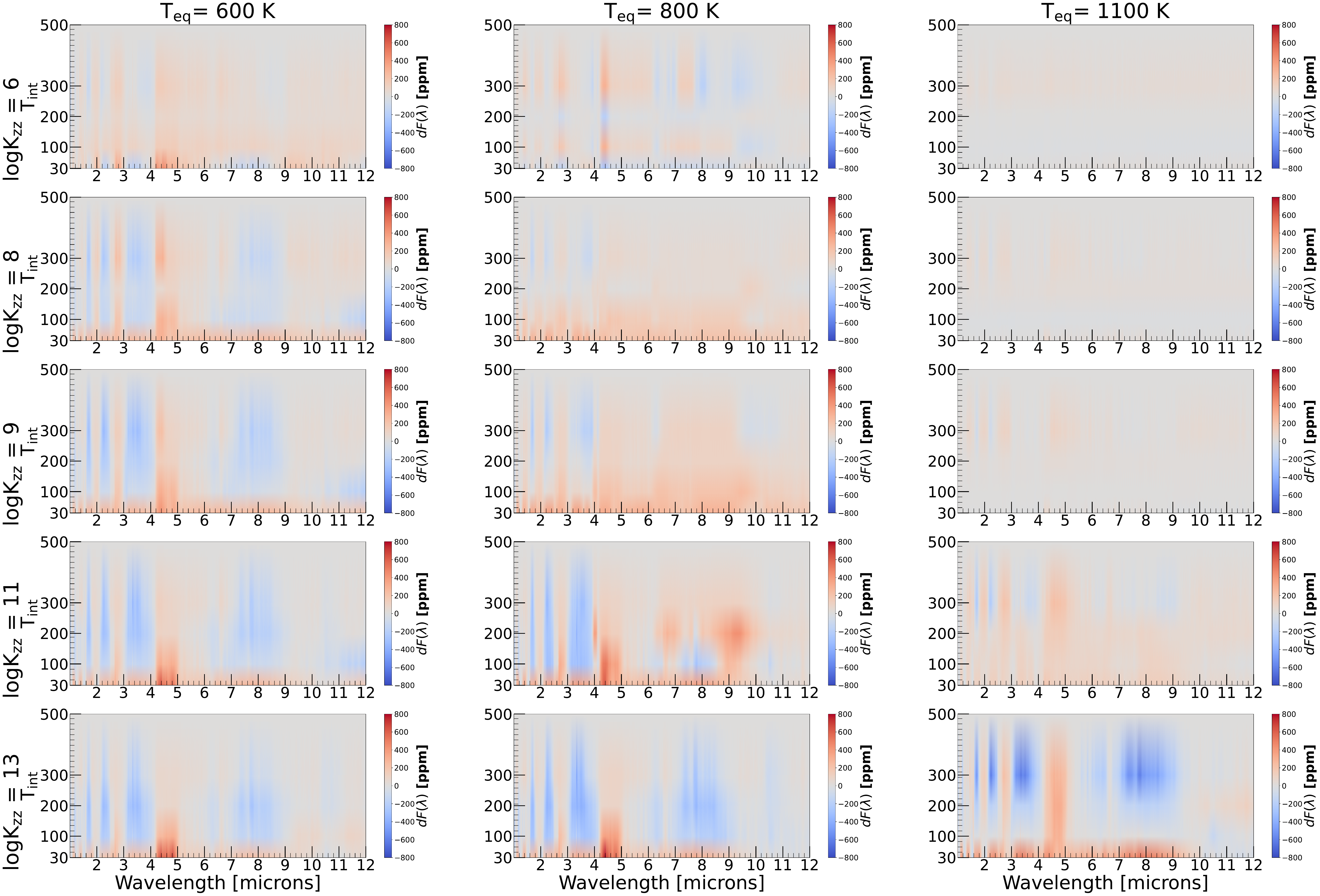}
  \caption{Sensitivity of the transmission spectra to changing {\tint} is shown here as a function of {\tint} and wavelength. Each column corresponds to a different {\teq} value whereas each row shows models for different {\kzz} values. All models shown here have C/O=1$\times$solar. A brighter red or a brighter blue color indicates that the transmission spectra is very sensitive to changing {\tint} at that {\tint} and wavelength value.}
\label{fig:TD_tint_teq_cto1pt1}
\end{figure*}

Figure \ref{fig:TD_tint_teq800_cto} presents the sensitivity of the transmission spectra to changing {\tint} when the {\tint} changes within our five {\tint} grid points of 30 K, 100 K, 200 K, 300 K, and 500 K. As our {\tint} grid points are non-uniformly spaced, unlike for {\kzz}, we plot the quantity $dF$($\lambda$) here, which is simply the difference in the renormalized transit depths between two {\tint} values. Like Figure \ref{fig:TD_kzz_teq800_cto}, each column of Figure \ref{fig:TD_tint_teq800_cto} also represents different C/O values. But unlike Figure \ref{fig:TD_kzz_teq800_cto}, each row of Figure \ref{fig:TD_tint_teq800_cto} represents a different value of {\kzz} with {\kzz} increasing from the top to the bottom row. It is clear that for the very O- rich case of C/O=0.1$\times$solar shown in the left columns, the sensitivity of the transmission spectra is
very subtle to changing {\tint}, especially for low {\kzz} values. But as the {\kzz} values increase, the {\meth} band between (3-4 $\mu$m) starts to show high sensitivity to changing {\tint}. 


The C/O=1$\times$solar models shown in middle column of Figure \ref{fig:TD_tint_teq800_cto} show a relatively higher level of sensitivity to changing {\tint} value compared to the C/O=0.1$\times$solar models, especially for {\kzz}$\ge$10$^{9}${\cms}. Most of the sensitivity here is shown in the {\meth} bands (3-4 $\mu$m and others) and the sensitivity of the {\meth} bands to changing {\tint} increases with increasing {\kzz} at this C/O value. The {\cotwo} absorption bands also show high sensitivity to changing {\tint} values, especially for higher {\kzz} values. For the C/O=2$\times$solar models shown in Figure \ref{fig:TD_tint_teq800_cto}, most of the sensitivity of the transit depth is mainly seen in the {\meth} bands (3-4$\mu$m and others) and the the {\cotwo} band between 4.2-4.3 $\mu$m.

Figure \ref{fig:TD_tint_teq_cto1pt1} shows the sensitivity of the transmission spectra at three different {\teq} values shown in each column for 600 K, 800 K, and 1100 K from left to right, respectively. The C/O ratio has been fixed to 1$\times$solar for this purpose. It is clear that the overall sensitivity of the transmission spectra to changing {\tint} diminishes with increasing {\teq}. For {\teq}=600 K, the {\meth} and {\cotwo} bands show the most sensitivity to changing {\tint}. This sensitivity increases progressively from low {\kzz} to high {\kzz} at {\tint}=600 K. The {\teq}=1100 K models shown in the right column show much lower sensitivity of the transit depths to changing {\tint} than the {\teq}=600 K and {\teq}=800 K models. This is especially true for {\kzz}$\le$10$^{9}${\cms}. However, when {\kzz} is higher than this value, some {\tint} sensitivity starts to appear in the spectrum. This sensitivity appears primarily in the {\meth}, {\cotwo} and {\co} bands (4.5-4.8 $\mu$m) in the {\kzz}=10$^{11}$ and 10$^{13}${\cms} cases.

Our sensitivity analysis of the transmission spectra reveals that the different wavelength regions in the transmission spectra show complex sensitivity with changing {\kzz} and {\tint} in different parts of the C/O, {\teq}, {\kzz}, and {\tint} parameter space. This suggests that the precision to which parameters like {\tint} and {\kzz} can be constrained for giant transiting planets depends on where that planet lands in this multi-dimensional parameter space. This kind of analysis can also be used to carefully design the wavelength region and the required signal-to-noise to achieve a certain science goal (like constraining {\tint} to a certain precision) with instruments like {\it JWST}. We avoid presenting similar analysis for eclipse spectroscopy as we chose to ignore the disequilibrium chemistry processes self-consistently within our {\tp} profile calculation and that would matter for emission spectra of these planets to a much larger extent than the transmission spectra analysis presented here.  

\subsection{Precursors to Soot/Haze}\label{sec:haze}

\begin{figure*}
  \centering
  \includegraphics[width=1\textwidth]{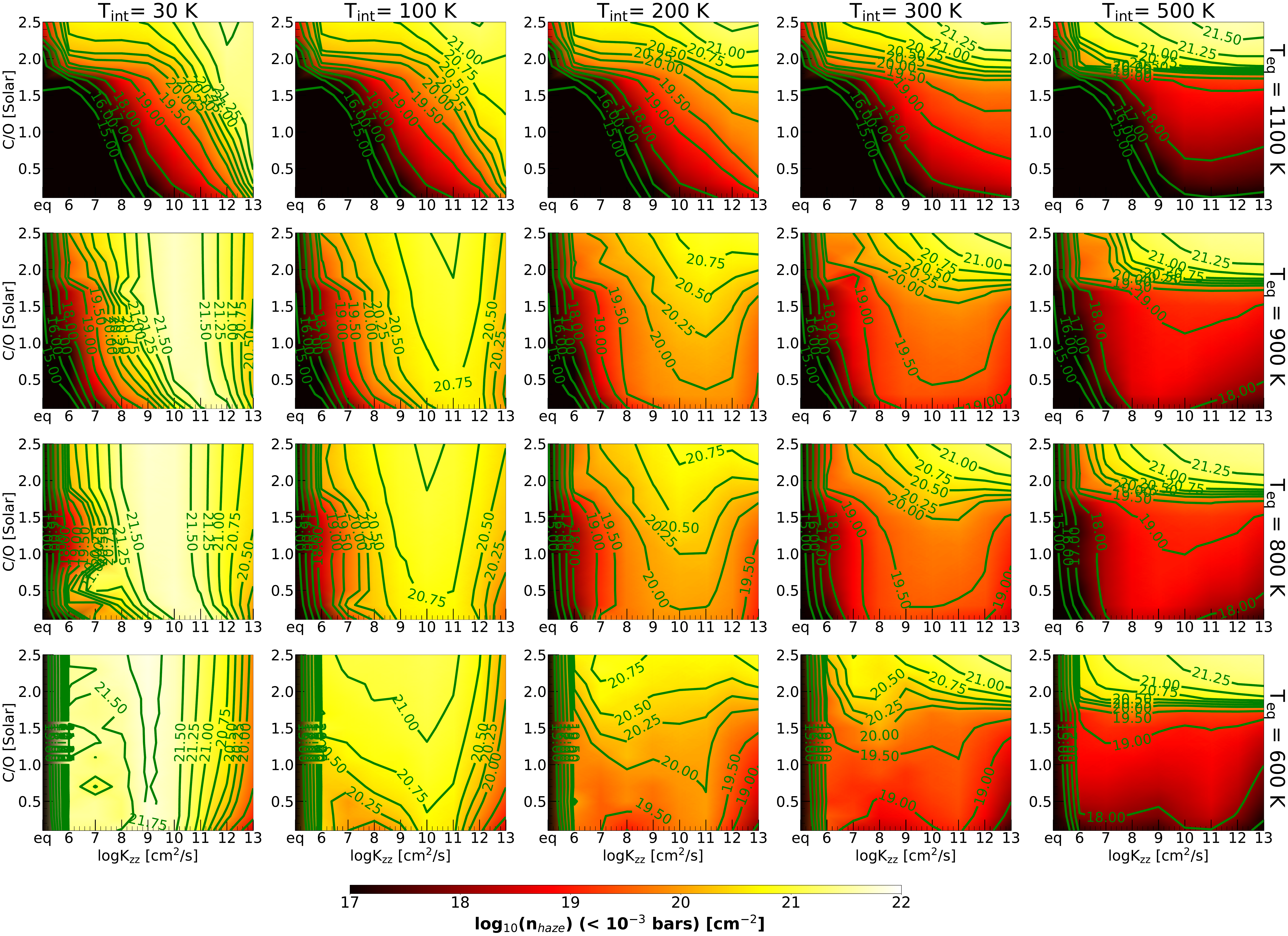}
  \caption{The column density of haze precursors at pressures smaller than 10$^{-3}$ bars has been shown as a heat map as a function of C/O and {\kzz} in each panel. Each row corresponds to a different {\teq} value from 1100 K to 600 K from top to bottom. Each column correspond to a different {\tint} value between 30 K and 500 K from left to right. The abundance of gases-- C$_2$H$_2$, C$_2$H$_4$, C$_2$H$_6$, C$_4$H$_2$, HCN, CH$_3$CN, and CS$_2$ were used to compute the column density of the haze precursor molecules.}
\label{fig:haze_cbyo}
\end{figure*}

\begin{figure*}
  \centering
  \includegraphics[width=0.85\textwidth]{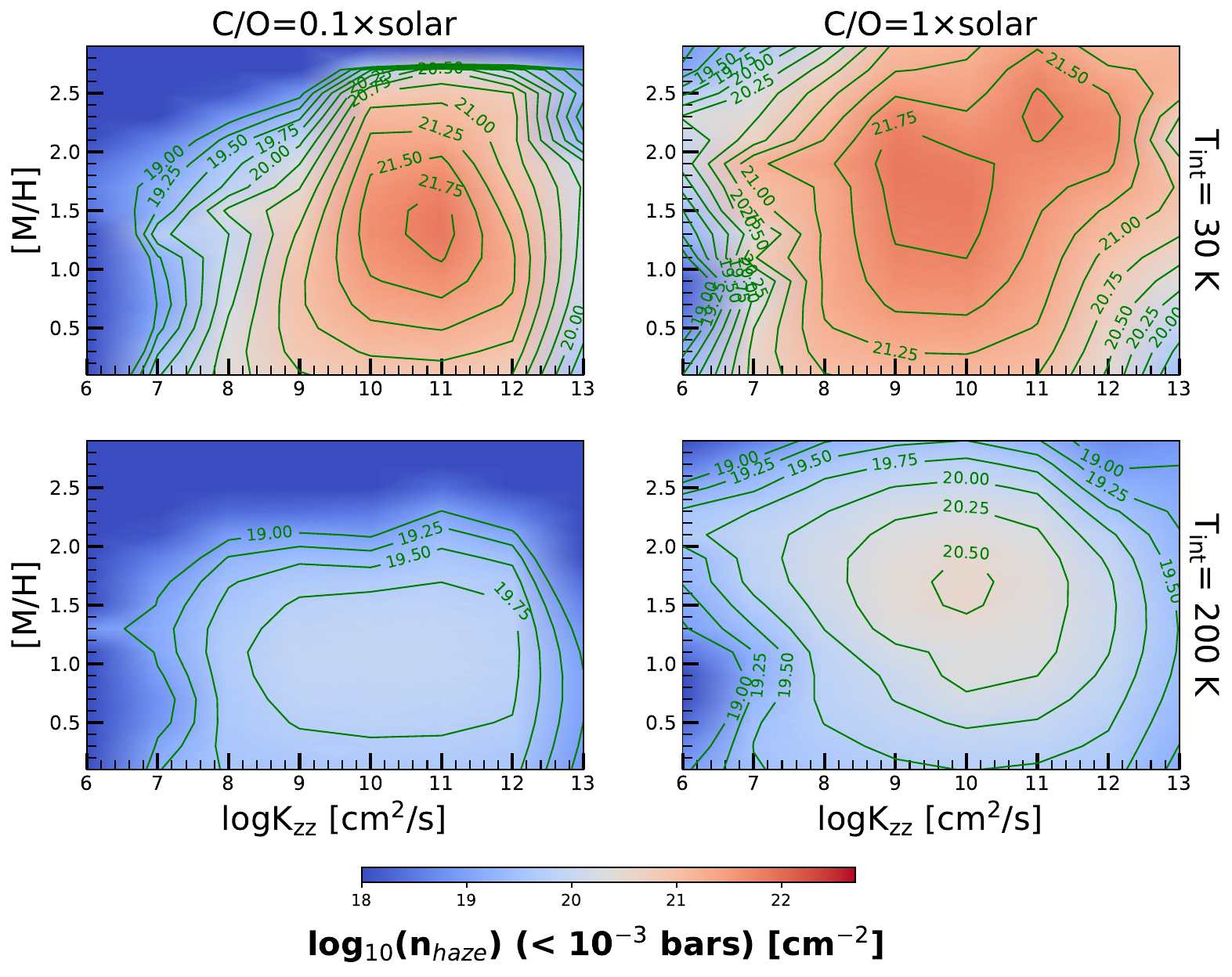}
  \caption{The column density of haze precursors at pressures smaller than 10$^{-3}$ bars has been shown as a heat map as a function of [M/H] and {\kzz} in each panel. The top row corresponds to {\tint}= 30 K, while the bottom row shows models with {\tint}= 200 K. The {\teq} has been set to 800 K for all the panels. The left column shows models for C/O=0.1$\times$solar while the right column shows models at C/O=1$\times$solar. The abundance of gases-- C$_2$H$_2$, C$_2$H$_4$, C$_2$H$_6$, C$_4$H$_2$, HCN, CH$_3$CN, and CS$_2$ were used to compute the column density of the haze precursor molecules.}
\label{fig:haze_met}
\end{figure*}

Photolysis of C- bearing gases (e.g., {\meth}) in the upper atmospheres of gas giants can lead to formation of gases like C$_2$H$_2$, C$_2$H$_4$, C$_2$H$_6$, etc, which in turn can act as precursors to haze/soot creation, if the conditions are favorable to further polymerization of these gases \citep[e.g.,][]{fortney13,morley13,morley15}. This polymerization can be enabled by the reducing nature of the upper atmosphere of these planets. Hazes can have large impact on the observed transmission, reflection, and emission spectra of exoplanets \citep[e.g.,][]{morley13,fortney13,morley15,ohno20,sing16}. While the {\it Photochem} kinetics model doesn't track the growth of these hydrocarbons to large polymers, we can still quantitatively estimate the abundances of these haze precursors across the vast parameter space explored in this work. We adopt the methodology presented in \citet{morley15} to estimate the column density of haze precursors above 10$^{-3}$ bars. Following \citet{morley15} and \citet{tsai21}, we include the following C- bearing molecules to estimate the abundance of haze/soot precursors--  C$_2$H$_2$, C$_2$H$_4$, C$_2$H$_6$, C$_4$H$_2$, HCN, CH$_3$CN, and CS$_2$. We calculate the column density of these gases at pressures smaller than 10$^{-3}$ bars across our model grid.

Each panel of Figure \ref{fig:haze_cbyo} shows the logarithm of the column density of these precursors as a function of C/O and {\kzz}. Similar to Figure \ref{fig:ch4_profiles}, each column represents a value of {\tint} while each row represents a different value of {\teq}. Figure \ref{fig:haze_cbyo} shows that the abundance of haze precursors can be very strongly dependant on {\kzz} in planets with {\tint}$\le$100 K. This sensitivity of the haze precursor column density to {\kzz} diminishes with increasing {\tint} at all {\teq} values. For {\teq}$\le$900 K models with {\tint}=30 K, the abundance of haze precursors increases first as a function of increasing {\kzz} before reaching a peak, beyond which it decreases with increasing {\kzz}, for a fixed C/O. The value of {\kzz} where this peak in haze precursor abundance occurs is a strong function of {\teq}. For {\teq}=1100 K and {\tint}=30 K, this peak happens at or beyond {\kzz}=10$^{13}$cm$^2$/s, whereas at {\teq}=800 K and {\tint}=30 K, this peak occurs at {\kzz}=10$^{10}$cm$^2$/s. At {\teq}=600 K, the peak occurs near {\kzz}=10$^{8}$cm$^2$/s, when {\tint}=30 K. Similar to the findings of \citet{fortney13,morley15}, we also find that the abundance of haze precursors shows an increase with decreasing {\teq} between {\teq}=1100 K and {\teq}= 900 K for a fixed C/O and {\kzz}. Below {\teq}=900 K, for models with {\tint}$\le$100 K, whether the column density of precursors increases or decreases with decreasing {\teq} depends largely on the {\kzz} and C/O. However, when the {\tint}$\ge$200 K, the column density of the precursors do not show much change with decreasing {\teq} below {\teq}= 900 K, for a given C/O and {\kzz} value. 

Above {\tint}$\ge$100 K, C/O  starts to influence the column density of haze precursors significantly, in addition to {\kzz}. Figure \ref{fig:haze_cbyo} shows that the influence of C/O on the precursors becomes stronger with increasing {\tint} for all {\teq}. For planets with {\tint} $\ge$ 300 K, the precursor column density shows a sharp increase when C/O $\ge$ 2$\times$solar. In these very C- rich atmospheres with hot interiors, the abundance of the precursors mostly remain independent of {\kzz}. For a given value of C/O and {\kzz}, Figure \ref{fig:haze_cbyo} also shows that the abundance of the precursors show an overall decline with increasing {\tint} for all {\teq} values. This can be seen in the overall fading of the brightness of the panels in Figure \ref{fig:haze_cbyo} as one goes from left to right. Figure \ref{fig:haze_cbyo} proves that the precursors to larger haze molecules in the upper atmosphere of H$_2$/He rich atmospheres show very complex dependence on {\kzz}, {\tint}, C/O, and {\teq}. For planets with hotter interiors both C/O and {\kzz} influence the availability of haze precursors in the upper atmosphere, but for planets with low {\tint}, {\kzz} almost solely influences the column density of these precursors in the upper atmosphere. This suggests that for low mass planets at old ages, C/O might not be very correlated to the abundance of hazes in their atmospheres. However, these results are only for 10$\times$solar metal enrichment.

 Figure \ref{fig:haze_met} shows the how the column density of the haze precursors vary with metallicity and {\kzz}. The top rows show models with {\tint}=30 K while the bottom row shows models with {\tint}= 200 K. The left and right columns show models at C/O= 0.1$\times$solar and C/O=1$\times$solar, respectively. All the models shown in Figure \ref{fig:haze_met} are for {\teq}=800 K. Figure \ref{fig:haze_met} shows that the amount of haze precursors in the upper atmosphere generally depends both on [M/H] and {\kzz}, to varying degrees, in all the four panels. When {\tint}=30 K and C/O=0.1$\times$solar (top left panel), the amount of haze precursor starts to show some [M/H] dependence when the {\kzz} is between 10$^{9}$ and 10$^{12}${\cms}. At {\kzz} values lower or higher than this range, the abundance of haze precursors shows strong dependence on {\kzz}. For C/O=1$\times$solar and {\tint}=30 K models shown in the top right panel, the abundance of haze precursors show [M/H] dependence when $10^{8}{\le}${\kzz}${\le}10^{10}${\cms}. For {\kzz} outside this range, the haze precursor column density becomes very sensitive to mixing with little to no dependence on [M/H]. When the {\tint} is higher (bottom panels), the sensitivity of the haze precursors to [M/H] and {\kzz} are also different. 

At {\tint}=200 K, the C/O=0.1$\times$ cases show dependence of the precursor column density to [M/H] when  10$^8\le${\kzz}$\le$10$^{12}$ {\cms}. But even when {\kzz} lies in between these values, the variation of the precursor abundances with changing [M/H] is much less rapid in this hotter interior model compared to the top panel. However, outside this range of {\kzz} values, the precursor show strong dependence on {\kzz}. When the atmosphere becomes more C- rich in these hotter interior models (bottom right panel), the haze precursor abundance shows greater metallicity dependence. For these models, the haze precursors become sensitive to {\kzz} for {\kzz}$\ge$10$^{12}$ {\cms} and {\kzz}$\le$10$^{8}$ {\cms}.  

The maximum abundance of haze precursors occurs between +1.0$\le$[M/H]$\le$+1.5 for the top left panel, while the bottom left panel shows maximum precursors between +0.5$\le$[M/H]$\le$+1.5. For the top and bottom right panels, maximum precursor abundances are reached between +1.3$\le$[M/H]$\le$+2.5 and +1.5$\le$[M/H]$\le$+2.0, respectively. The top right panel, with  C/O=1$\times$solar and {\tint}= 30 K,  also shows that the maximum haze precursor abundance extends to higher metallicities (e.g., [M/H]=+2.5), if {\kzz} is high. All the panels show that the haze precursor abundances can be very low if mixing is not vigorous enough. Comparing the top to the bottom panel in both the columns of Figure \ref{fig:haze_met} shows that planets with lower {\tint} show enhanced levels of haze precursors at all [M/H], {\kzz}, and C/O values compared to planets with hotter interiors. Both Figure \ref{fig:haze_cbyo} and Figure \ref{fig:haze_met} therefore suggest that the atmospheres of lower mass planets at older ages might have a higher haze content than higher mass planets. However, this only holds true if their interiors have not been heated through external perturbations.

\section{Discussion}\label{sec:disc}

\subsection{Radiative-convective-Photochemical-Equilibrium}

\begin{figure*}
  \centering
  \includegraphics[width=1\textwidth]{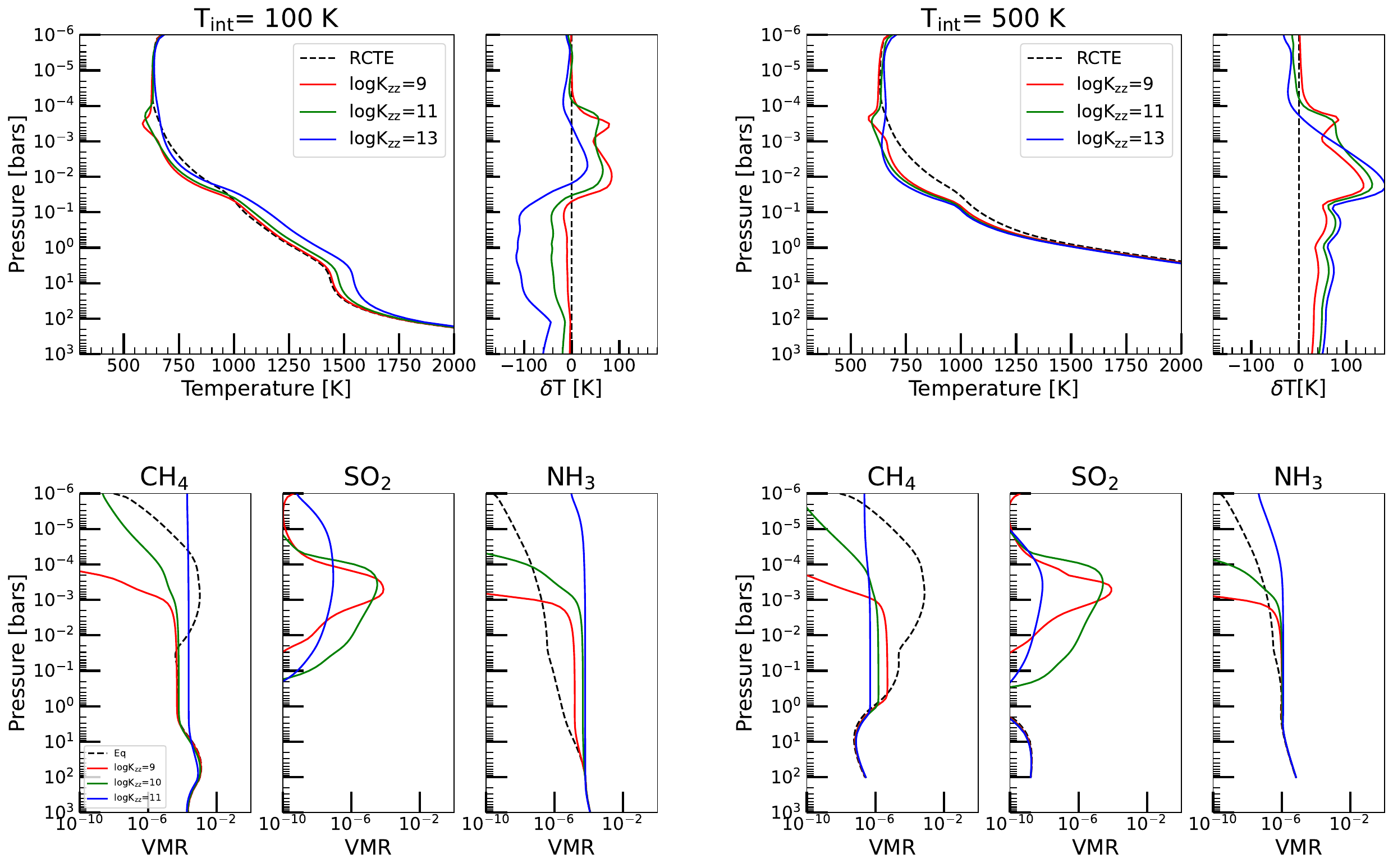}
  \caption{Difference between the {\tp} profiles calculated using the RCPE model for three different {\kzz} values and the RCTE model is shown in the top left panel for a {\tint}=100 K model. The same differences are shown for a {\tint}=500 K model in the top right panel. The three bottom left panels show the abundance profiles for each model for {\meth}, {\sotwo}, and {\amon}  at {\tint}=100 K. The same quantities for the {\tint}= 500 K model are shown in the three bottom right panels. All the models shown here are for {\teq}=800 K, [M/H]=+1.0, and C/O=1$\times$solar.}
\label{fig:rcpe_kzz}
\end{figure*}

Figure \ref{fig:rcpe} has already established that there can be significant differences in the {\tp} structure calculated using RCTE and RCPE models. This has been shown in previous studies as well \citep[e.g.,][]{drummond16,bell23,welbanks24}. This self-consistency between disequilibrium chemistry and {\tp} profiles has been found to be  important for atmospheric models used to interpret the emission spectra of brown dwarfs and directly imaged planets \citep{Mukherjee22,Mukherjee22a,lacy23,Philips20,karilidi21,mukherjee24}.  For the high signal-to-noise spectra of these objects, these atmospheric details matter.

However, we have ignored the self-consistent approach for the bulk of this work as it is not expected to matter much in a parameter space exploration and identification of major trends in chemistry, which is the main focus of this work. The correction of {\tp} profile due to RCPE calculations relative to RCTE models will typically be dependant on {\kzz}. Figure \ref{fig:rcpe_kzz} shows the {\tp} profiles computed with RCPE modeling for three different {\kzz} values for a {\teq}=800 K object with [M/H]=+1.0 and solar C/O. We calculate these RCPE profiles for two {\tint} values of 100 K and 500 K, which are shown in Figure \ref{fig:rcpe_kzz} along with the RCTE {\tp} profiles computed in each case. The chemical abundances in each model for {\meth}, {\amon}, and {\sotwo} are also shown. It is clear that the correction on the {\tp} profiles relative to RCTE models is quite a strong function of both {\kzz} and {\tint}. For example, the absolute $\delta{T}$ in the {\tint}=500 K cases are larger but not very {\kzz} dependant, whereas the absolute $\delta{T}$ is relatively lower but much more {\kzz} dependant in the {\tint}=100 K. Moreover, we also find that the RCPE models have slightly different radiative-convective boundaries from the RCTE models too. This sensitivity can have modest effects on the absolute abundances presented in this work but does not appear to be strong enough to alter the trends presented here. 

We have also made this simplification because here we only explore the effect of the planetary parameters on \emph{transmission spectra} of planets. \citet{drummond16} has shown that the effect of the difference between the RCTE and RCPE calculations is much stronger for \emph{eclipse spectroscopy}. Figure \ref{fig:rcpe_kzz} reiterates this point and suggests that RCPE models should be used when eclipse spectroscopy data of warm and cold transiting planets are interpreted.

\subsection{Dependence of Photochemical Calculations on the UV spectra of host stars}

\begin{figure*}
  \centering
  \includegraphics[width=1\textwidth]{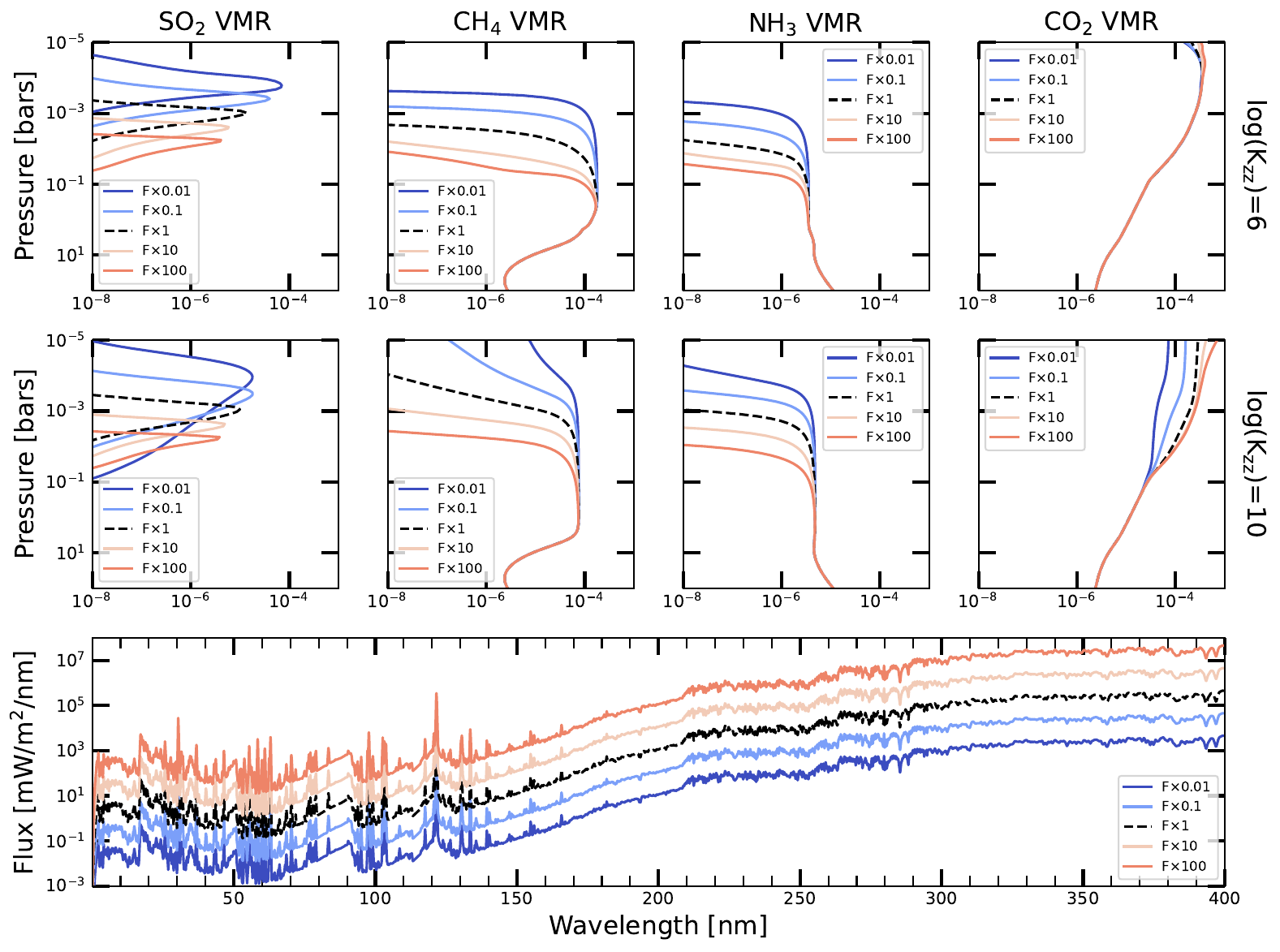}
  \caption{The upper set of panels show the change in abundance profile of key molecules due to change in the X-ray/UV flux incident on the planet when {\kzz}=10$^6${\cms}. Profiles for {\sotwo}, {\meth}, {\amon}, and {\cotwo} are shown from left to right, respectively. The black-dashed line shows the profile for the nominal case. The middle row compares the abundances when {\kzz}=10$^{10}${\cms}. The lower panel shows the X-ray/UV spectrum used to calculate each of these models.}
\label{fig:XUV_flux_whole}
\end{figure*}

\begin{figure*}
  \centering
  \includegraphics[width=1\textwidth]{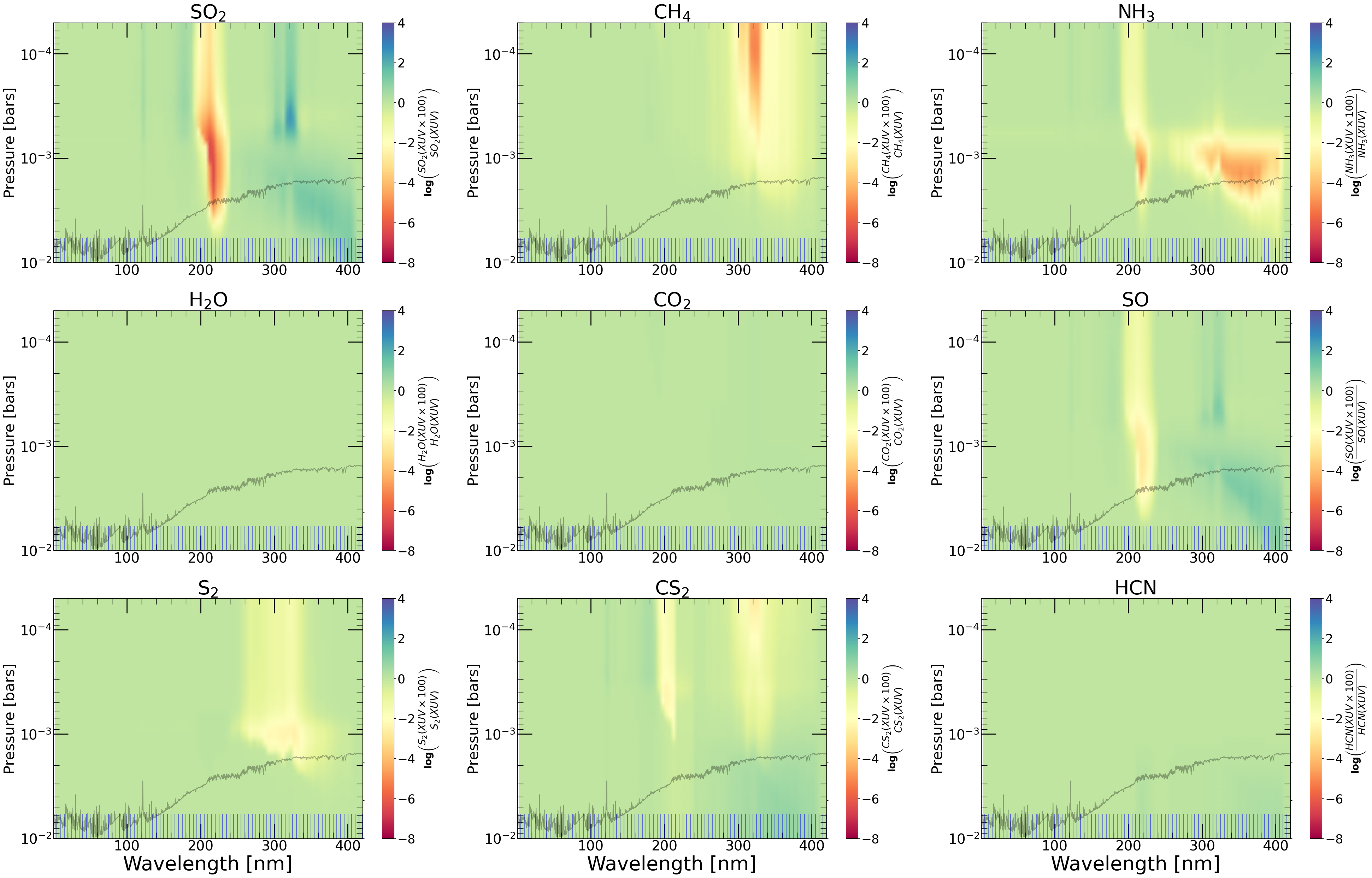}
  \caption{This plot shows how the abundance of a certain gas changes at a given pressure when the UV flux at a given wavelength bin is increased by a factor of $\times$100. This has been quantified with $Y$ in Equation \ref{eq:eq_xuv}. This $Y$ quantity has been plotted as a heat map in each panel for {\sotwo}, {\meth}, {\amon},{\water}, {\cotwo}, SO, S$_2$, CS$_2$, and HCN. The X-axis in each panel is the wavelength of the UV flux and the Y-axis is pressure in the atmosphere. The wavelength bins used for boosting the UV fluxes are also shown as vertical lines at the bottom of each panel along with the nominal UV spectra incident on the planet. A positive value for $Y$ at a given pressure and wavelength means that the molecular abundance has increased relative to the nominal case when UV flux at those wavelengths were boosted. A negative $Y$ value corresponds to a decrease instead. }
\label{fig:XUV_flux}
\end{figure*}

Figure \ref{fig:chemical_profiles} shows that photochemistry shapes a significant portion of the abundance profiles of gases like {\meth} and {\sotwo} in/near the pressures typically probed by transmission spectroscopy. Therefore, another planetary parameter which can be influential on the photospheric abundances of gases in irradiated giant planets is the host star X-ray/UV spectra incident on the planetary atmosphere.

The black-dashed line in the bottom panel of Figure \ref{fig:XUV_flux_whole} shows the nominal UV spectra used for the parameter space exploration presented in \S\ref{sec:results}. The other colored lines show the same UV spectra multiplied with different factors of 100$\times$, 10$\times$, 0.1$\times$, and 0.01$\times$. The top panels in Figure \ref{fig:XUV_flux_whole} show the chemical abundance profiles calculated for the same planet  but with these different incident UV fluxes on the planet. We use the {\teq}= 800 K, {\tint}= 300 K, {\kzz}= 10$^6$cm$^2$s$^{-1}$, and C/O= 1.1$\times$solar case for all results presented in this section. Abundance profiles for {\sotwo}, {\meth}, {\amon}, and {\cotwo} are shown from top left to right, respectively. It is clear that changes in the whole UV flux by five orders of magnitude does not cause the peak {\sotwo}  abundance to change much, but instead has a significant impact on the pressure where it is produced, which can change by two orders of magnitude. For {\meth} and {\amon}, that same change in UV flux can cause the pressure at which they are photochemically depleted to change by a factor of 10. On the other hand, the UV flux has  almost no effect on the {\cotwo} abundance profiles. The middle panels in Figure \ref{fig:XUV_flux_whole} show the effect of UV flux on the abundances when {\kzz}=10$^{10}${\cms}. The effect of the UV flux on the abundances are more pronounced in this case compared to the lower {\kzz} case shown in the top panel. Additionally, {\cotwo} shows much stronger variation in the upper atmosphere with changing UV flux when the  mixing is strong. This exercise shows that having constraints on the overall flux levels of UV photons incident on giant planets is important for interpreting abundances of some gases like {\sotwo} or {\meth} from observations.

We further explore the sensitivity of atmospheric chemistry to different wavelength regions of the UV spectrum. We take the nominal UV spectra from Figure \ref{fig:XUV_flux_whole} and divide it into a large number of wavelength bins. Then, we boost the flux at each wavelength bin by a factor of 100$\times$ and recalculate the chemistry while keeping the rest of the fluxes in other wavelength bins the same. Figure \ref{fig:XUV_flux} shows the results from this exercise. The nominal UV spectra are shown in the three panels in light gray, and the many vertical lines at the bottom of each panel depict the wavelength bins that we use. Each panel shows the quantity,
\begin{equation}\label{eq:eq_xuv}
    Y= log\left(\dfrac{X(F(w)\times100)}{X(F(w))}\right)
\end{equation}
as a heat map, where $X(F(w)\times100)$ is the abundance of a gas $X$ calculated when the UV flux at wavelength $w$ has been boosted by a factor of 100. $X(F(w))$ is the abundance of the same gas calculated using the nominal UV flux. When $Y$ in Equation \ref{eq:eq_xuv} is greater/lower than zero, it means that the boost in UV flux in that wavelength bin causes an increase/decrease in the abundance of molecule $X$ relative to the abundance calculated using the nominal UV flux. 

The top left most panel in Figure \ref{fig:XUV_flux} shows $Y$ for {\sotwo}. It shows that the {\sotwo} abundance near 0.1 to 3 mbar shows a decrease if the UV flux between 200 to 240 nm is boosted by a factor of 100. This sensitivity is due to the high UV cross-section of {\sotwo} between 200-240 nm shown in Figure \ref{fig:xsec1} in Appendix \S\ref{sec:xsec}. It also shows that the {\sotwo} abundance is slightly enhanced between 1-10 mbar if the UV flux near 350-400 nm is boosted. The {\sotwo} also shows a small increase near the 0.1 mbar level if the UV flux between 300-340 nm is increased. Both of these enhancements in {\sotwo} are linked with the additional destruction of S$_2$ and CS$_2$ when the flux between 300-400 nm is increased. This can be seen in the bottom left and bottom middle panels of Figure \ref{fig:XUV_flux}. This enhanced destruction of S$_2$ and CS$_2$ is due to the strong UV cross-sections of S$_2$ and CS$_2$ between 300-400 nm and 300-350 nm, respectively. Interestingly, the $Y$ map for {\water} in the bottom left panel shows the opposite behavior than {\sotwo} when fluxes in the same wavelength ranges are boosted. But this is not visible in the plot as we choose to keep the same color scale across all panels in Figure \ref{fig:XUV_flux} so that the sensitivity of each gas to changing UV flux can be compared with the rest. 

The top middle panel in Figure \ref{fig:XUV_flux} shows $Y$ for {\meth}. {\meth} abundance shows a small decrease near 0.1-3 mbar region when fluxes between 350-400 nm are boosted. {\meth} also shows a sharp decrease near the 0.1-0.6 mbar level when flux between 300-350 nm is increased by a factor of 100. Like {\sotwo}, {\amon} abundance also shows a sharp decrease between  0.1-3 mbar when flux between 200-240 nm is increased. This is also related to the high UV cross-section of {\amon} (see Figure \ref{fig:xsec1}) between 200-240 nm. Between 0.8-2 mbar,  {\amon} shows a depletion when UV fluxes at wavelengths higher than 300 nm are boosted. Comparing Figure \ref{fig:XUV_flux_whole} and Figure \ref{fig:XUV_flux} shows that determining the level of UV flux incident on the whole planet in very broad wavelength bands is much more important than determining the very detailed nature of the UV SED to interpret the chemical abundances in irradiated planet atmospheres and connecting them to planetary bulk properties. However, we note that photochemical kinetics calculations are highly non-linear in nature and our calculations have several simplifications. For example, a more comprehensive radiative--transfer approach like Monte--Carlo radiative transfer and $T-P$ dependant UV cross-sections might provide additional insights.

\subsection{Effect of Condensation on Chemical Kinetics Calculations}

Our photochemical model does not track alkali metals and other condensibles relevant to the deep portions of gas-giant atmospheres. In reality, species like Na$_2$S and MnS, which we do not model, should condense at relatively high pressures and temperatures which may prevent some amount of sulfur from reaching the upper atmosphere \citep[e.g.,][]{morley2012neglected}. Therefore, our omission of these species should cause our simulations to over-predict the abundance of sulfur gases (e.g., SO$_2$) in the upper atmosphere. However, we expect our over-prediction to be small and inconsequential, because sulfur should be far more abundant than the alkali metals. For example, the Sun has $\sim 10 \times$ more sulfur atoms than sodium atoms \citep{lodders09}, meaning Na$_2$S condensation could, at most, only sequester about 5\% of the available sulfur.
In summary, we expect the lack of condensibles in our photochemical model to cause our results to over-estimate the sulfur in the upper atmosphere by an amount that does not effect our overall conclusions which is based on order-of-magnitude trends in sulfur bearing species.

\subsection{Steady-State Convergence of Chemical Kinetics Model}\label{sec:convergence}

To achieve a photochemical steady-state, the {\it Photochem} code integrates the photochemical equations forward in time until the atmosphere ceases to change. Our code identifies convergence to such a steady-state with the following criteria:
\begin{equation}
  \Delta y_\text{max} = \max \left|\frac{f_{ij}(t_n) - f_{ij}(t_{n}/2)}{f_{ij}(t_n)} \right|
\end{equation}
\begin{equation}
  \frac{\Delta y_\text{max}}{\Delta t} = \frac{\Delta y_\text{max}}{t_n/2}
\end{equation}
Here, $f_{ij}$ is the mixing ratio of species $i$ in layer $j$, and $t_n$ is the current integration time, making $\Delta y_\text{max}$ the maximum relative change over the last $t_n/2$ seconds. We assume a steady-state is achieved when $\Delta y_\text{max} < 0.05$ and $\frac{\Delta y_\text{max}}{\Delta t} < 10^{-4}$. From testing, these criteria appear appropriate over a wide parameter space. Furthermore, the VULCAN photochemical code \citep{tsai17} uses similar convergence requirements that have proved reliable for warm gas-rich planets (e.g., \citet{tsai23}).

{\it Photochem}, like many other chemical kinetics models (e.g., VULCAN \citep{tsai21}), computes the atmospheric chemistry in a grid of atmospheric altitudes instead of the pressure space used in the \texttt{PICASO} climate model. As a result, as the chemistry of the atmosphere evolves in the chemical kinetics model, the mean molecular weight of the atmosphere also changes. So, when the altitudes are converted back to pressures after the kinetics models have converged, the {\tp} profile effectively becomes ever so slightly different than the input {\tp} profile. To avoid this, the {\it Photochem} code adaptively re-grids and adjusts the temperature profile during integration as to ensure the {\tp} profile matches the input \texttt{PICASO} {\tp} profile. This approach is particularly very important for the RCPE models we have presented in this work, where the chemical kinetics model is iteratively coupled directly with the climate model.

\section{Conclusions}\label{sec:conc}
We have explored the variation of atmospheric chemistry in close-in giant transiting exoplanets across a vast parameter space of bulk planetary properties like {\teq}, {\tint}, [M/H], and C/O. Our exploration includes chemical disequilibrium processes such as vertical mixing and photochemistry and is applicable for a wide variety of planets in terms of their mass, age, and proximity to their host stars. We have arrived at the following conclusions from this exploration,

\begin{enumerate}
    \item We have shown that the the photospheric abundances of gases like {\meth} can be sensitive to just the {\kzz} parameter or just the C/O ratio or can also be quite degenerate between the two depending on the bulk properties of the planet like {\tint} or {\teq} and also atmospheric properties like the vigor of vertical mixing. We have identified these trends in this work at different parts of the parameter space.

    \item We have shown that {\co} is a good tracer of the C/O ratio and remains largely independent of {\kzz}, unless the {\tint} of the planet is $\le$ 100 K. For these cold interior planets, {\co} can become much better tracers of {\kzz} than C/O.

    \item {\cotwo} on the other hand shows strong sensitivity to {\kzz} if the {\teq}$\le$900 K. This sensitivity is particularly enhanced for cold interiors. We also show that {\amon} is an excellent tracer of {\kzz}, without much dependence on C/O, as expected.

    \item Photochemically produced {\sotwo} on the other hand depends quite strongly on the C/O of the planet unless the {\kzz} exceeds high values like $\sim$ 10$^{11}${\cms}. We also show that {\sotwo} abundance shows a very sharp decline once the {\teq}$\le$600 K. For these colder {\teq} planets, we find that the S- content of the photosphere is carried by other gases like CS, CS$_2$, and S$_8$. We also find that atomic S is a major carrier of S- near the photosphere irrespective of {\teq}.

    \item We also explore how the abundances of some key gases in the atmosphere behave in the [M/H] vs. {\kzz} space. Among other trends, we find that other S- carriers like OCS and H$_2$S can show a rapid increase with {\kzz} in the photosphere, especially for metal enriched objects.

    \item In order to explore the major chemical transitions in the atmosphere with {\teq}, we  explore the {\teq} vs. abundance trends of gas combinations like {\co}-{\meth}, {\amon}-{\ntwo}, and {\sotwo}-H$_2$S-S. Our findings on these key transitions reflect the findings from previous work  \citep{fortney20,ohno23,ohno232,baxter21,hobbs21,polman23}. Interestingly, we find that even in the photosphere, atomic S can carry a significant fraction of the S- inventory along with H$_2$S and {\sotwo}.

    \item In order to explore how sensitive is the transmission spectra to parameters like {\kzz} and {\tint} in different parts of the parameter space, we present a wavelength resolved sensitivity analysis of the transmission spectra to these parameters. We find that the sensitivity of the transmission spectra to these parameters is quite complex. In some parts of the parameter space, the {\meth} absorption bands might show great sensitivity to {\kzz}, while in some parts the {\cotwo} or {\sotwo} bands might show the highest sensitivity to changing {\kzz} or {\tint}.

    \item We calculate the column number density of molecules like C$_2$H$_{x}$, HCN, etc, which can act as precursors to photochemical hazes in the upper atmosphere. We find that these precursors molecules show very strong dependence on {\kzz} and almost no dependence on C/O for planets with cold interiors ({\tint}$\le$100 K) at 10$\times$solar metallicity. We also find that planets with low {\tint} are likely to have significantly higher abundance of haze precursors in the upper atmosphere compared to planets with hotter interiors. This is especially relevant to older low mass planets with no external heating.
\end{enumerate}

There are several aspects of our modeling of giant planet atmospheric chemistry which can be improved in the near future. These include inclusion of more elements in our chemical kinetics calculations and inclusion of condensation and rainout chemistry within our chemistry calculation as well. We think that the assumptions made in our modeling approach are sufficient for the typical quality of transmission spectra obtained with {\it JWST}, especially in the near-future. But these kind of improvements might lead to new avenues of characterizing the physics and chemistry of giant planet atmospheres and also their bulk properties.

\section{Acknowledgement}
SM acknowledges the Templeton TEX cross-training fellowship for supporting this work. JJF acknowledges the support of JWST Theory Grant JWST-AR-02232-001-A.
K.O. acknowledges support from the JSPS KAKENHI Grant Number JP23K19072. 
N.W. was supported by the NASA Postdoctoral Program. 
We acknowledge use of the \emph{lux} supercomputer at UC Santa Cruz, funded by NSF MRI grant AST 1828315.

{\it Software:} \texttt{PICASO 3.0} \citep{Mukherjee22}, \texttt{PICASO} \citep{batalha19}, pandas \citep{mckinney2010data}, NumPy \citep{walt2011numpy}, IPython \citep{perez2007ipython}, Jupyter \citep{kluyver2016jupyter}, matplotlib \citep{Hunter:2007}, the model grid will be formally released via Zenodo.

\section{Appendix}\label{sec:xsec}
\begin{figure*}[h!]
  \centering
  \includegraphics[width=1\textwidth]{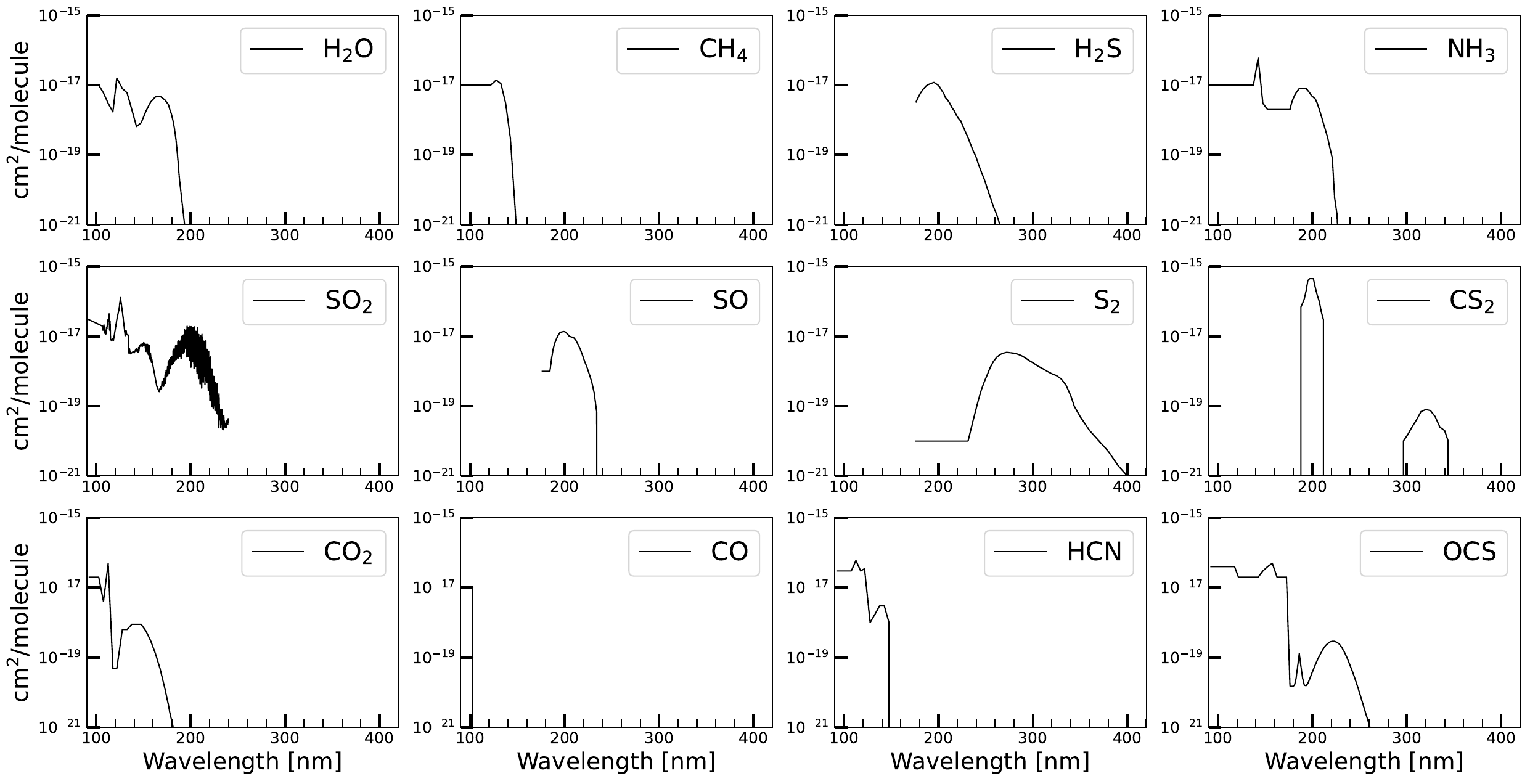}
  \caption{The UV cross section as a function of wavelength for different gases are shown in each panel. These cross-sections were used in our chemical kinetics modeling.}
\label{fig:xsec1}
\end{figure*}

Here we present the UV cross-sections of key gases used in the {\it Photochem} chemical kinetics model. Figure \ref{fig:xsec1} shows the UV cross-section used in this work for 12 major molecules. Each panel shows the cross-section for a different molecule.

\bibliography{sample631}{}
\bibliographystyle{aasjournal}

\end{document}